\documentclass[fleqn]{article}
\usepackage{pstricks}    
\usepackage{fancybox}    
\usepackage{amsmath}
\usepackage{amssymb}     %
\usepackage{pifont}      
\usepackage{booktabs}    
\usepackage{multirow}    
\usepackage{tabularx}
\usepackage{longtable}

\usepackage[english]{babel}     
\usepackage{graphicx}
\usepackage[figuresright]{rotating}
\usepackage{axodraw}
\usepackage[T1]{fontenc}

\newrgbcolor{darkgreen}{0.3 0.6 0}
\newrgbcolor{darkred}{0.8 0 0}
\newrgbcolor{darkblue}{0 0 0.7}
\newrgbcolor{darkyellow}{0.6 0.6 0}
\newrgbcolor{darkcyan}{0 0.6 0.6}
\newrgbcolor{darkmagenta}{0.7 0 0.7}

\begin{document} 
{\centering
{\Large {\bf
Preliminary Investigation of a Waveform Analysis with the WASA 
and the ACQIRIS Readout Electronics.

\vspace{1.0cm}  
{\large R.\ Fabbri$^{a, b}$, R.\ Engels$^a$ }}\\

\vspace{0.3cm}      
{\large 
$^a$ Forschungszentrum J\"{u}lich (FZJ), J\"{u}lich \\
$^b$ Corresponding Author: r.fabbri@fz-juelich.de} \\
\vspace{1.5cm}
      \hspace{3.7cm} 
\centering{\today}
}
}

\vspace{1.5cm}
\begin{abstract}
  The Group for the development of neutron and gamma detectors in the 
  Central Institute of Engineering, Electronics and Analytics 
  \mbox{(ZEA-2)} at Forschungszentrum J\"{u}lich (FZJ) is developing
  a fast Anger Camera prototype for improving the rejection of the 
  gamma contamination during the detection of neutrons.
  The prototype is based on a scintillating plate for neutron capture 
  and on the subsequent generation of scintillating light collected 
  by a matrix of 4x4 vacuum Photomultipliers R268 by Hamamatsu.
  According to the impinging point position of the incoming neutrons 
  the light is collected by different PMTs, and via dedicated algorithms  
  the $x$ and $y$ coordinates can be calculated.
  In this note the WASA and ACQIRIS readout electronics are compared while 
  performing a waveform analysis of the signals generated by using 
  both an analogue pulse generator and an LED+PMT system. Different 
  options of pre-amplifiers and amplifiers are considered, and the 
  results are here presented and commented. 
  At this stage of the prototype development, 
  systematical studies were not performed while the scope of this 
  work was only to validate the principle of operations by using
  both readout systems.  
\end{abstract}

{\large \vspace{-18cm} \hspace{-4cm} 
  Forschungszentrum J\"{u}lich\\

  \hspace{-4.cm}
  Internal Report No. FZJ\_2013\_02893
}

\newpage  

  \tableofcontents
  \vspace{-20cm} \hspace{6cm} 

\newpage  
\section{Introduction}
%
An Anger prototype made of a 4x4 matrix of vacuum 
photomultipliers has been built at the ZEA-2 Institute 
in the FZJ.
Two different options for the readout electronics were analyzed,
the ACQIRIS DC282~\cite{ACQIRIS} and the WASA~\cite{WASA} QCD 
electronics. 

In this paper the proof of principle to perform a waveform 
analysis with both systems is presented, without, at the moment, 
further investigating 
the systematical uncertainty involved. In combination with
an offline analysis~\cite{WAVEFORM_ANALYSIS} to disentangle the 
contamination of gammas from neutrons, the waveform 
analysis should help to perform experiments with a larger
purity of signals induced by neutrons.

The proposed prototype is expected to undergo a full series of measurements 
at the research reactor FRM-II~\cite{FRM-II} to show 
the capability to cope with the high demanding performance.
%

\section{\mbox{WASA Electronics: Description and Properties}}
%
\label{sec:WASA}
For the WASA experiment a specific QDC card was developed . 
The design of the electronics comprises a fast signal and data 
processing by usage of modern technologies like field-programmable 
gate arrays (FPGAs). 

\begin{figure*}[b!]
  \includegraphics[height=6cm,width=12.cm]{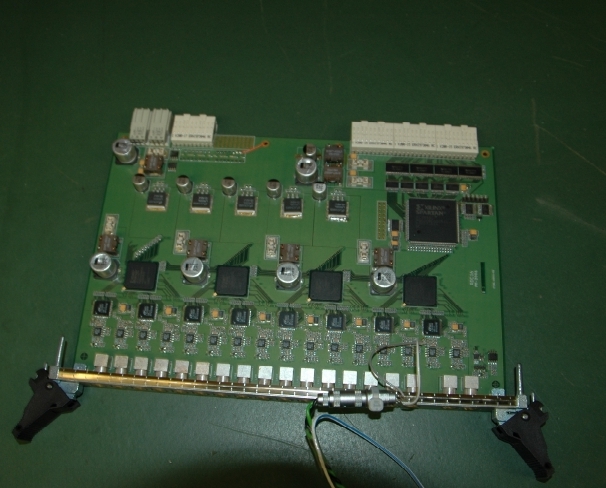} \\
  \vspace{-0.4cm}
  \caption{QDC module of the WASA electronics with the sixteen 
           $50$ Ohm Lemo connectors for the input signals. 
           Visible at the front-end side (on the right) is the ad-hoc 
           made connection to the ADC to test the processing of 
           differential signals.}
  \label{fig:WASA_schematics}
\end{figure*}

A detailed description of the electronics can be found in~\cite{WASA}. 
Here we address only its main features: the front-end, 
Fig.~\ref{fig:WASA_schematics}, is equipped with
LEMO $50$ Ohm connectors to handle up to $16$ single-ended signals.
Each FPGA handles four input channels and stores in a buffer
the values provided by the individual ADCs, which can sample 
their corresponding signal at a rate of either $160$ MHz 
(every $6.25$ ns) or $80$ MHz. The data acquisition can 
be configured to store either the entire waveform in a selectable 
gate window (up to $6.4$ $\mu$s) or its peaking amplitude. 
A pre-(after-)window for the DAQ can be set to up $\pm6.4$ $\mu$s.

In this work the WASA electronics has been used in the so-called 
fast mode ($160$ MHz sampling rate). In this mode the signal 
is initially processed by the internal differential amplifier 
AD8132, and then sampled by the $12$ bit ADC MAX1213
using only $11$ bits. 
Each FPGA handles four channels, while in the 
the main FPGA register several cuts can be set to select events 
in specific amplitude and time windows. 
The setting of the FPGA is performed via API functions provided
by the driver developers.

\section{\mbox{Analysis with an analogue Voltage Pulse Generator}}
\subsection{Test-Bench Description}
\begin{figure*}[b!]
   \hspace{-0.6cm}
      \includegraphics[height=6.0cm,width=13.cm]{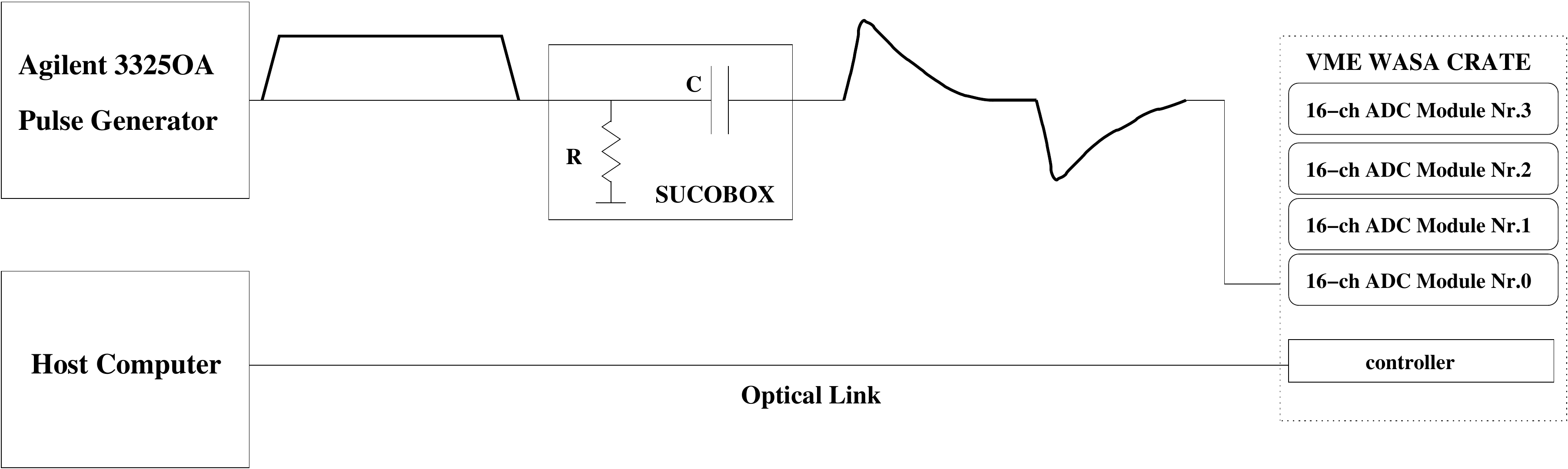}
  \caption{Test-bench setup used for the waveform analysis with the 
           WASA readout system at FZJ. A similar experimental setup 
           is used when using the ACQIRIS system: here network twisted cables
           are used for transmitting the data to the host computer.} 
  \label{fig:TestBench_Pulser}
\end{figure*}

%
The characterization measurements of the readout electronics were 
performed at the laboratories in ZEA-2.
The typical setup of the test-bench is shown in 
Fig.~\ref{fig:TestBench_Pulser}. 
No generator of tailed pulses with fast rise time 
(at the order of a few nanoseconds)
was available; thus the analogue pulse generator Agilent 3325OA
was used to generate voltage pulse with rise and fall time values 
set at $5$ ns (the minimum allowed configuration). 

To generate a current pulse an $RC$ high-pass filter was used,  
maintaining a fast rise time as expected in a real experiment
(large rise time values were observed to give rise to undershoots of 
the signal tail).
The chosen filter, built up in a SUCOBOX, has the following parameters
\begin{eqnarray}
    R & = & 50 \ \Omega  \nonumber \ , \\
    C & = & 10 \ nF \ ,
    \label{eq:RC}
\end{eqnarray}
and generates two pulses with opposite polarity (corresponding 
to the rising or falling edge of the voltage pulse) with 
characteristic decay time of $\tau_{c} = 500 \ ns$. 
The distance between the two pulses is given by the width of the incoming 
voltage pulse.
The readout systems were properly configured to sample only one of the
two peaks generated by the RC filter.

Using the same experimental configuration the signal was 
injected through a Lemo cable with $50\ \Omega$ impedance alternatively 
in one of the $16$ input channels of the investigated WASA board, and 
in one of the four DC282 ACQIRIS channels (set to $50$ Ohm input impedence), 
and the sampled waveforms 
were shipped to an external computer and there recorded for a later 
offline analysis. The approximately $6$ m long $50$ 
$\Omega$ Lemo cable used resulted 
in a signal (as measured at the oscilloscope) attenuation of a factor 
$0.58$ using the amplitude of the output voltage set to $20$ mV. 
This value is then used during the entire scan to rescale all the 
remaining amplitude values used.
Note that this measurement at the scope could be quite un-precise, 
and thus might be affected by a strong systematical bias.
%

\subsection{Measurements with the WASA System}
\begin{figure*}[t!]
\hspace{-1cm} \vspace{-0.5cm}
      \includegraphics[height=5cm,width=7.cm]{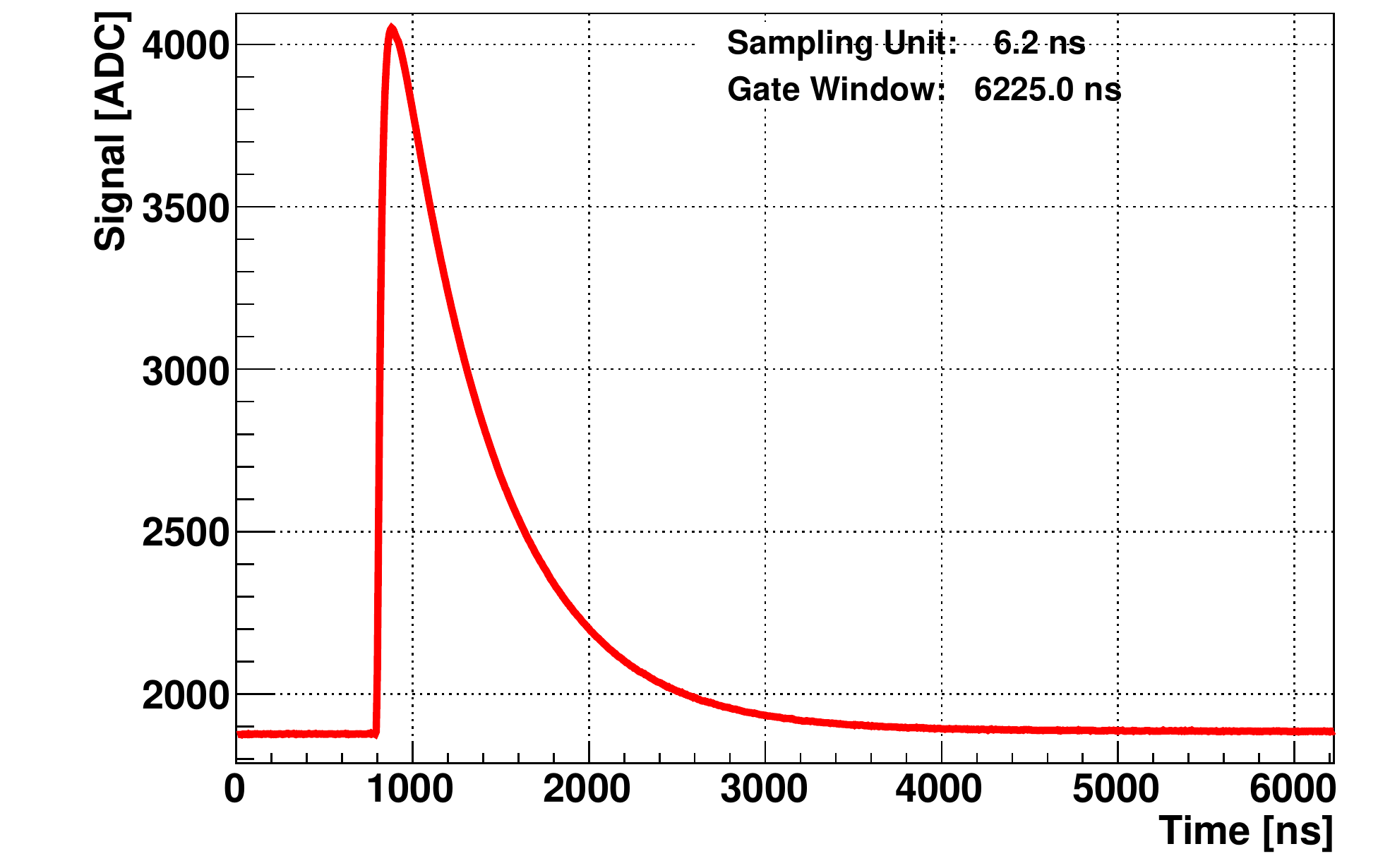}
      \includegraphics[height=5cm,width=7.cm]{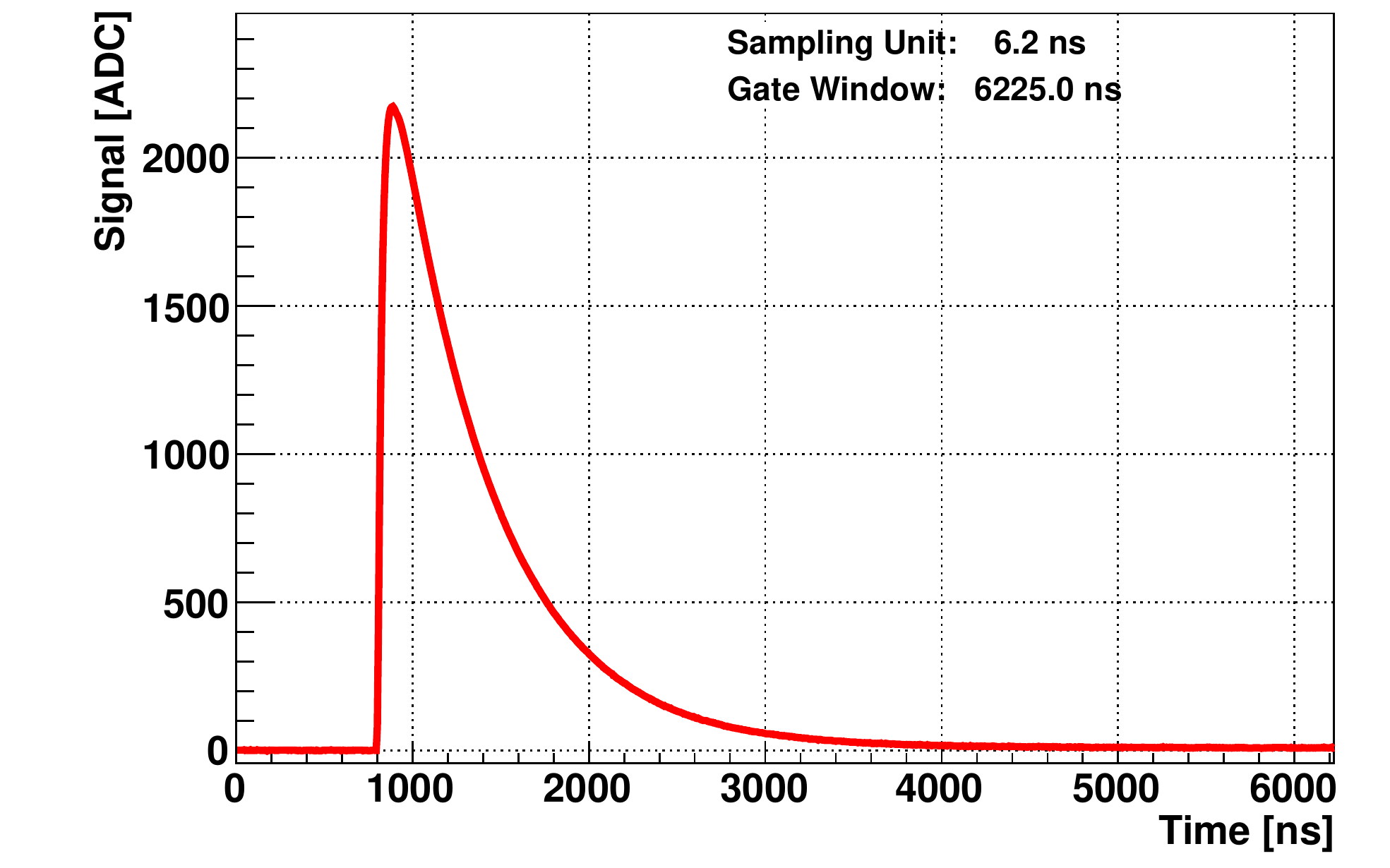}\\

\hspace{-1cm} \vspace{-0.5cm}
      \includegraphics[height=5cm,width=7.cm]{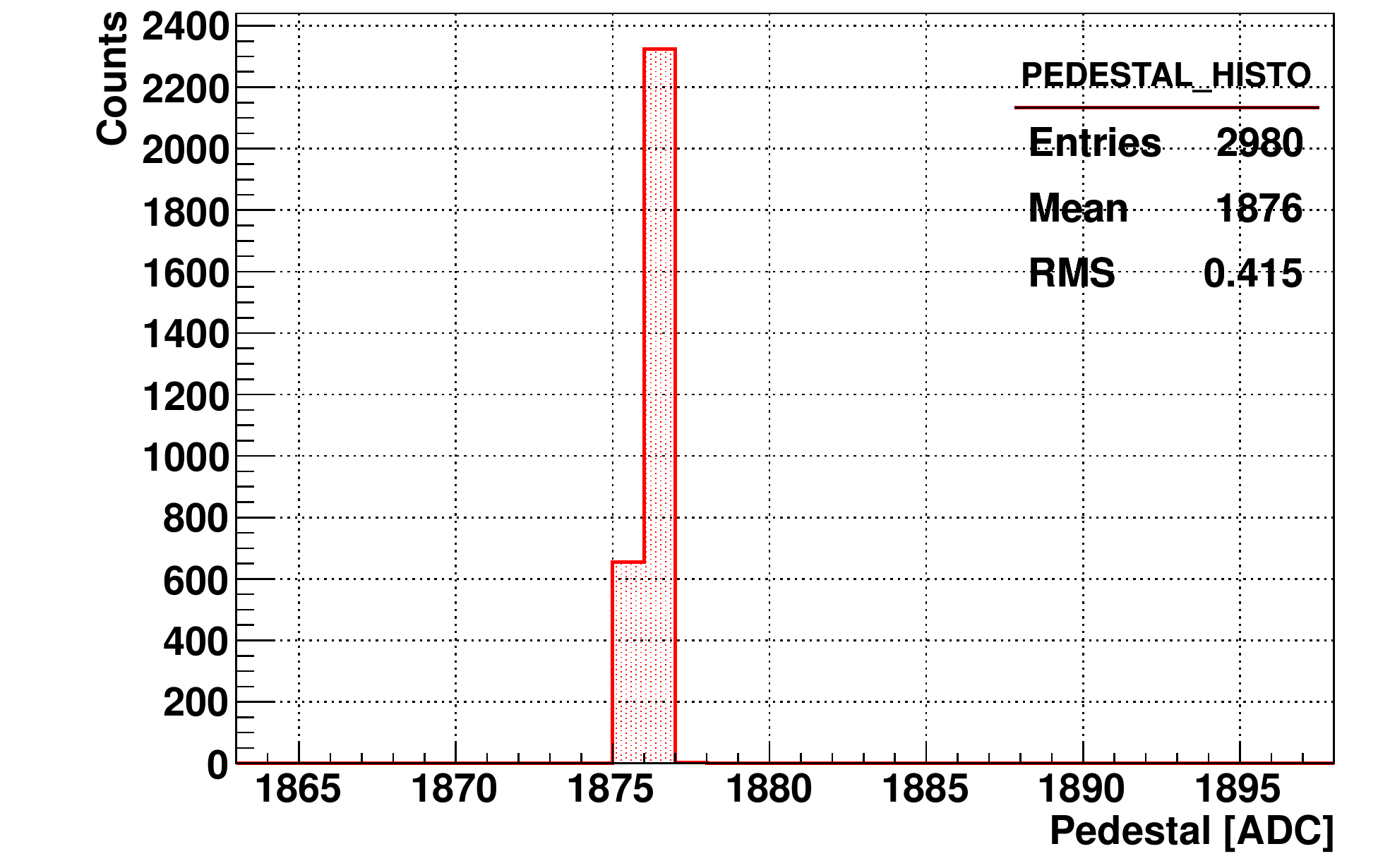}
      \includegraphics[height=5cm,width=7.cm]{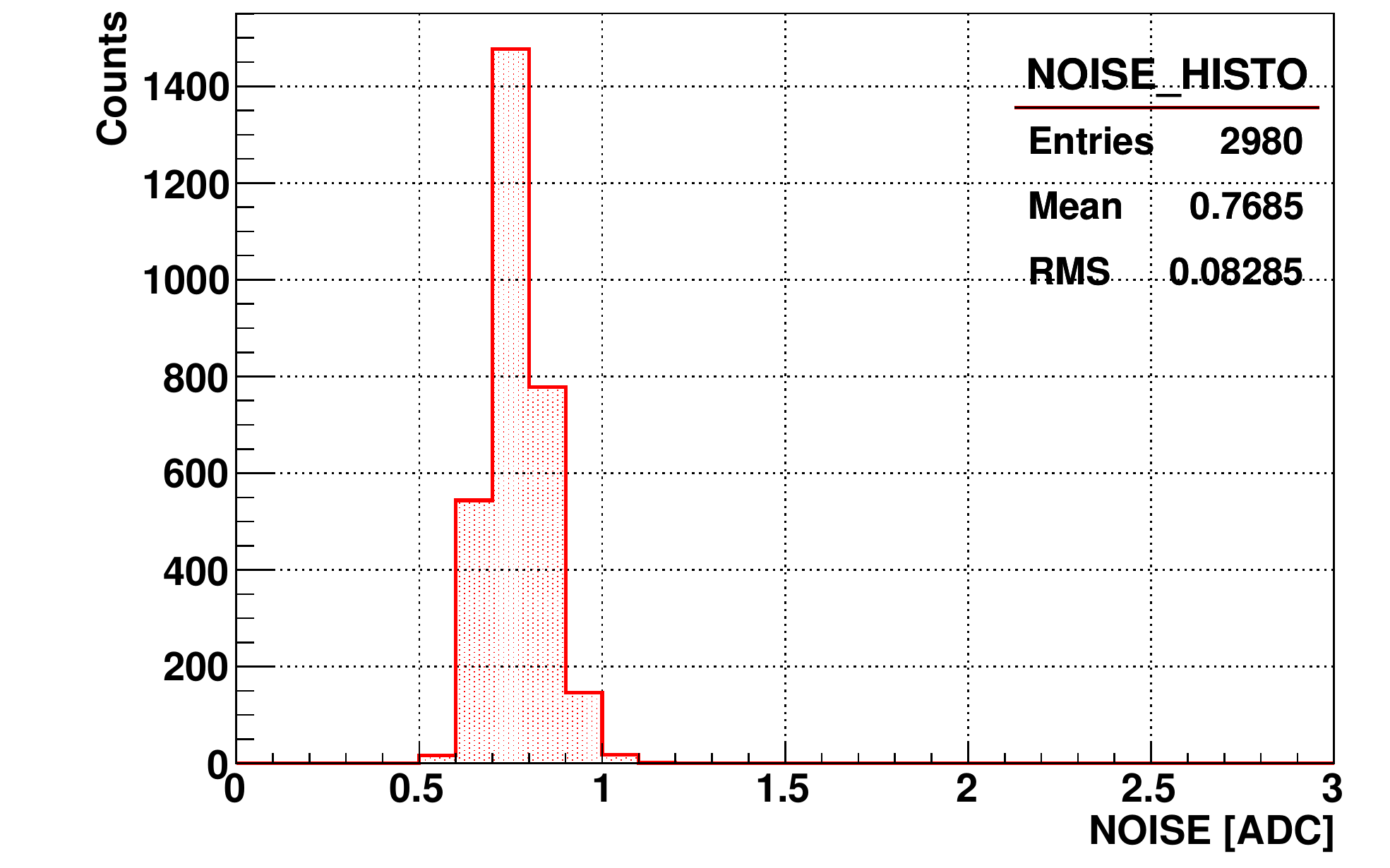}\\

\hspace{-1cm} 
      \includegraphics[height=5cm,width=7.cm]{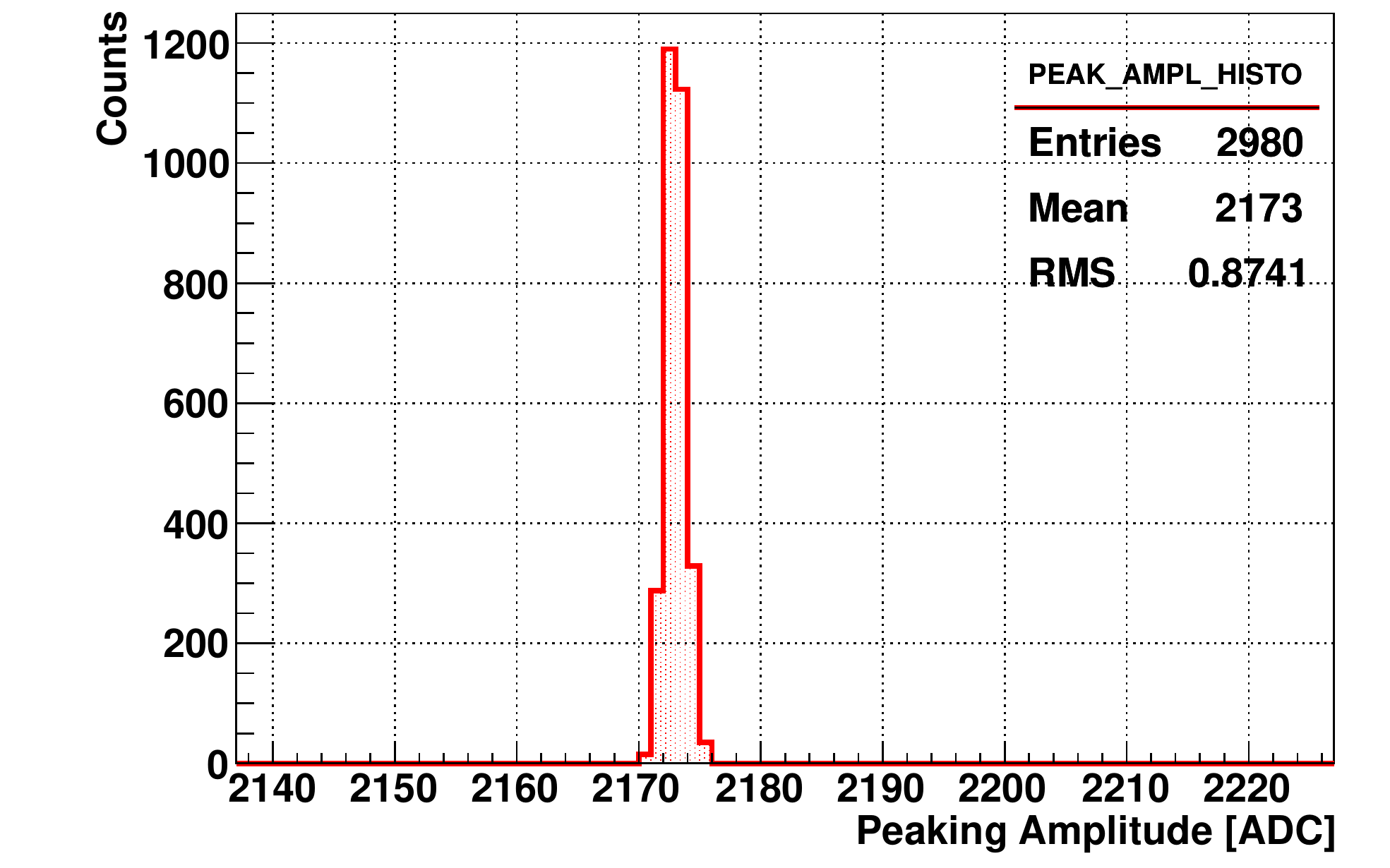}
      \includegraphics[height=5cm,width=7.cm]{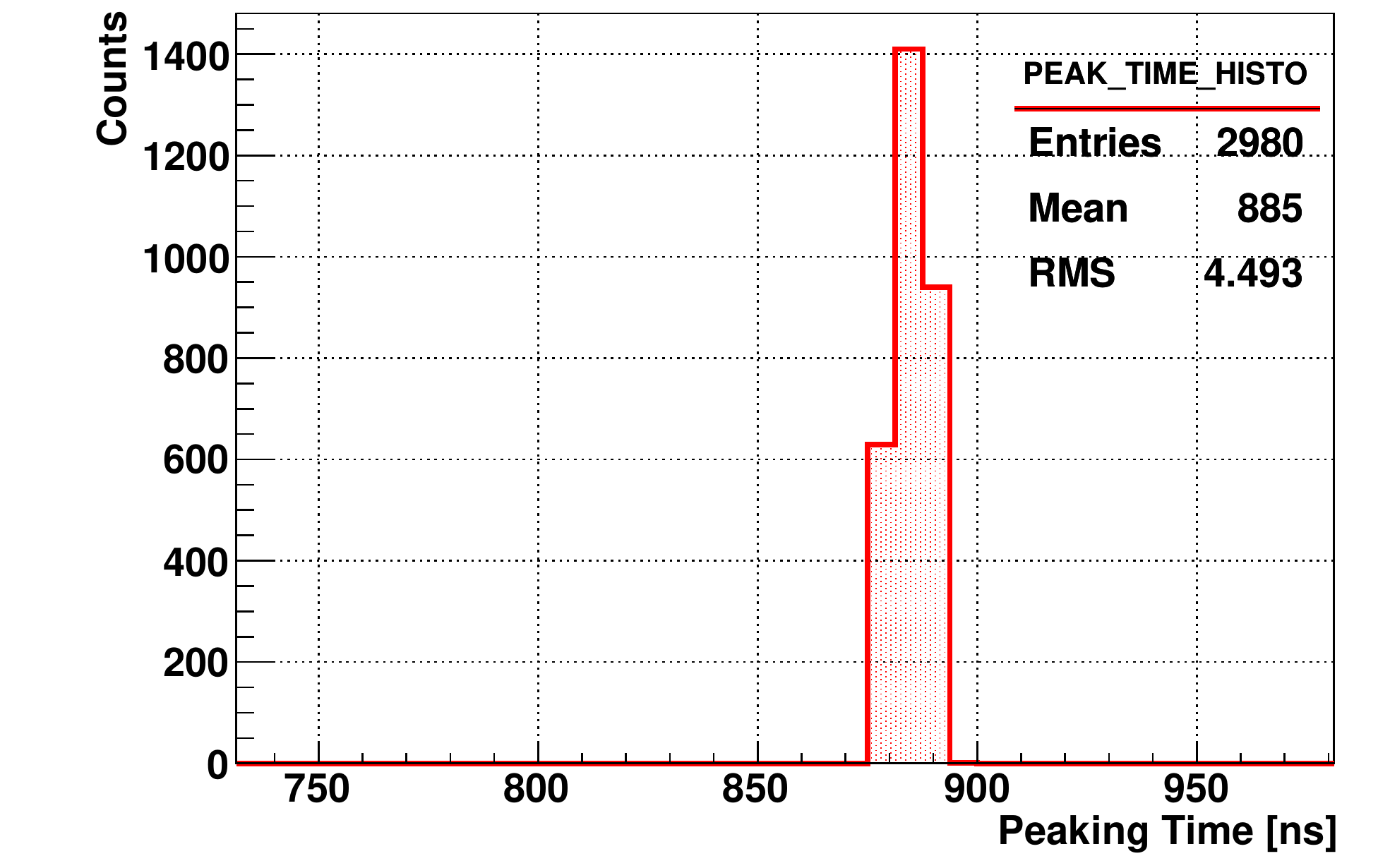}\\

\vspace{-0.5cm}
  \caption{Typical histograms obtained during the waveform 
           analysis with the WASA electronics. These examples were obtained 
           with $1900$ mV voltage amplitude provided by the pulse generator.}
  \label{fig:WASA_PULSER_EXAMPLE}
\end{figure*}

%
To measure the ADC resolution of the WASA system a linear 
scan in the input voltage was performed. The results of a typical 
measurement is presented in Fig.~\ref{fig:WASA_PULSER_EXAMPLE}
(here shown for the input voltage amplitude of 
$1900$ mV).

In the upper panels it is shown the effect of the pedestal calculation and 
its subtraction to the data in the waveform analysis. The acquisition 
system (and with ACQIRIS as well) allows to record also a certain 
amount of data accumulated before the trigger for the acquisition is 
generated. In this example, for each waveform event from the initial 
$100$ points (corresponding to \mbox{$100\cdot 6.25 = 625$ ns}) the mean value 
and RMS are calculated and saved in the corresponding
pedestal and noise histograms, shown in the middle panels. 

For each event the calculated pedestal is subtracted to the waveform 
sampling data bringing, as expected, the baseline to zero, as 
visible in the right upper panel. Note the absence of undershoot
in the signal tail region within this experimental configuration.  

The test-bench plus readout system has a low noise, well below
one ADC unit.

In principle, the deposited charge can be calculated out of the 
waveform integral (after pedestal subtraction). Unfortunately at 
this stage the ADC resolution is not known yet, and should be 
extracted out of the data. The measurement to calculate the ADC 
resolution (mV/ADC) is presented in the next subsection.
%

\subsection{Linearity of the WASA system}
%
A linear scan is performed, injecting into the WASA  system 
waveforms with different values of the input voltage pulse amplitude, 
as shown in the Tab.~\ref{tab:WAS_SCAN_PULSER}.
\begin{table}[t!]
  \begin{center}
\begin{tabular}{||c|c|c||}
\hline 
\hline
\multicolumn{2}{|c|}{ \textbf{WASA System: Linear Scan} }\\
\hline
\hline
   \bf{Output Pulser [mV]:} & \bf{Measured Peaking Amplitude [ADC]:} \\
\hline 
\hline 
     20 &     23.4 \\
     40 &     46.1 \\
     80 &     91.6 \\
    120 &    136.9 \\
    140 &    159.6 \\
    180 &    204.8 \\
    220 &    250.5 \\
    300 &    341.5 \\
    340 &    386.7 \\
    420 &    478.0 \\
    500 &    569.2 \\
    580 &    660.0 \\
    660 &    751.4 \\
    740 &    842.6 \\
    820 &    934.1 \\ 
    900 &   1025.0 \\
    980 &   1117.0 \\
   1060 &   1212.0 \\
   1140 &   1304.0 \\
   1220 &   1395.0 \\
   1300 &   1487.0 \\
   1380 &   1578.0 \\
   1460 &   1670.0 \\
   1540 &   1761.0 \\
   1620 &   1853.0 \\
   1700 &   1944.0 \\
   1780 &   2036.0 \\
   1820 &   2082.0 \\
   1860 &   2127.0 \\
   1900 &   2173.0 \\
\hline
\hline
\end{tabular}
\caption{Linear scan performed with the WASA readout system. 
         The total error in the measured peaking amplitude is 
         conservatively considered as being unity.
         For a correct comparison between the two sets of data 
         the correction factor of $0.58$ should be applied to 
         the output pulser values.}
\label{tab:WAS_SCAN_PULSER}
  \end{center}
\end{table}

\begin{figure}[t!]
  \hspace{-2cm}
  \includegraphics[height=6.cm,width=8cm]
                   {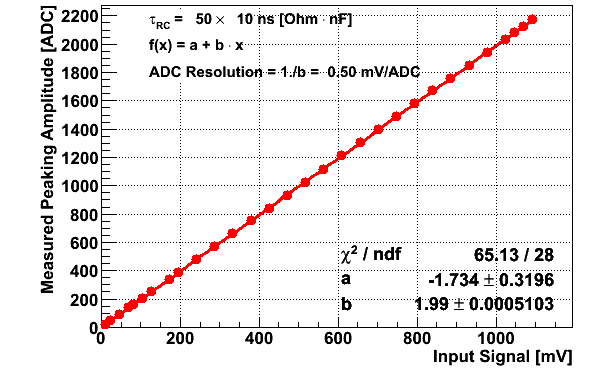}
  \includegraphics[height=6.cm,width=8cm]
                   {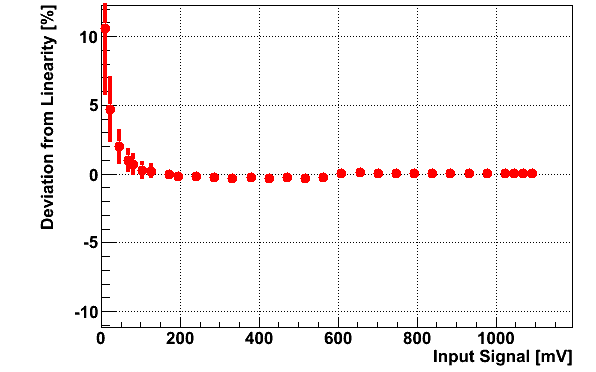}\\

  \hspace{-2cm}
  \includegraphics[height=6.cm,width=8cm]
                   {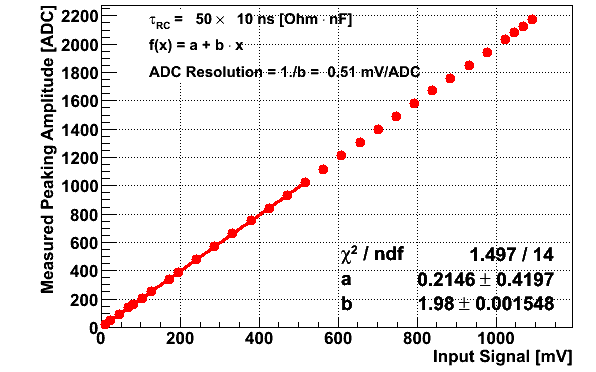}
  \includegraphics[height=6.cm,width=8cm]
                   {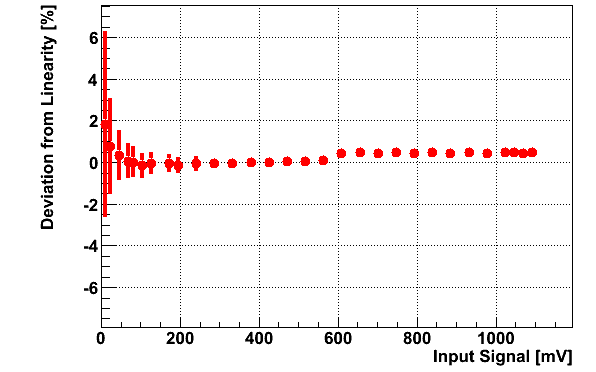}
  \caption{The ADC resolution is here extracted via a linear fit 
           to the data of the scan with respect to the amplitude of the 
           input voltage pulse.}
  \label{fig:WASA_LINEARITY_PULSER}
\end{figure}

The above data are presented in Fig.~\ref{fig:WASA_LINEARITY_PULSER}. 
A linear function is fitted to the data in two different measurement 
regions, resulting in an ADC resolution of approximately $0.5$ mV per ADC.

Out of the fit results, the deviation to linearity is calculated in 
percent, and is presented in the right panels of the picture. 
The deviation is typically well below $1\%$, although unknown systematical 
effects (not investigated yet) appear. 

Using the extracted ADC resolution, the total injected charge can be
calculated for each waveform and the corresponding mean value can be 
measured for each input voltage amplitude. The results will be compared 
in the next section with the data obtained in the same experimental 
conditions with the ACQIRIS readout system. 

\subsection{Waveform Analysis: WASA vs \mbox{ACQIRIS} Comparison}
\label{sec:WAVEFORM_PULSER}
The ACQIRIS readout electronics can store the waveform sampling 
values already in millivolt units, and performs a calibration 
at startup. We can thus measure the injected charge directly 
from the data, without any need of extra calibration. 

This type of measurement was performed for three 
amplitude values for the output voltage, $1900$, $1220$ and $580$ mV, 
thus allowing for an estimation of the systematical evaluation 
of the total injected charge as measured by both readout systems. 
Please note that to the systematical effects should be also included 
possible effects due to the offline analysis. 
 
\begin{figure*}[t!]
\hspace{-1cm} \vspace{-0.5cm}
      \includegraphics[height=5cm,width=7.cm]
               {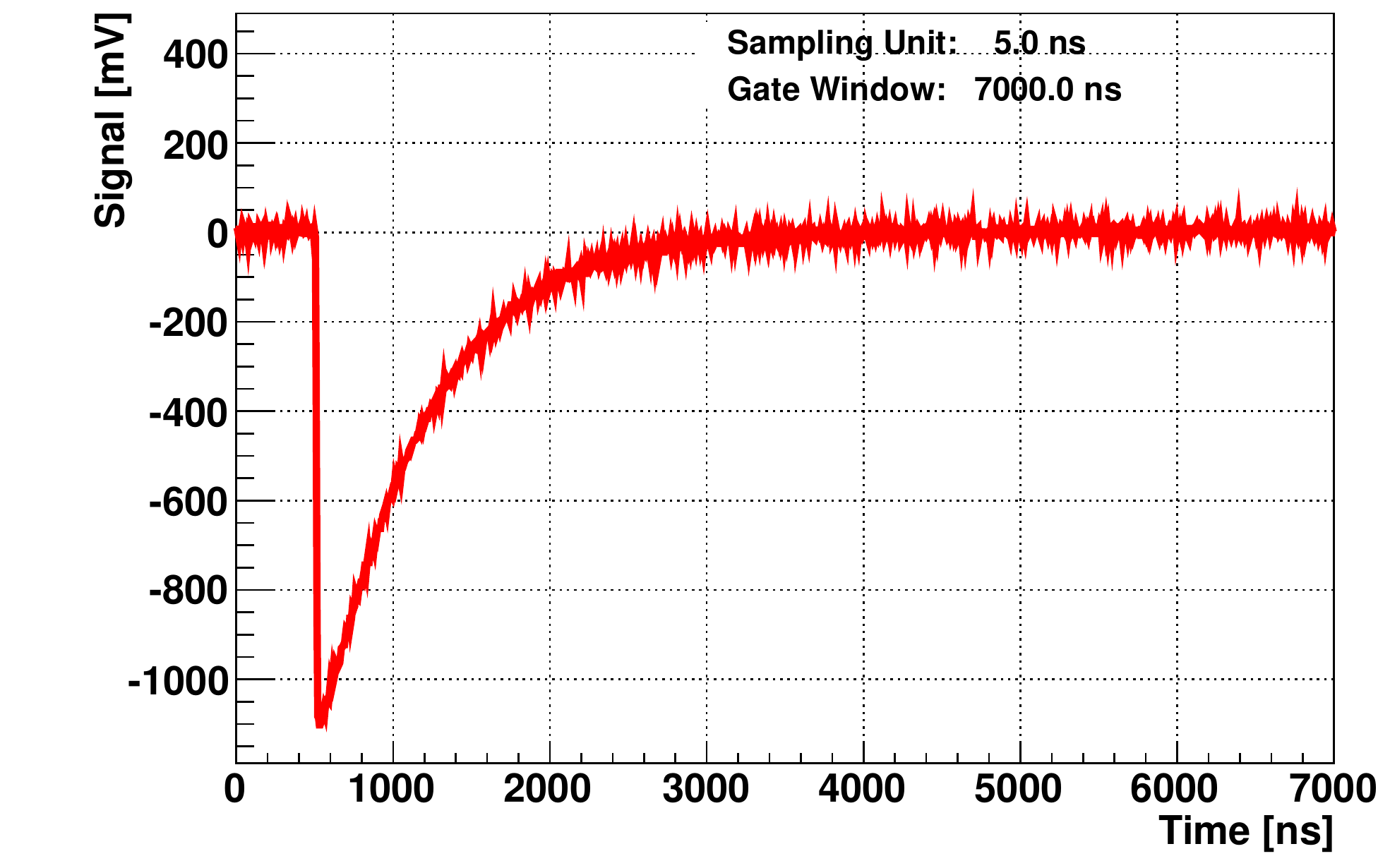}
      \includegraphics[height=5cm,width=7.cm]
               {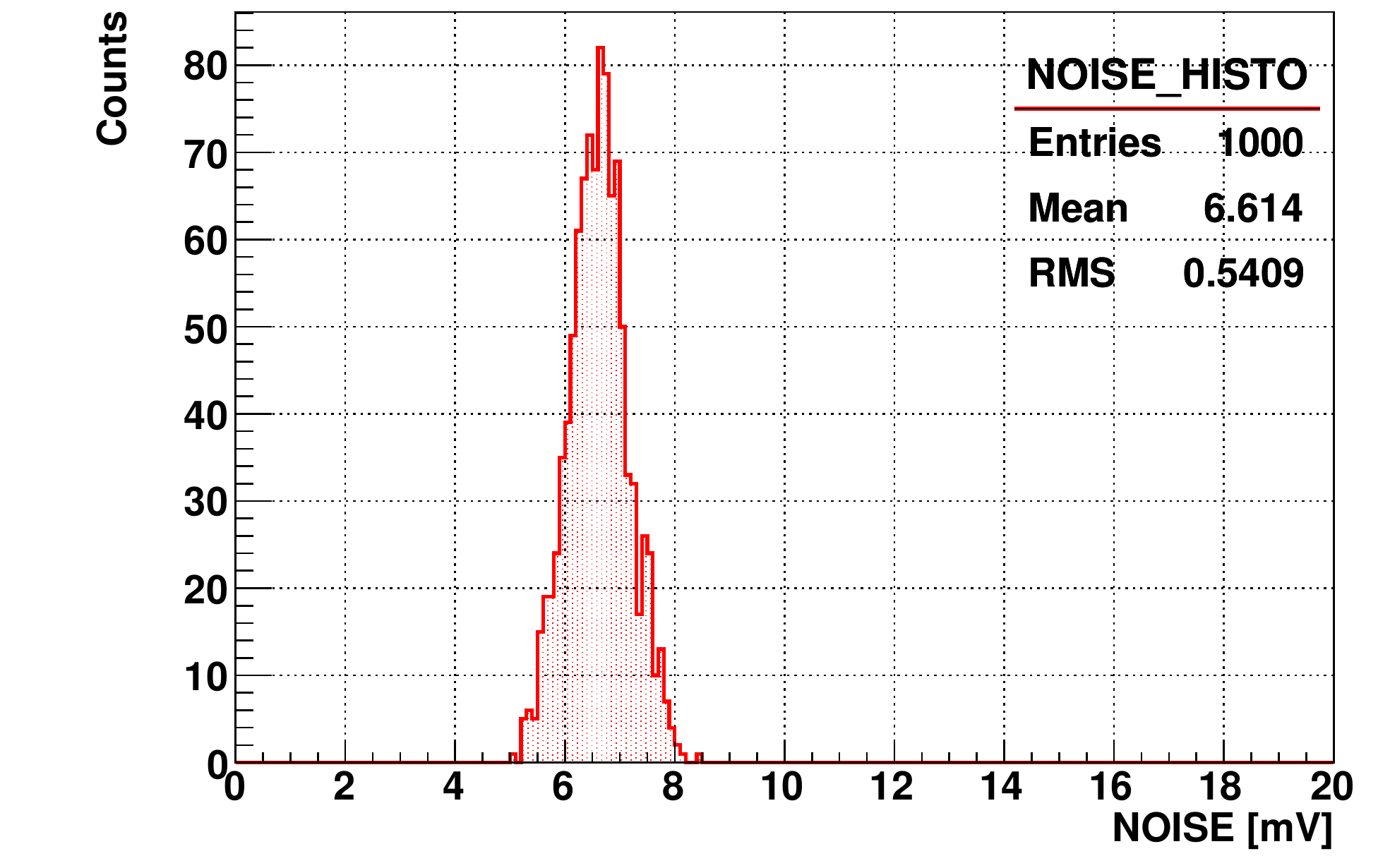}\\

\hspace{-1cm}
      \includegraphics[height=5cm,width=7.cm]
               {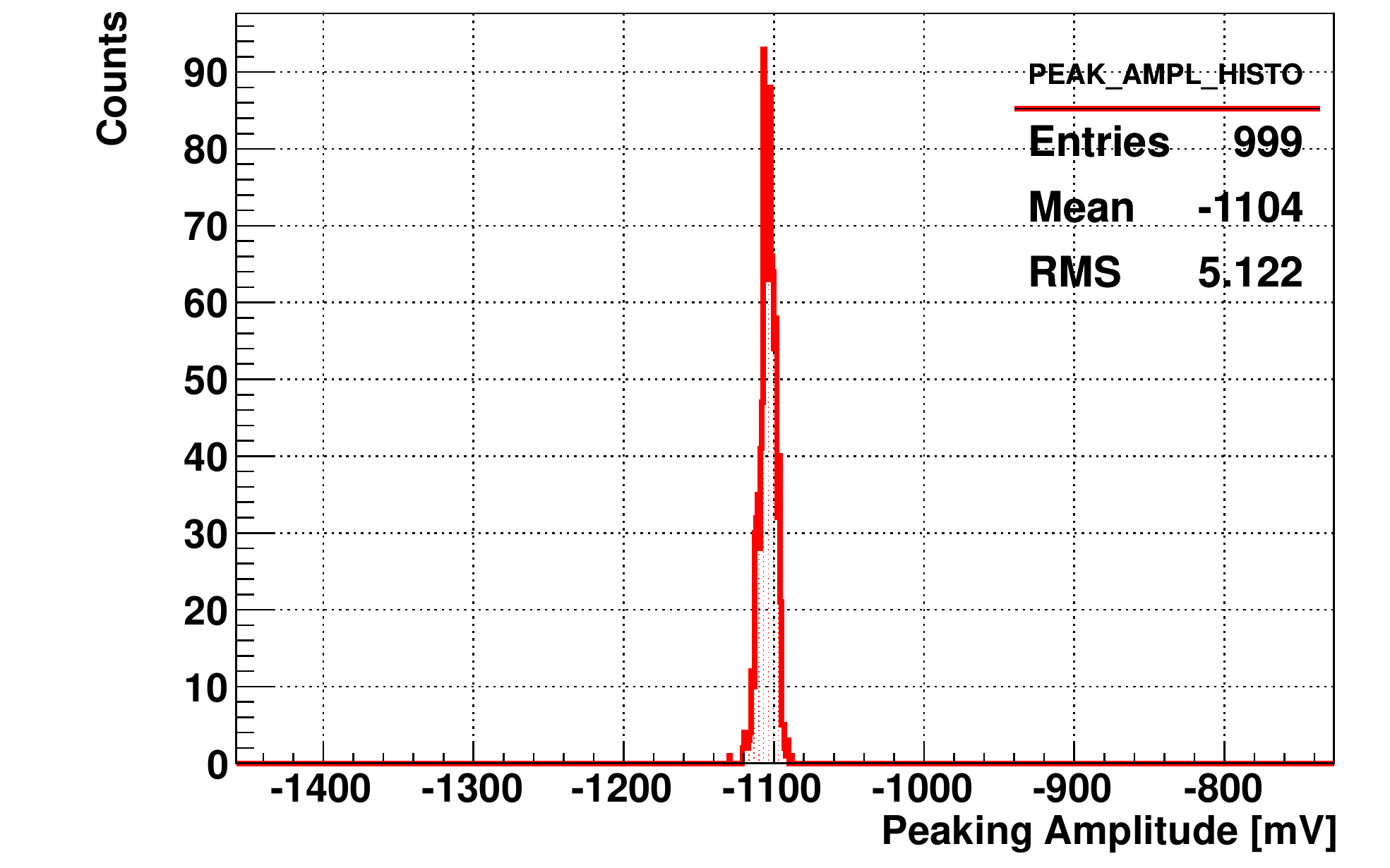}
      \includegraphics[height=5cm,width=7.cm]
               {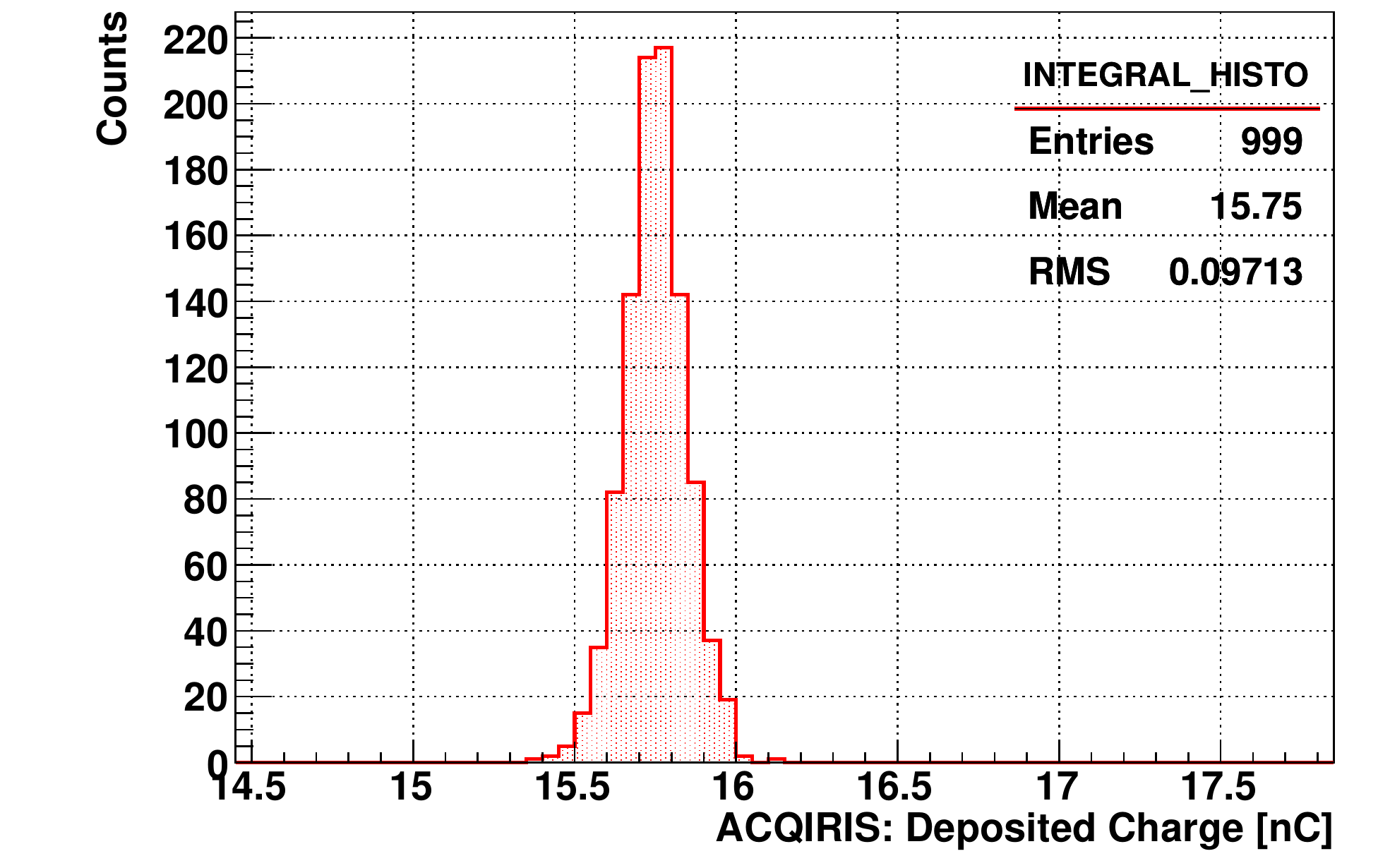}\\

  \vspace{-0.7cm}
  \caption{Typical histograms obtained during the waveform 
           analysis with the ACQIRIS electronics. These examples were obtained 
           with $1900$ mV voltage amplitude provided by the pulse generator.}
  \label{fig:ACQIRIS_PULSER_EXAMPLE}
\end{figure*}

As an example, the results for the output voltage pulse with 
amplitude of $1900$ mV are presented in Fig.~\ref{fig:ACQIRIS_PULSER_EXAMPLE}. 

As above mentioned, the ACQIRIS system provides the data already in millivolts
units, thus allowing a fast calculation of the deposited charge. The formula
used is 
\begin{eqnarray}
     INTEGRAL = \sum_i [ Signal_i \cdot CONVERSION \cdot TIME\_SAMPLING\_UNIT ] \ , 
\end{eqnarray}
being the sum performed when the signal is found below a certain threshold
(in this case the threshold is given by $3\cdot \sigma_{noise}$), the 
TIME\_SAMPLING\_UNIT was $5$ ns during this data taking, and the  
CONVERSION factor is simply $1/1000/50$ (with $50\ \Omega$ impedance) 
to convert from millivolt to Coulomb.
 
\begin{figure}[b!]
 \hspace{-2cm}  \vspace{-0.5cm}
 \includegraphics[height=5.4cm,width=7.5cm]
                 {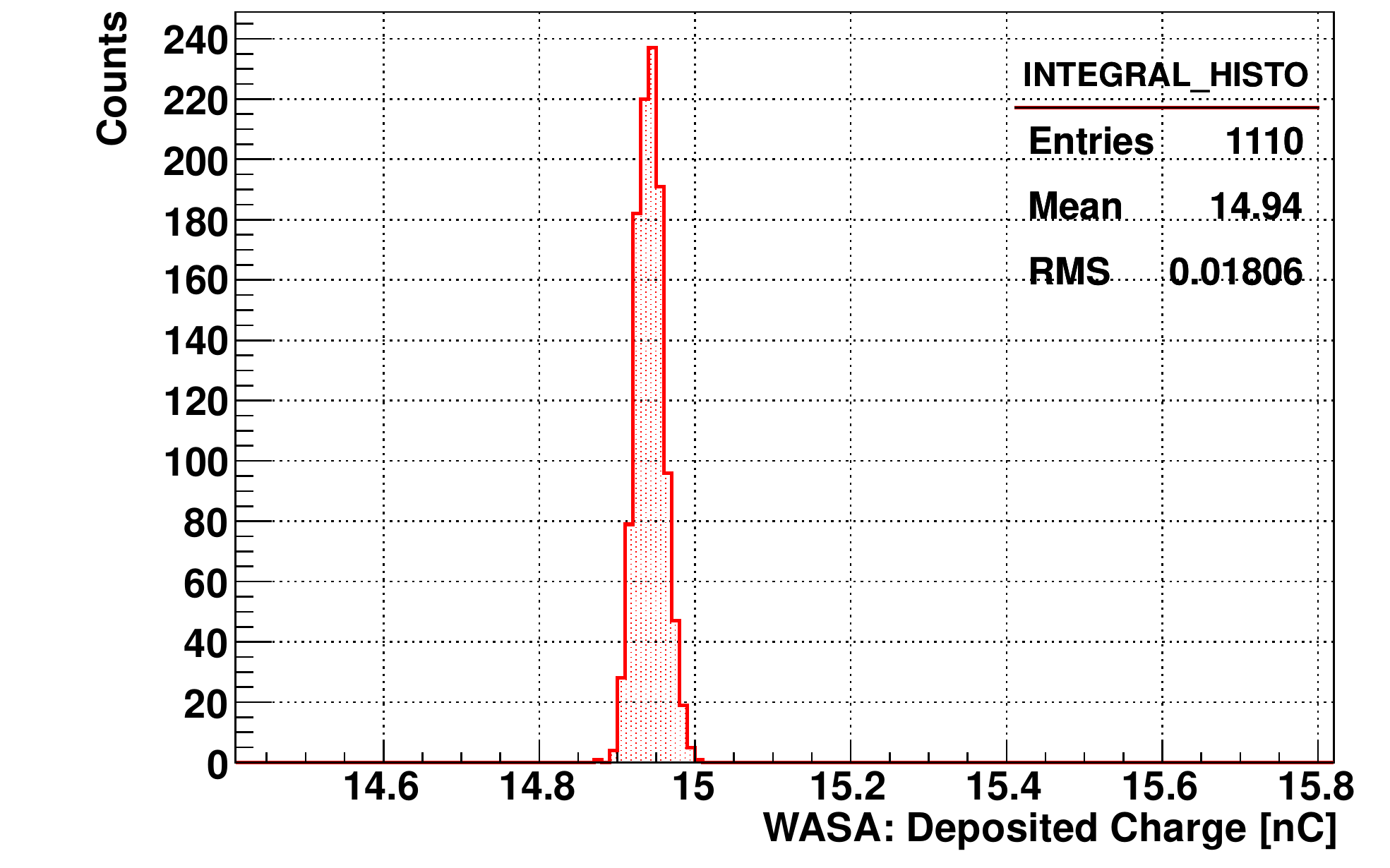}
 \includegraphics[height=5.4cm,width=7.5cm]
                 {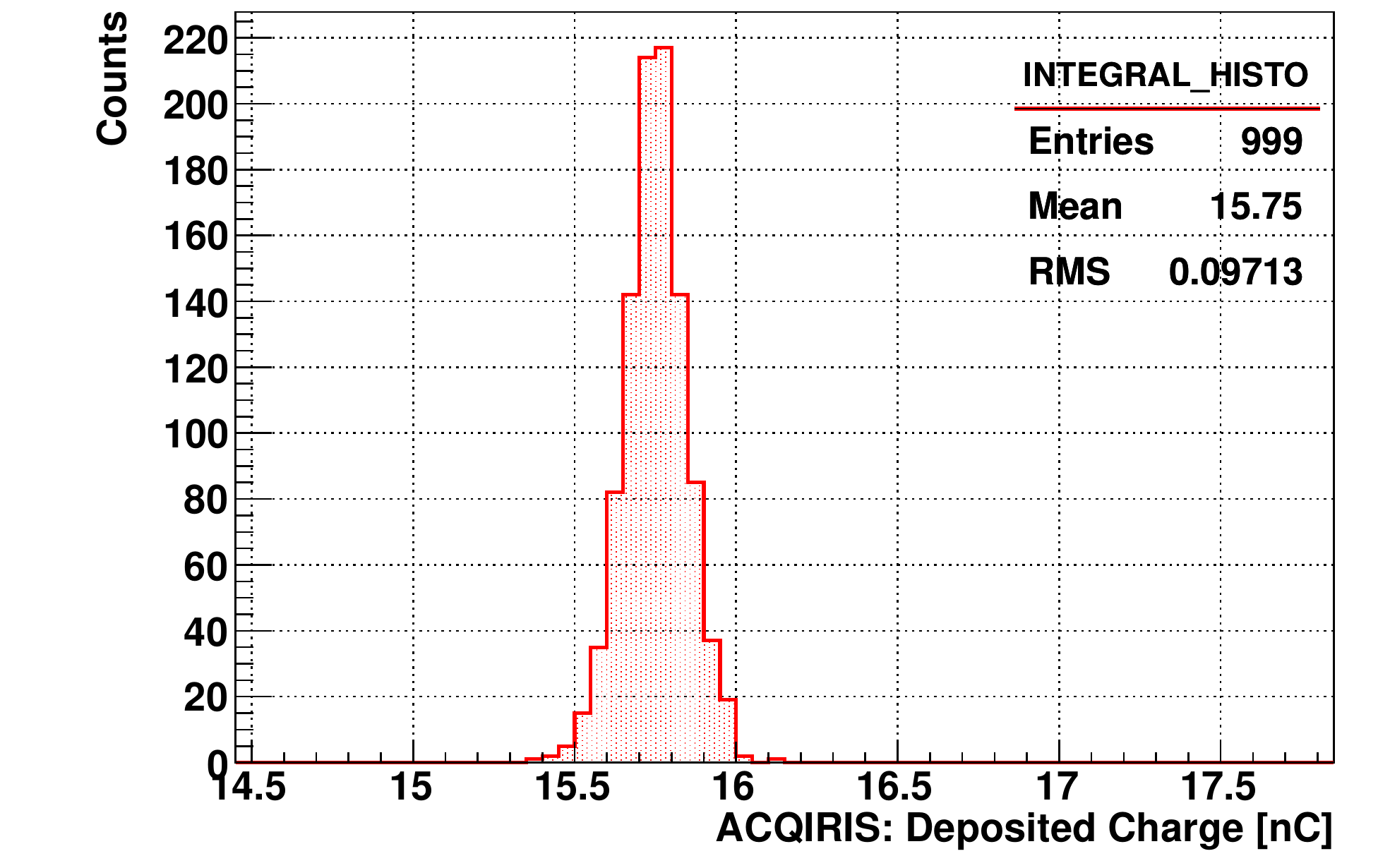}\\

 \hspace{-2cm} \vspace{-0.5cm}
 \includegraphics[height=5.4cm,width=7.5cm]
                 {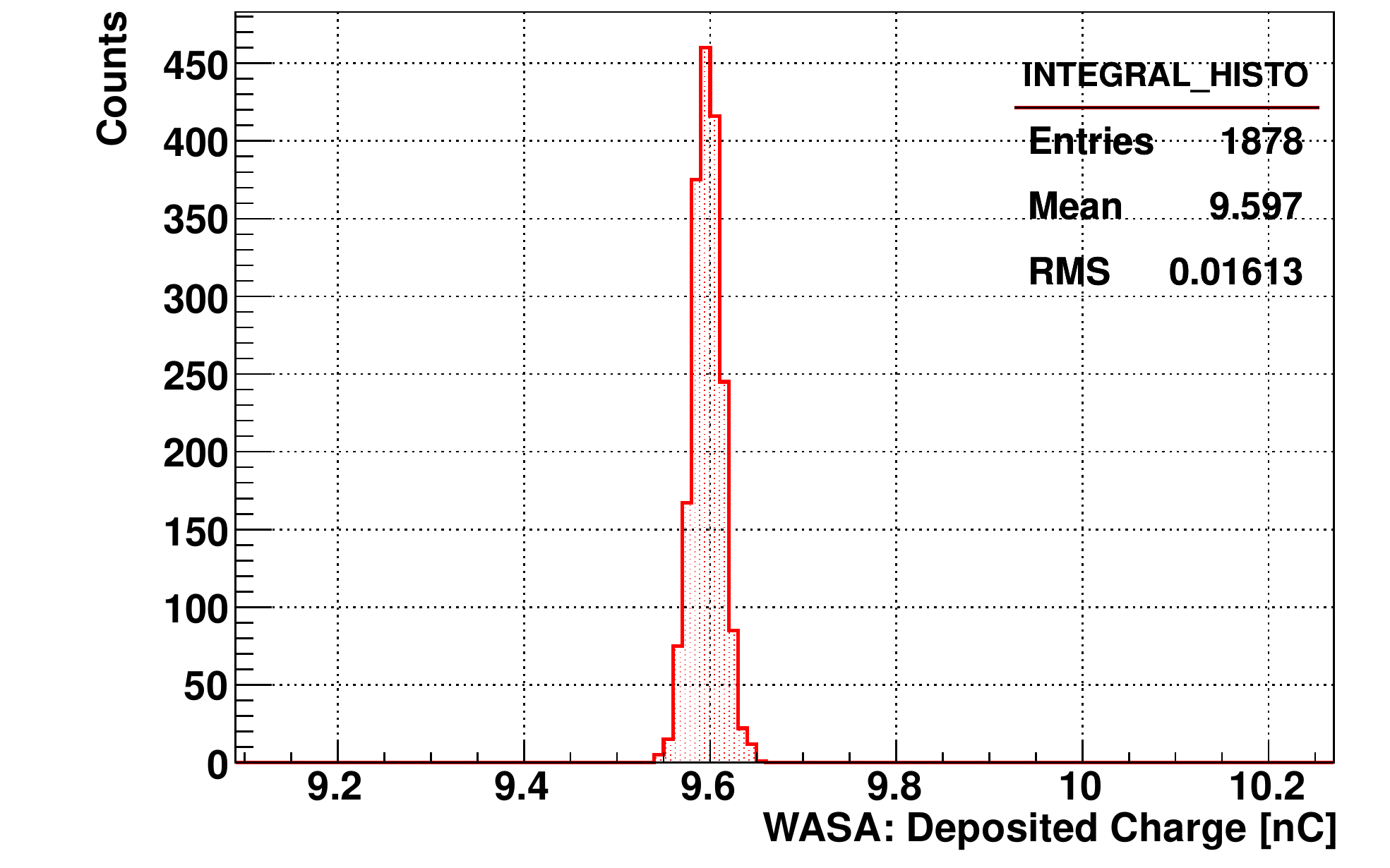}
 \includegraphics[height=5.4cm,width=7.5cm]
                 {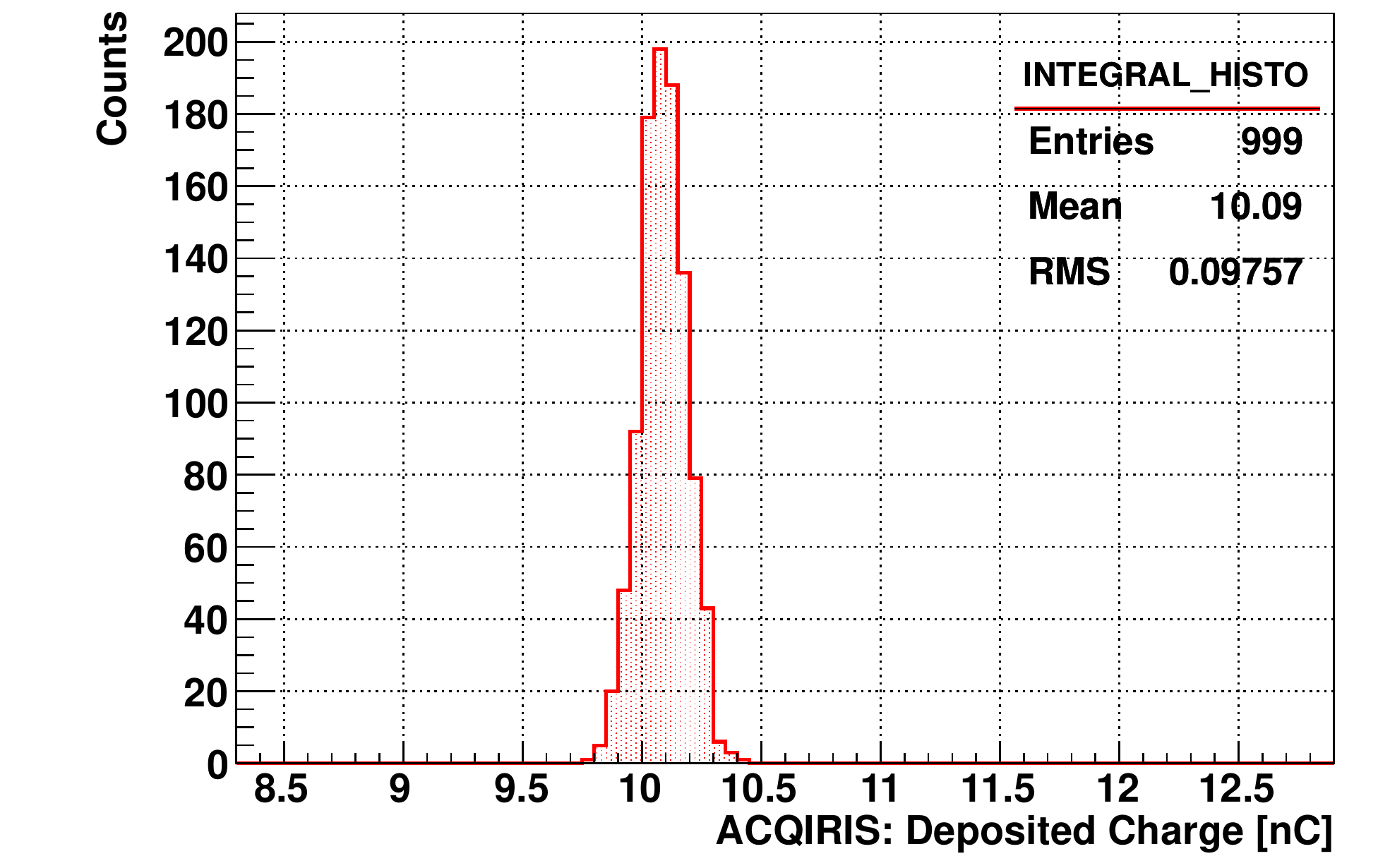}\\

 \hspace{-2cm}
 \includegraphics[height=5.4cm,width=7.5cm]
                 {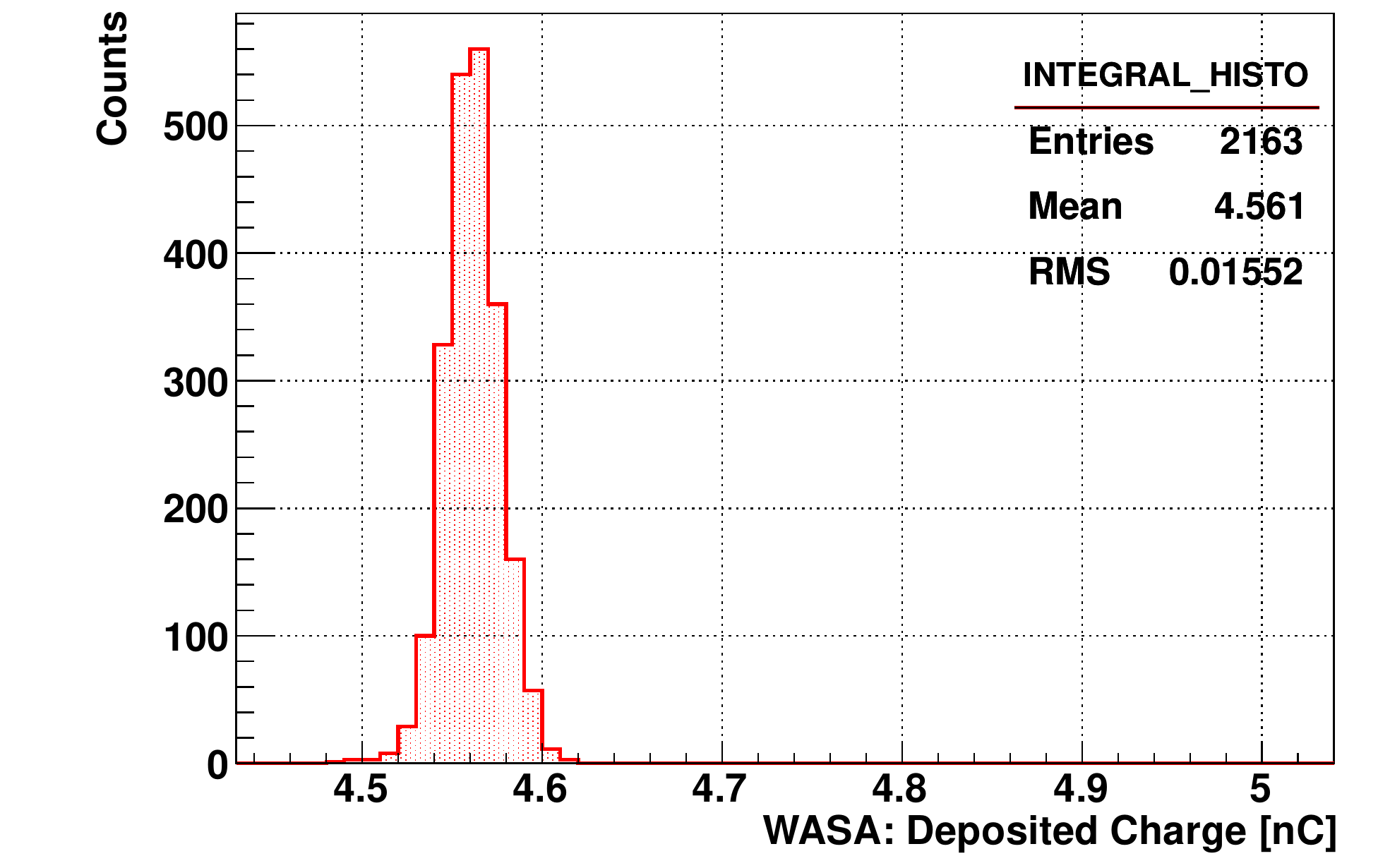}
 \includegraphics[height=5.4cm,width=7.5cm]
                 {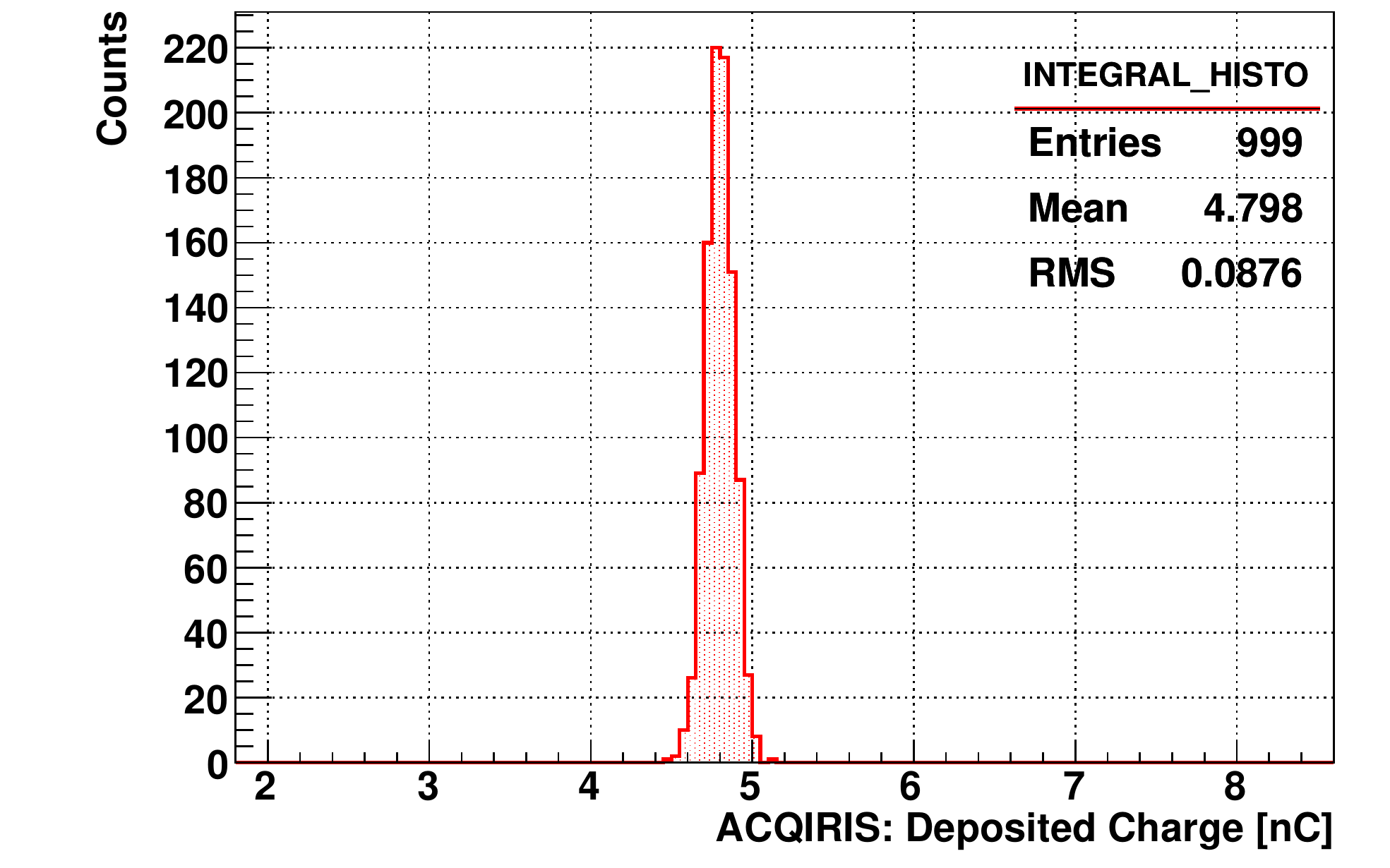}

 \vspace{-0.5cm}
  \caption{Integrated charge injected in the two readout systems 
           for three values of the voltage pulse amplitude.}
  \label{fig:COMPARE_CHARGE_PULSER}
\end{figure}

The comparison of the integrated charge as measured by the two systems is 
presented in Fig.~\ref{fig:COMPARE_CHARGE_PULSER} for the values 
$1900$, $1220$ and $580$ mV of the voltage pulse amplitude. 

For all three experimental setups a relative $5\%$ difference appears.
Preliminary, for the moment this value is considered as a contribution 
to the systematical
uncertainty of the measurement. It should be also mentioned that 
the ADC to mV conversion factor was derived by a direct observation in 
the oscilloscope with a large uncertainty. In the next measurements 
this conversion factor for the WASA system will be extracted by 
measuring the peaking amplitude with the ACQIRIS readout electronics 
in the same experimental conditions. Unfortunately, this way the results 
between the two devices will be not any more completely independent, although 
a large systematical scale uncertainty will be removed. 

\section{Precision of the ACQIRIS System with Respect to the Sampling Unit}
%
Up to now the ACQIRIS system was used at the sampling unit of $5$ ns 
which is the available setting closest to the $6.5$ ns unit used in the 
WASA system. Although the comparison between the two systems has given
promising consistent results, one should investigate how much could 
improve the measurement of the total injected charge when running 
the ACQIRIS system with finer sampling units. 

\begin{figure}[t!]
\hspace{-2cm}
  \includegraphics[height=6.cm, width=7.5cm]
                  {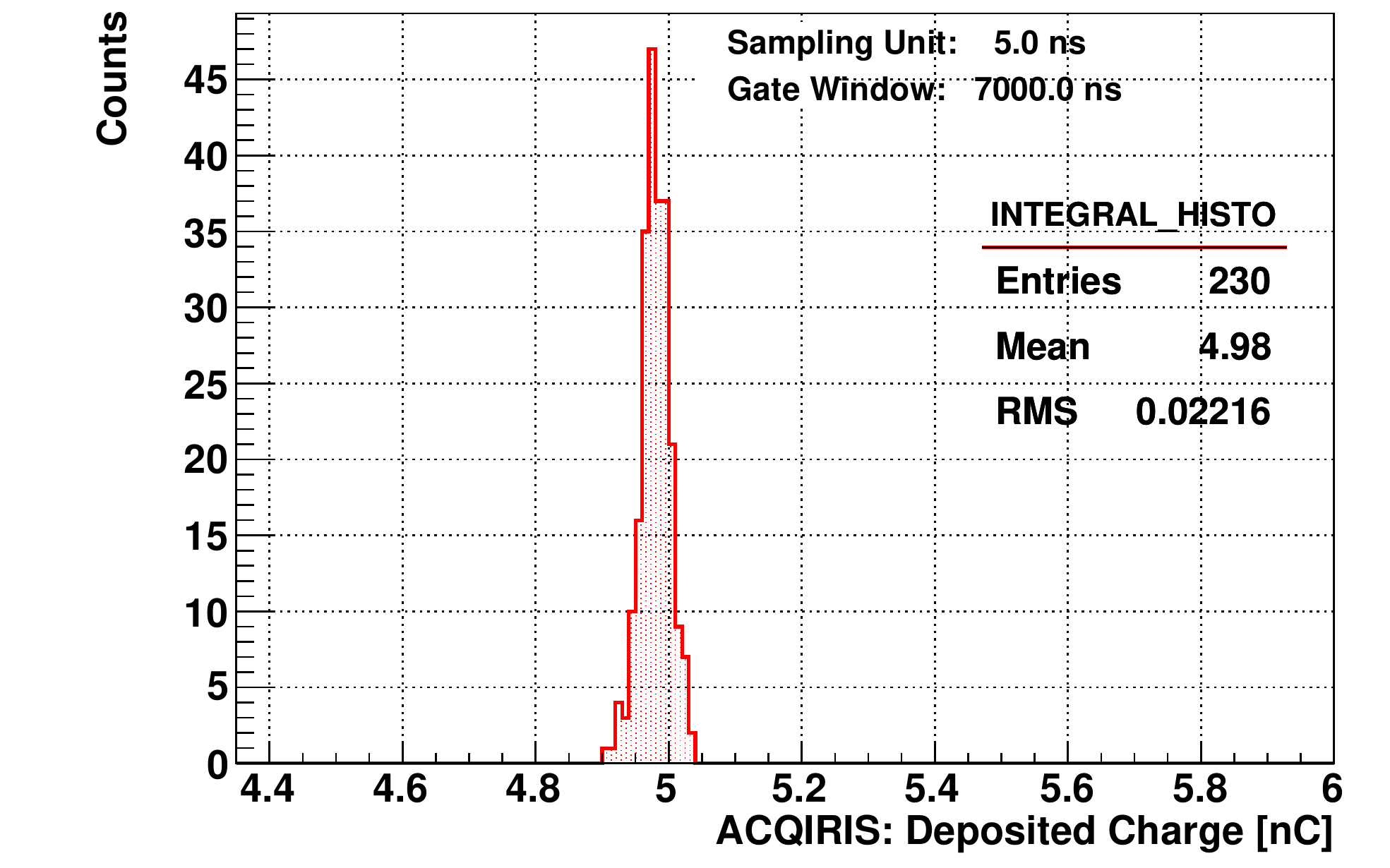} 
  \includegraphics[height=6.cm, width=7.5cm]
                  {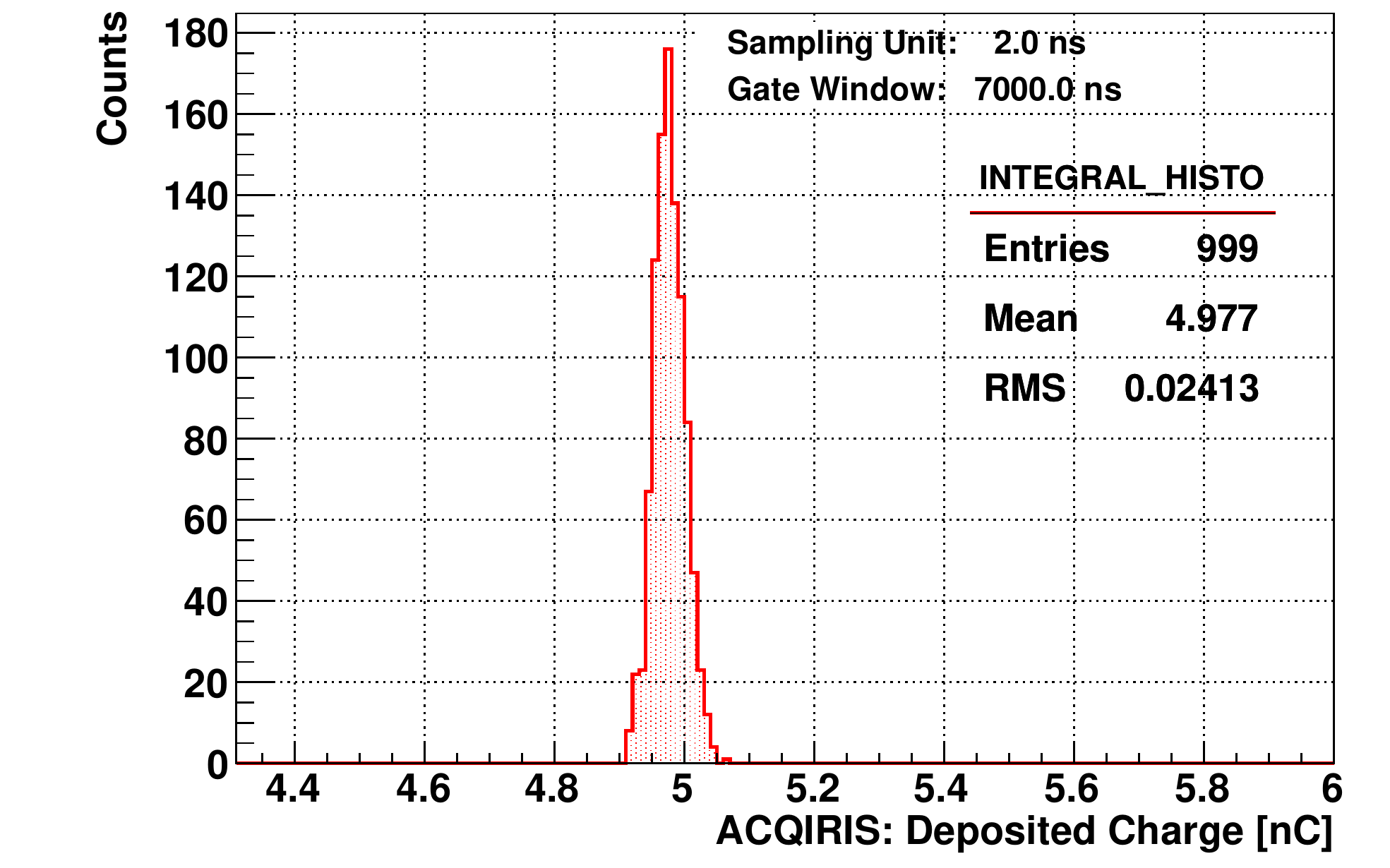} \\

\hspace{-2cm}
  \includegraphics[height=6.cm, width=7.5cm]
                  {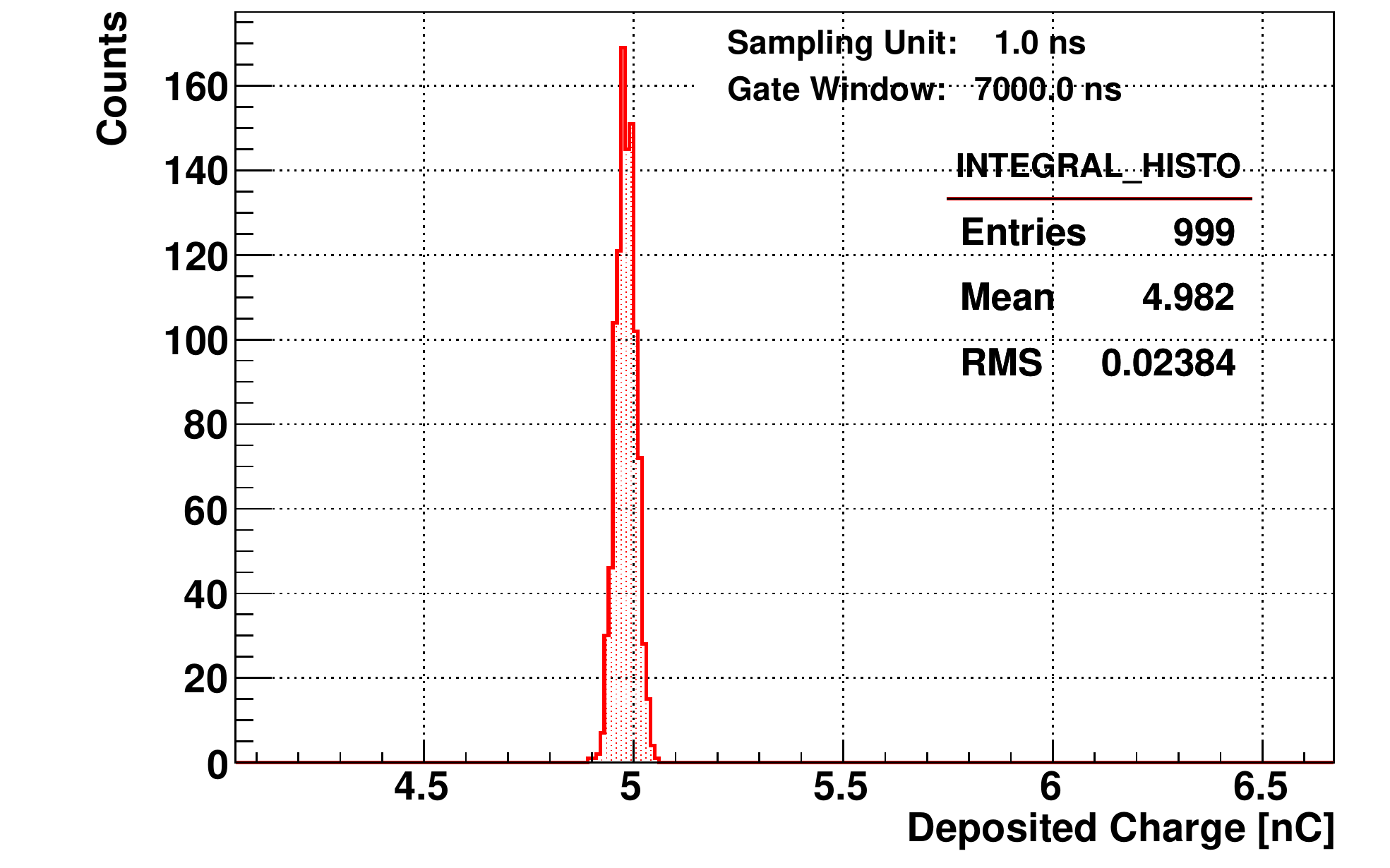} 
  \includegraphics[height=6.cm, width=7.5cm]
                  {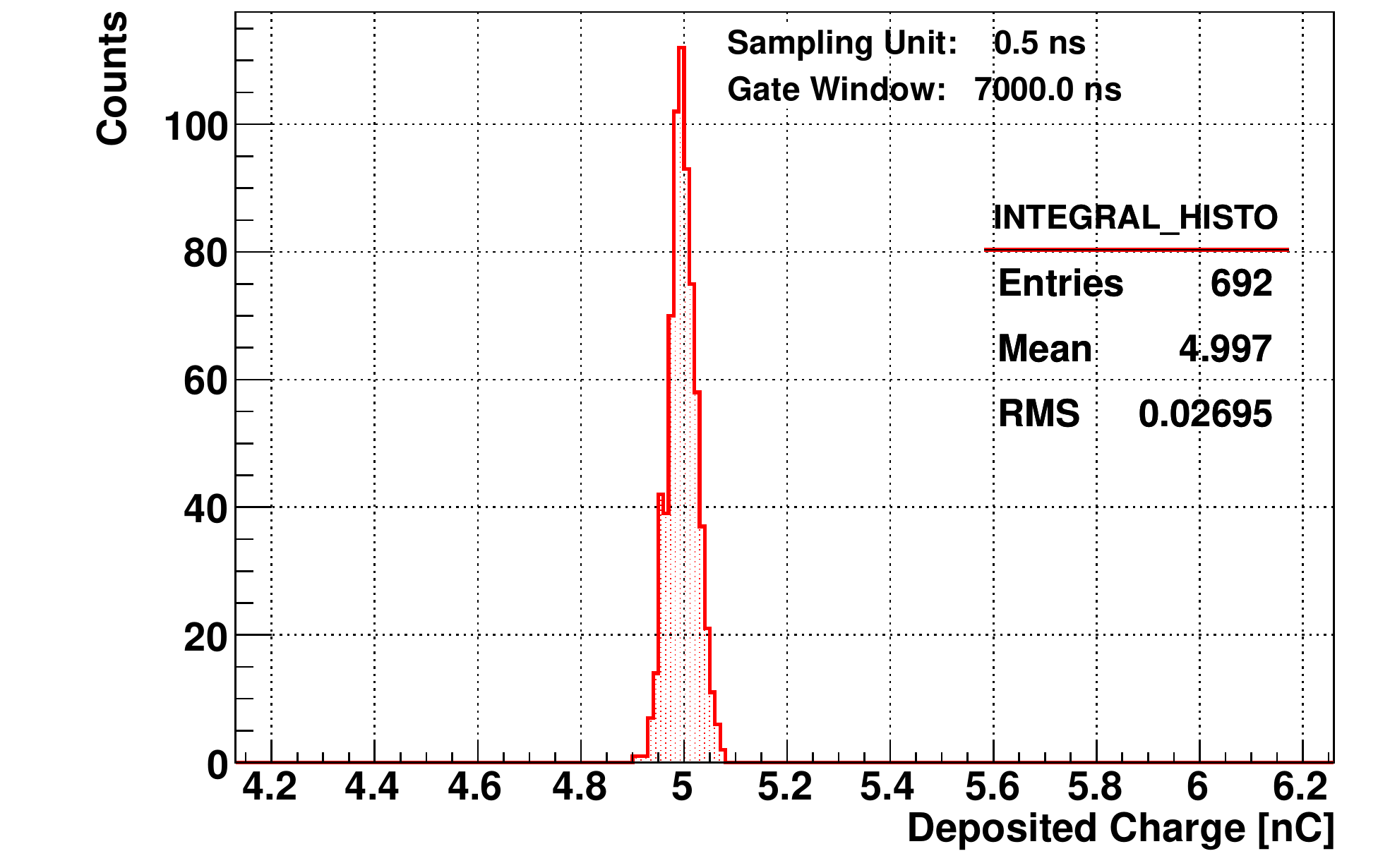} \\
  \vspace{-0.6cm}
  \caption{The integral of the injected charge is investigated 
           with the ACQIRIS system at different sampling unit values.}
  \label{fig:ACQIRIS_SAMPLING_UNITS}
\end{figure}

The results for $5.0$, $2.0$, $1.0$ and $0.5$ ns is presented in 
Fig.~\ref{fig:ACQIRIS_SAMPLING_UNITS} using a voltage pulse 
with amplitude $580$ mV.
The results show that for this type of analysis the variation 
in the calculation of the integral of the injected charge is at
per-mill level, thus negligible. Having shown this feature, in the 
following analyses presented in this note the ACQIRIS system will 
be run with the sampling unit of $5$ ns, as close as possible
to the finest WASA setup ($6.25$ ns).  

\section{Analysis with an LED coupled to Vacuum PMTs}
%
So far, waveforms generated by a pulse generator were 
investigated giving good agreement between the two readout system.  
The next natural step of the investigation was the waveform
analysis using signals generated directly by photomultipliers. 
In this study the vacuum photomultiplier 8850 by BURLE Electron 
Tubes was used. 

\subsection{Test-Bench Description}
\begin{figure}[t!]
   \hspace{-0.6cm}
      \includegraphics[height=6.0cm,width=13.cm]{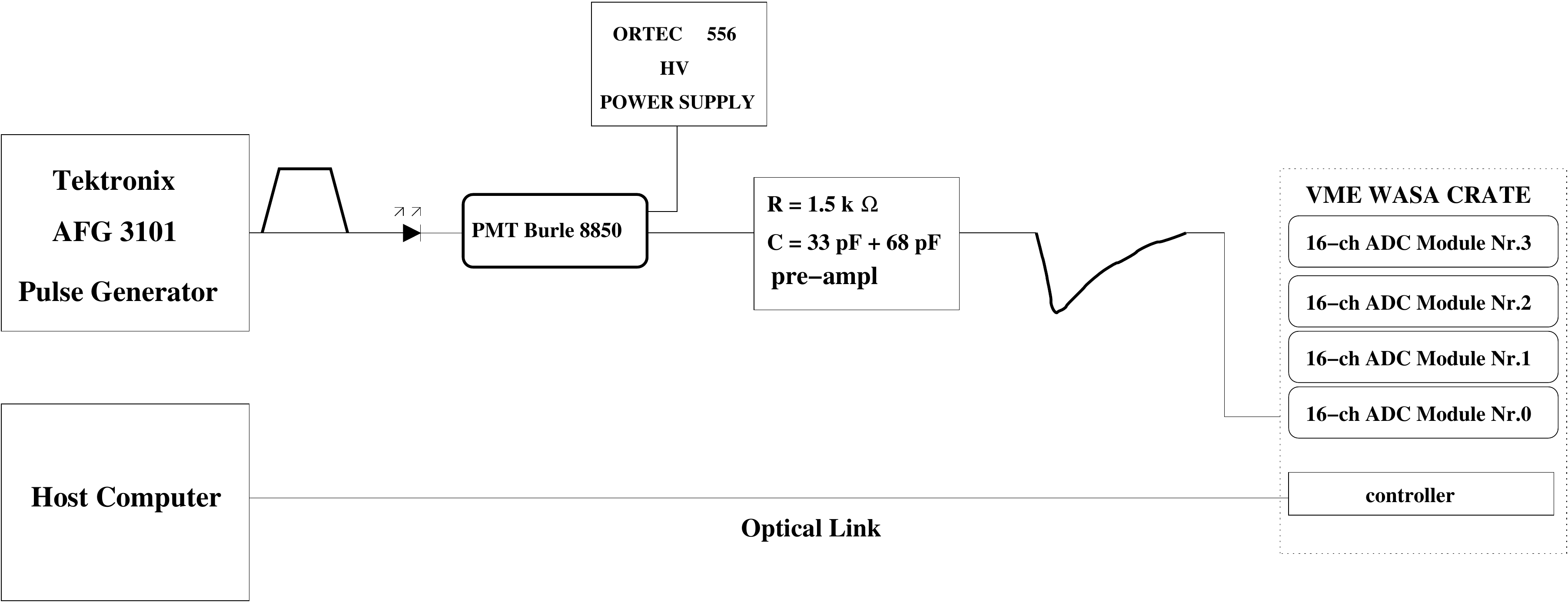}
  \caption{Test-bench setup used for the waveform analysis with the 
           WASA readout system at FZJ using the signal generated by a 
           photomultiplier. A similar experimental setup 
           is used when using the ACQIRIS system; here network twisted cables
           are used for transmitting the data to the host computer.} 
  \label{fig:TestBench_LED}
\end{figure}

%
The test-bench was modified to cope with signals from PMTs, 
and the experimental setup built to measure the properties 
of wave shifting fibers was used.
The setup is shown in Fig.~\ref{fig:TestBench_LED}. 

A $10$ ns wide voltage pulse with $5$ ns rise- and fall-time is sent to a
blue LED which highlights the photocathode of the photomultiplier.
Different intensities of the LED were used leaving the voltage pulse width 
unchanged, to avoid an increase of the rise-time of the current signal 
generated by the PMT, whose high-voltage was left to its nominal value
$1000$ V. Instead, to change the intensity of the LED light output 
the voltage amplitude of the pulse generator was changed. 

The pre-amplifier built for the Anger camera project~\cite{JUDIDT}
was used here, soon after the PMT, after modifying the RC stage to 
the following configuration
\begin{eqnarray}
    R & = & 1.5 \ k \Omega  \nonumber \ , \\
    C & = & 33 \ pF \ + 68 \ pF. 
    \label{eq:ANGER_PREAMPL}
\end{eqnarray}

With the above mentioned configuration of the voltage pulse generator
and of the pre-amplifier no undershooting of the signal was observed. 

\subsection{Measurements with the WASA System}
\label{sec:WASA_PMT}
To verify the linearity of the system in the new experimental 
configuration for the WASA system a  
scan in the input voltage to the LED was performed. 
The response of the LED is typically not proportional to the input 
voltage, thus we used the results obtained with the ACQIRIS system 
to verify the linearity of the system 
(see Sec.~\ref{sec:WAVEFORMS_LINEARITY}).

\begin{figure*}[h!]
  \hspace{-1cm}
  \includegraphics[height=5cm,width=7.cm]
           {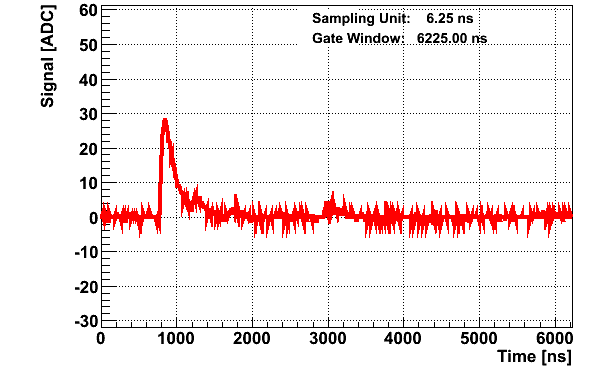}
  \includegraphics[height=5cm,width=7.cm]
           {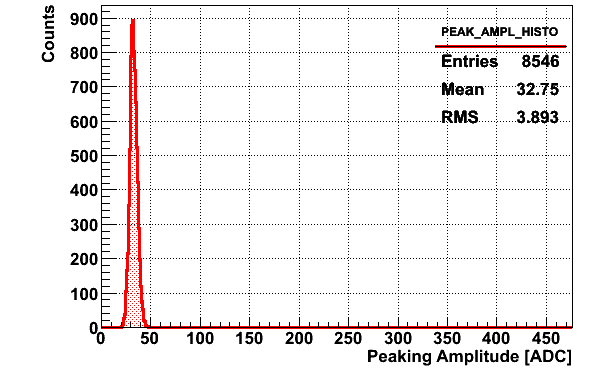}
  \vspace{-0.8cm}
  \caption{Waveform and peaking amplitude distribution measured with 
           WASA electronics for \underline{$5.4$ V} voltage amplitude 
           sent to the LED.}
  \label{fig:WASA_LED_5.4}
\end{figure*}

\begin{figure*}[h!]
  \hspace{-1cm}
  \includegraphics[height=5cm,width=7.cm]
           {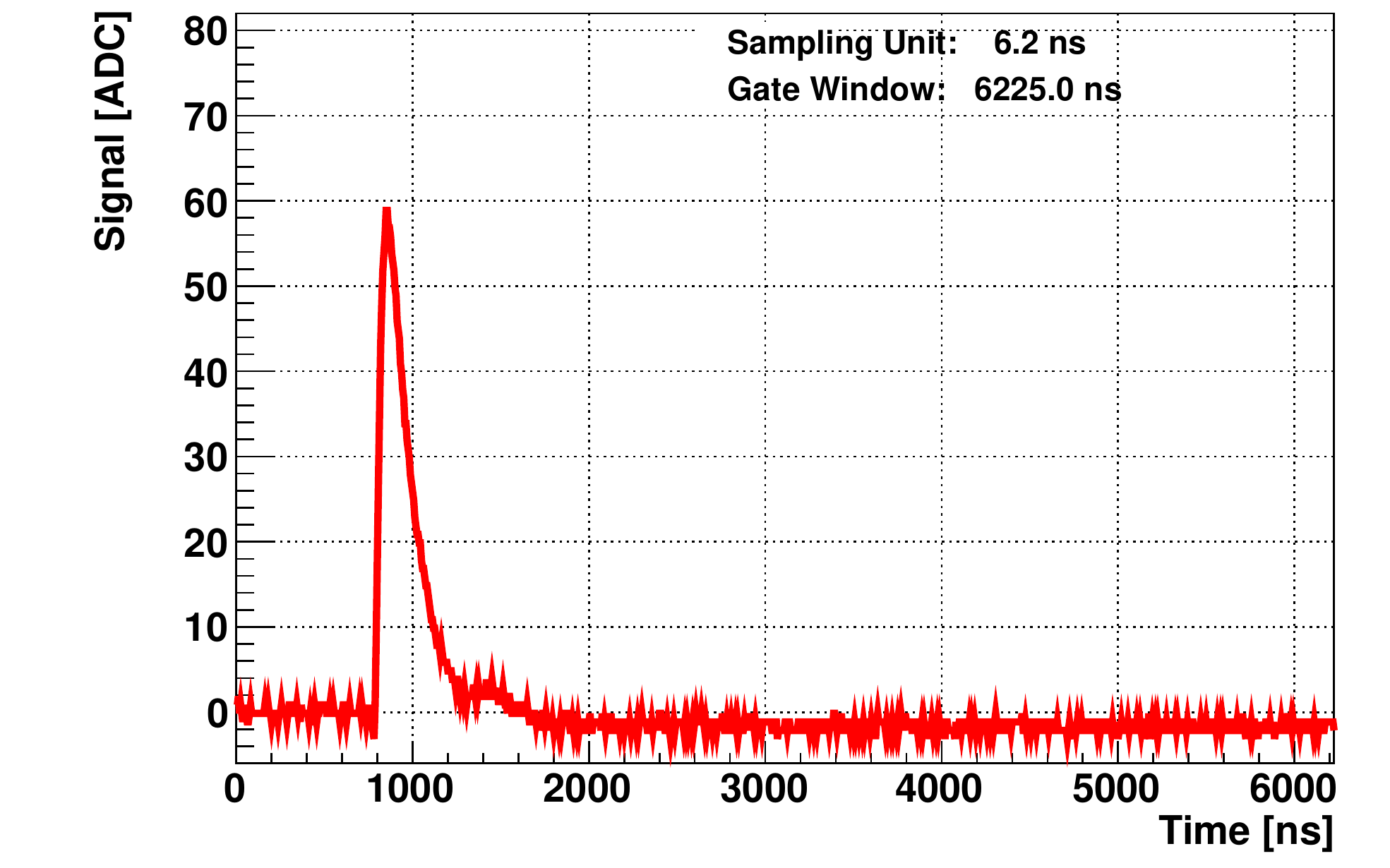}
  \includegraphics[height=5cm,width=7.cm]
           {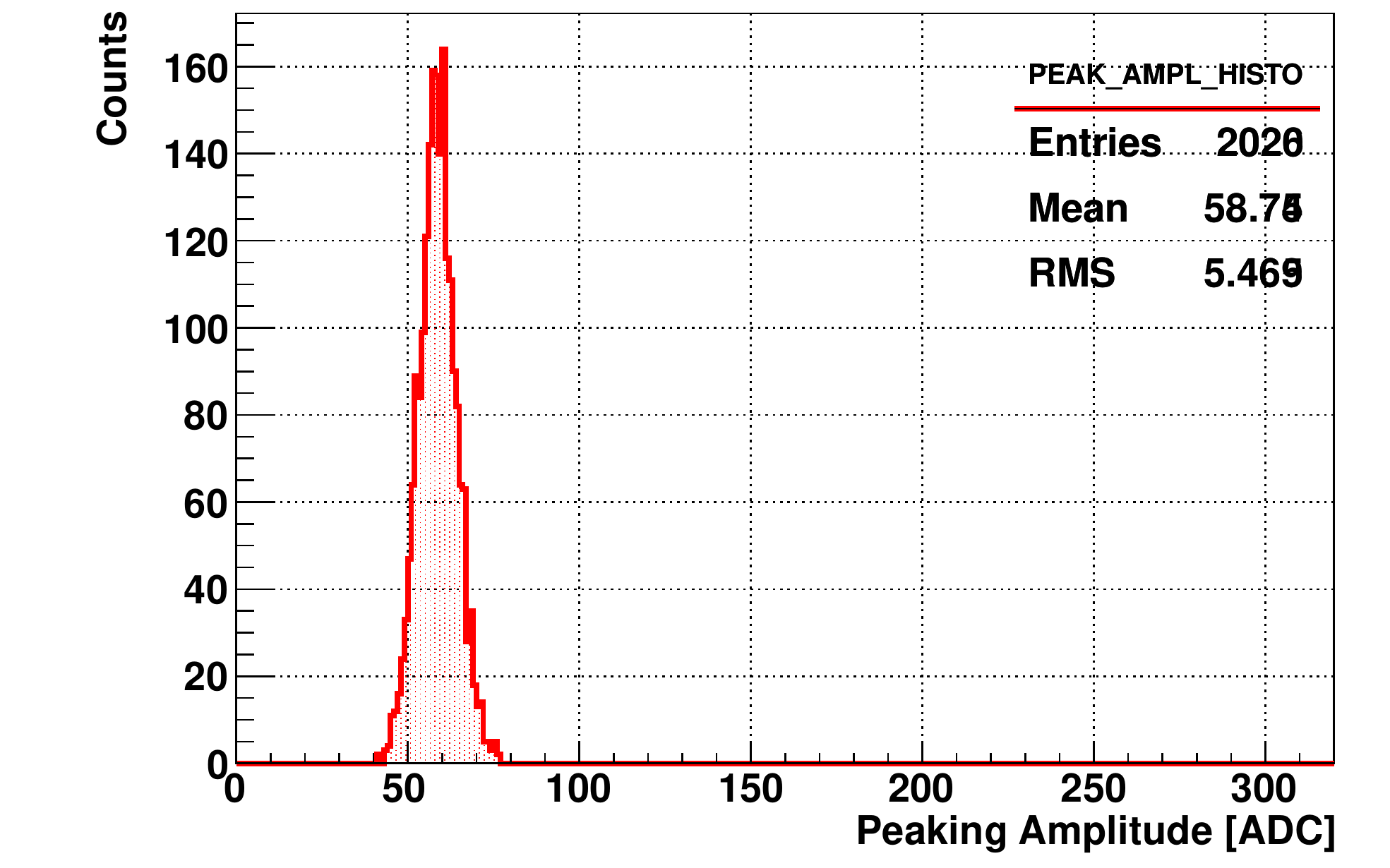}
  \vspace{-0.8cm}
  \caption{Waveform and peaking amplitude distribution measured with 
           WASA electronics for \underline{$6.3$ V} voltage amplitude 
           sent to the LED.}
  \label{fig:WASA_LED_6.3}
\end{figure*}

\begin{figure*}[h!]
  \hspace{-1.0cm}
  \includegraphics[height=5cm,width=7.cm]
           {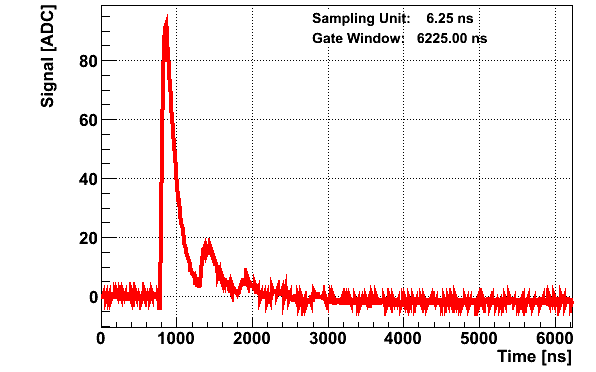}
  \includegraphics[height=5cm,width=7.cm]
           {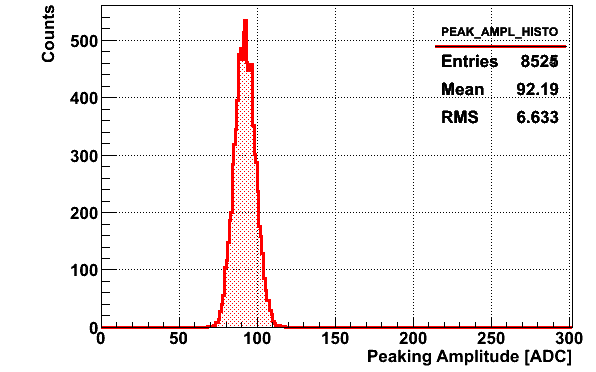}
  \vspace{-0.8cm}
  \caption{Waveform and peaking amplitude distribution measured with 
           WASA electronics for \underline{$7.2$ V} voltage amplitude 
           sent to the LED.}
  \label{fig:WASA_LED_7.2}
\end{figure*}

\begin{figure*}[h!]
  \hspace{-1.0cm}
  \includegraphics[height=5cm,width=7.cm]
           {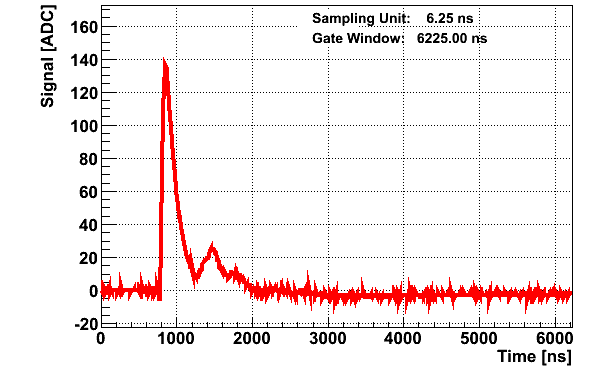}
  \includegraphics[height=5cm,width=7.cm]
           {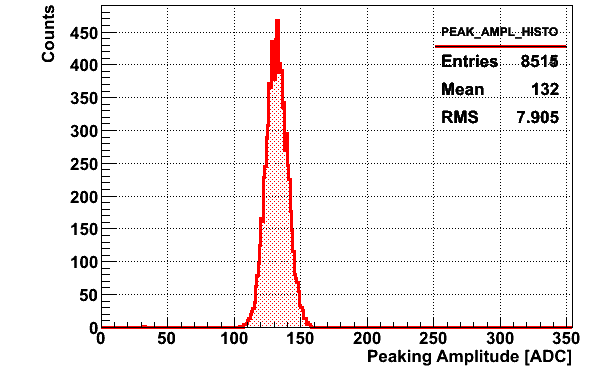}
  \vspace{-0.8cm}
  \caption{Waveform and peaking amplitude distribution measured with 
           WASA electronics for \underline{$8.1$ V} voltage amplitude 
           sent to the LED.}
  \label{fig:WASA_LED_8.1}
\end{figure*}

\begin{figure*}[h!]
  \hspace{-1.0cm}
  \includegraphics[height=5cm,width=7.cm]
           {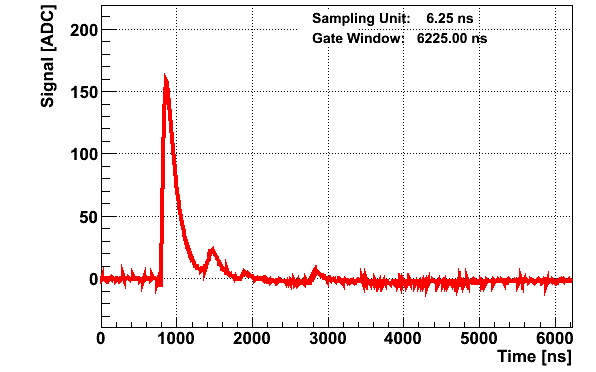}
  \includegraphics[height=5cm,width=7.cm]
           {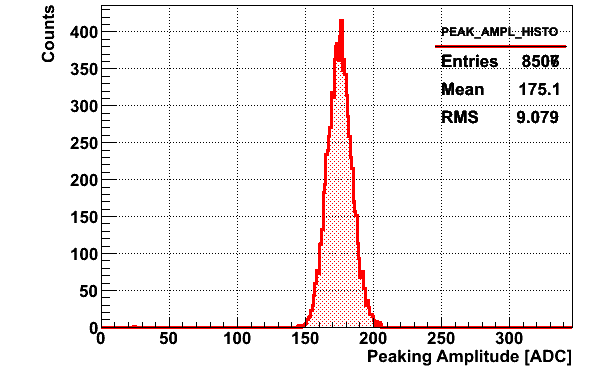}
  \vspace{-0.8cm}
  \caption{Waveform and peaking amplitude distribution measured with 
           WASA electronics for \underline{$9.0$ V} voltage amplitude 
           sent to the LED.}
  \label{fig:WASA_LED_9.0}
\end{figure*}

The peaking amplitude distributions and 
a typical waveform (after pedestal subtraction) are presented 
for different values of the voltage input to the LED in the 
Fig.~\ref{fig:WASA_LED_5.4}~-~\ref{fig:WASA_LED_9.0},
and the results
are shown in Tab.~\ref{tab:WASA_SCAN_LED}.
\begin{table}[b!]
  \begin{center}
\begin{tabular}{||c|c|c||}
\hline 
\hline
\multicolumn{2}{|c|}{ \textbf{WASA System: LED Intensity Scan} }\\
\hline
\hline
   \bf{Voltage Amplitude to LED [V]:} &\bf{Measured Peaking Amplitude [ADC]:}\\
\hline 
\hline 
    5,4 &     33 \\
    6.3 &     59 \\
    7.2 &     92 \\
    8.1 &    132 \\
    9.0 &    175 \\
\hline
\hline
\end{tabular}
\caption{Scan performed with the WASA readout system increasing 
         the LED intensity by changing the amplitude of the voltage 
         pulse sent to the LED.}
\label{tab:WASA_SCAN_LED}
  \end{center}
\end{table}
Due to the limitations of the voltage pulse generator, values 
not larger than $10$ V could be used, thus limiting the investigated 
ADC region of the WASA system.
 
The conversion to millivolt of the values obtained in ADC units 
is provided by the similar measurements performed with the 
ACQIRIS readout system, shown in the next section. For the same 
amplitude of the voltage pulse to the LED a relation between 
the peaking amplitude value measured by WASA (in ADC units) 
and by ACQIRIS (in millivolt) was obtained. 
%

\subsection{Measurements with the ACQIRIS  System}
\label{sec:ACQIRIS_LED}
\begin{figure*}[h!]
  \hspace{-1cm}
  \includegraphics[height=5cm,width=7.cm]
           {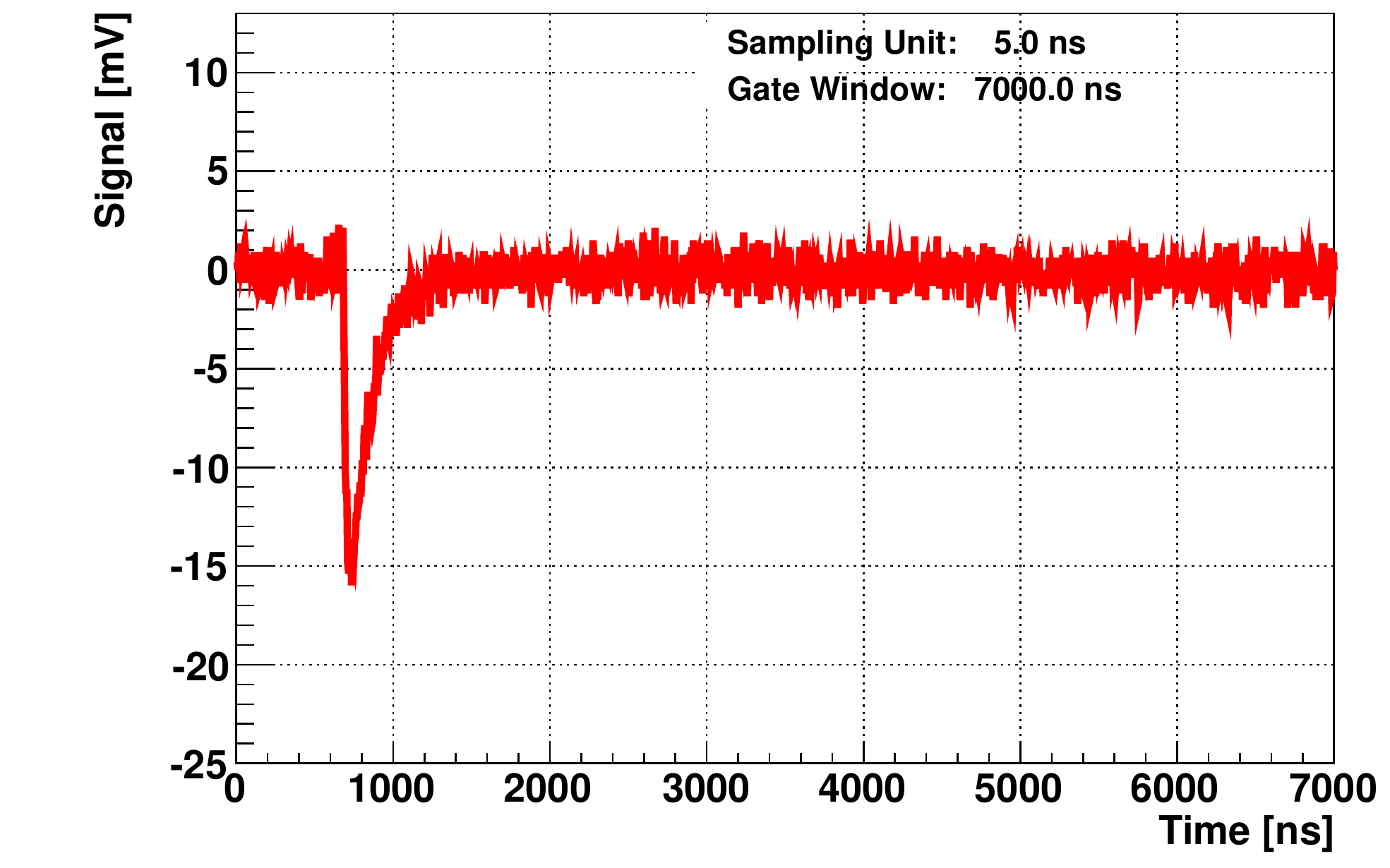}
  \includegraphics[height=5cm,width=7.cm]
           {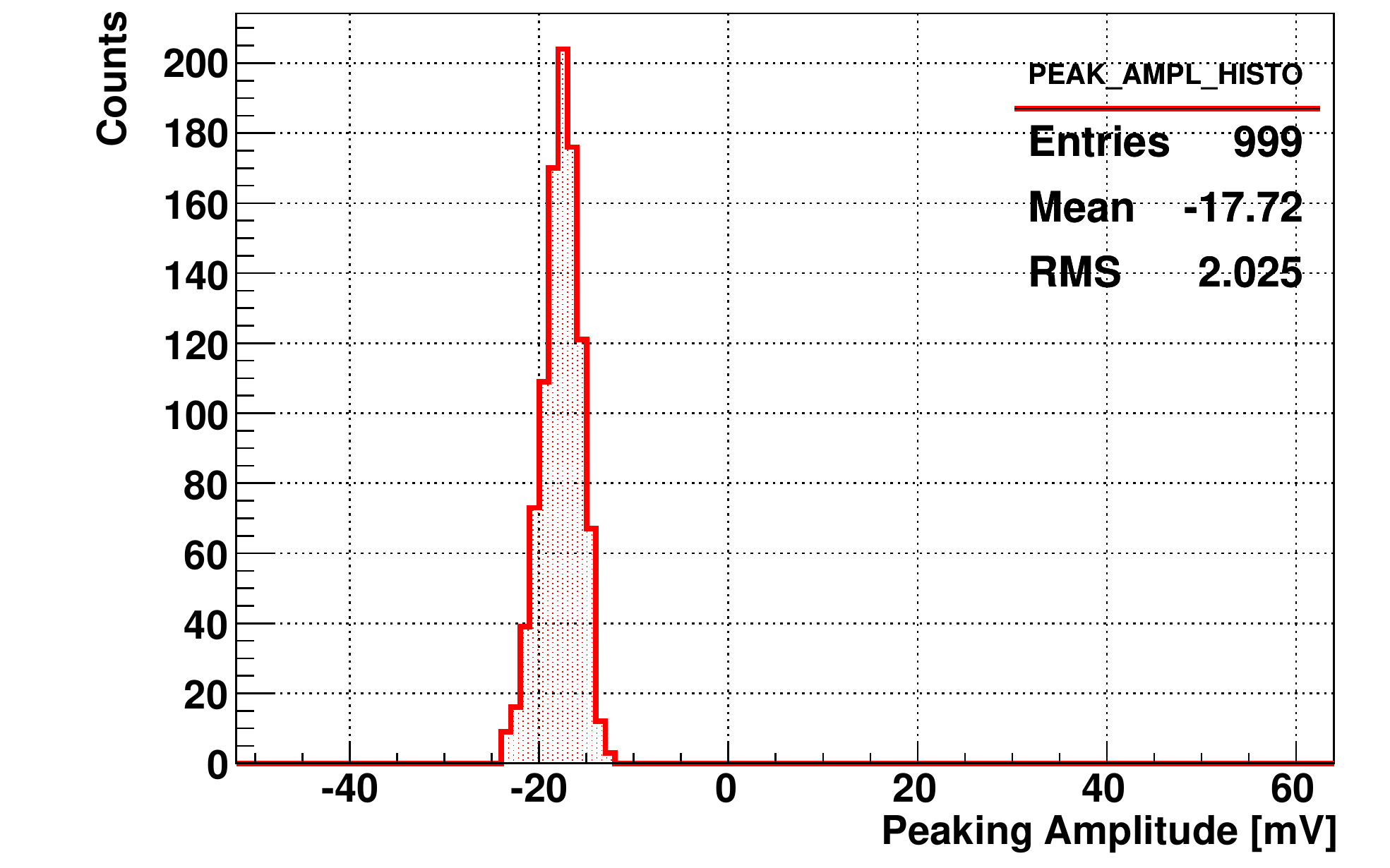}
  \vspace{-0.8cm}
  \caption{Waveform and peaking amplitude distribution measured with 
           \mbox{ACQIRIS} electronics for \underline{$5.4$ V} voltage 
           amplitude sent to the LED.}
  \label{fig:ACQIRIS_LED_5.4}
\end{figure*}

\begin{figure*}[h!]
  \hspace{-1cm}
  \includegraphics[height=5cm,width=7.cm]
           {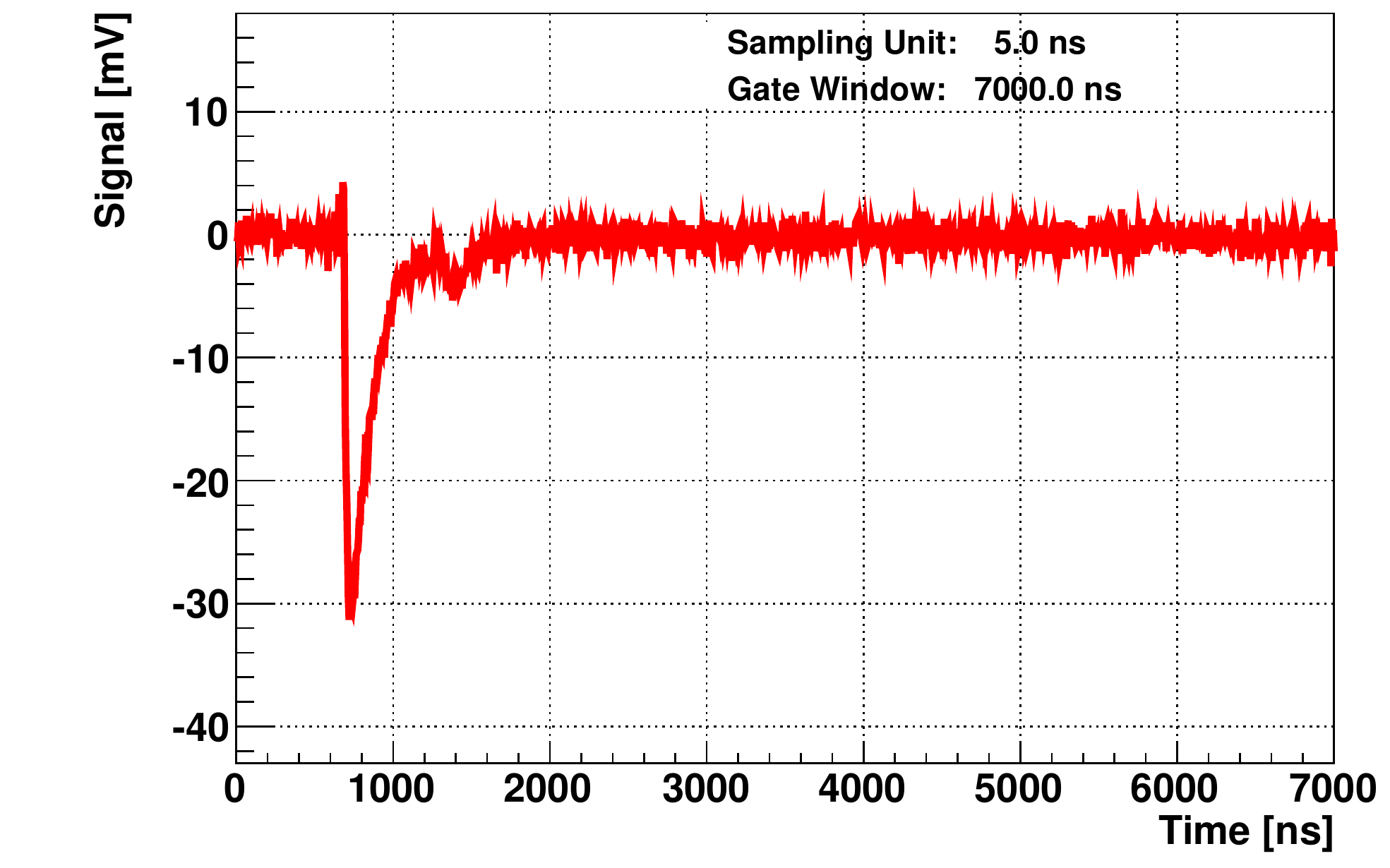}
  \includegraphics[height=5cm,width=7.cm]
           {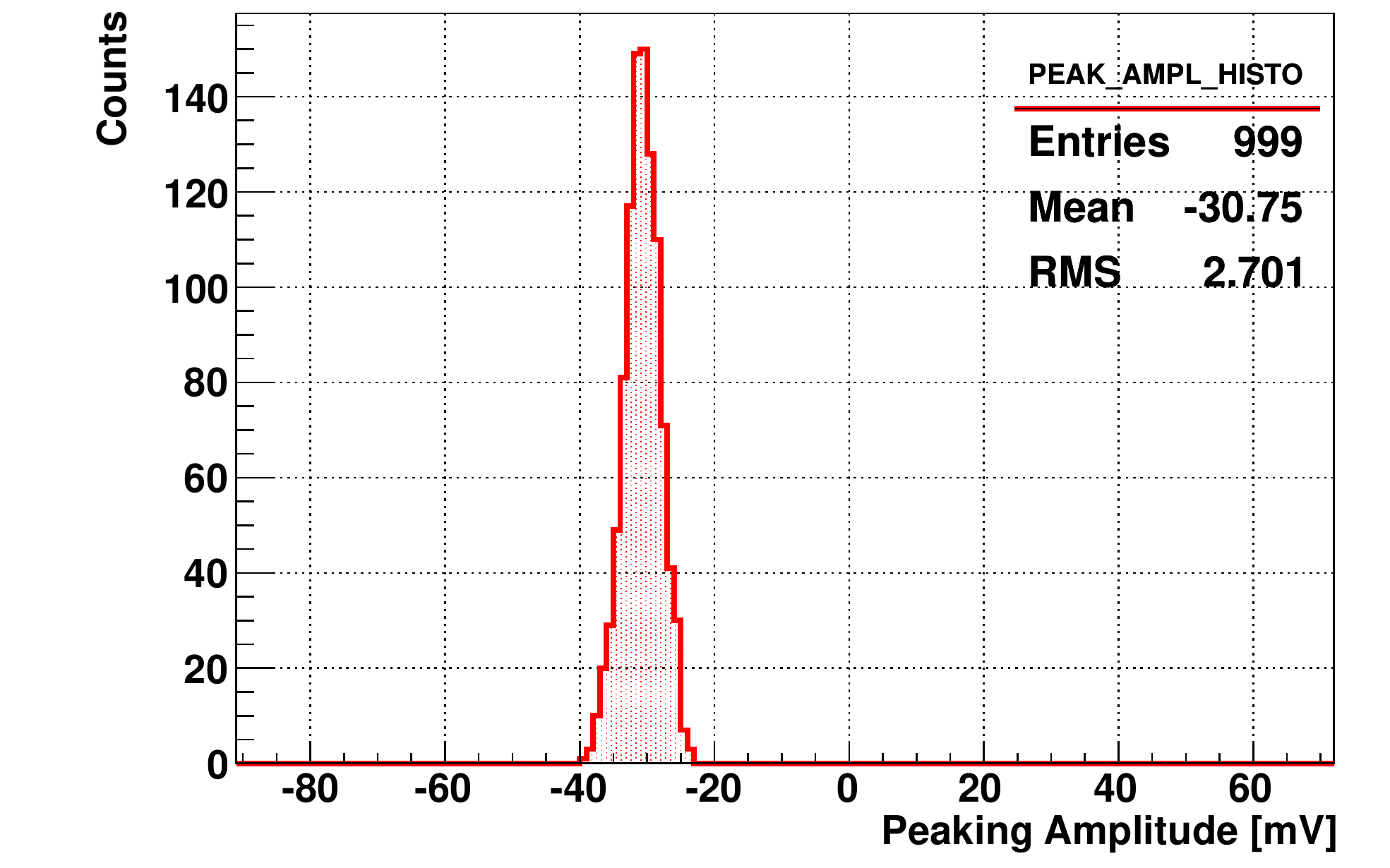}
  \vspace{-0.8cm}
  \caption{Waveform and peaking amplitude distribution measured with 
           \mbox{ACQIRIS} electronics for \underline{$6.3$ V} voltage 
           amplitude sent to the LED.}
  \label{fig:ACQIRIS_LED_6.3}
\end{figure*}

\begin{figure*}[h!]
  \hspace{-1.0cm}
  \includegraphics[height=5cm,width=7.cm]
           {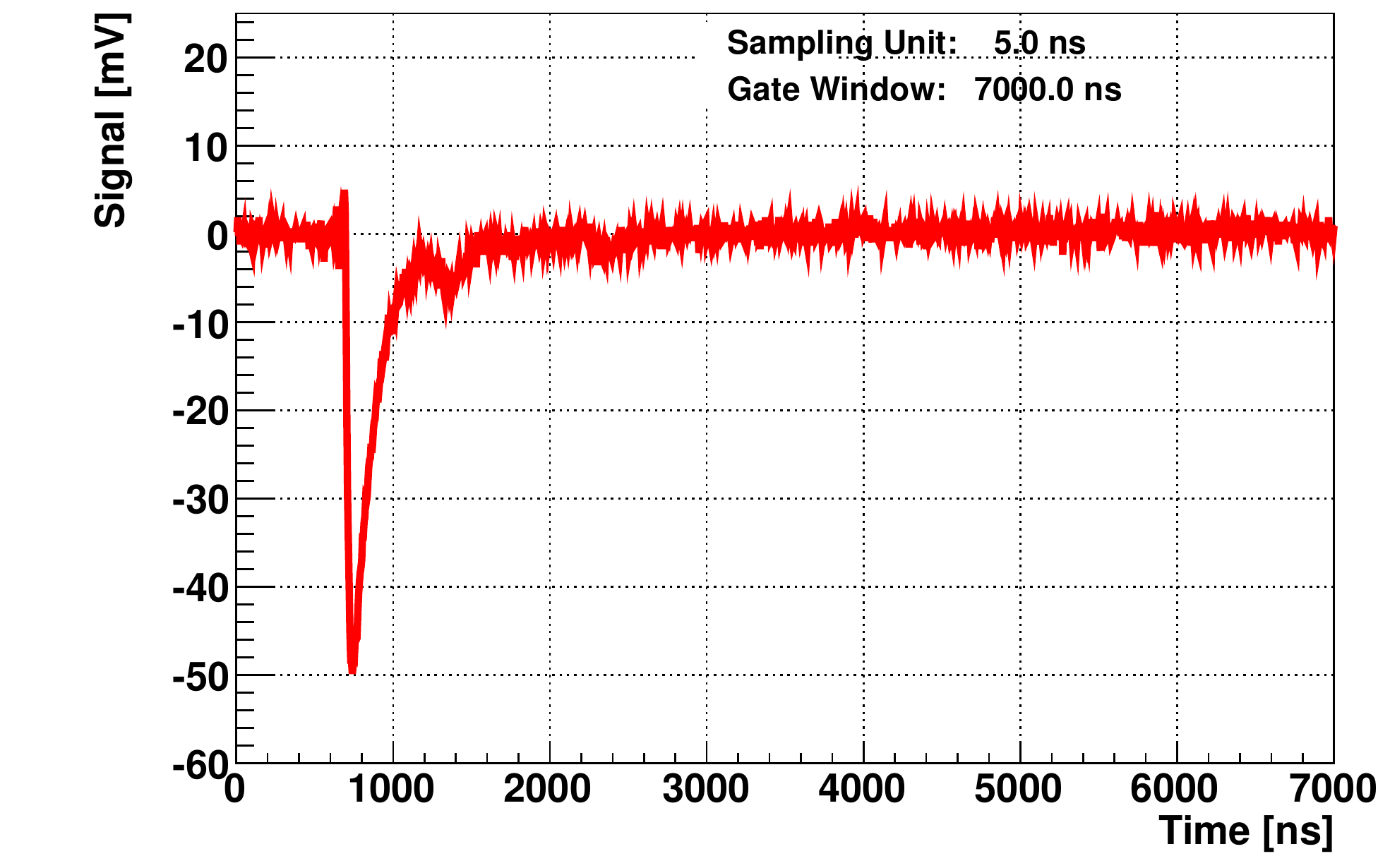}
  \includegraphics[height=5cm,width=7.cm]
           {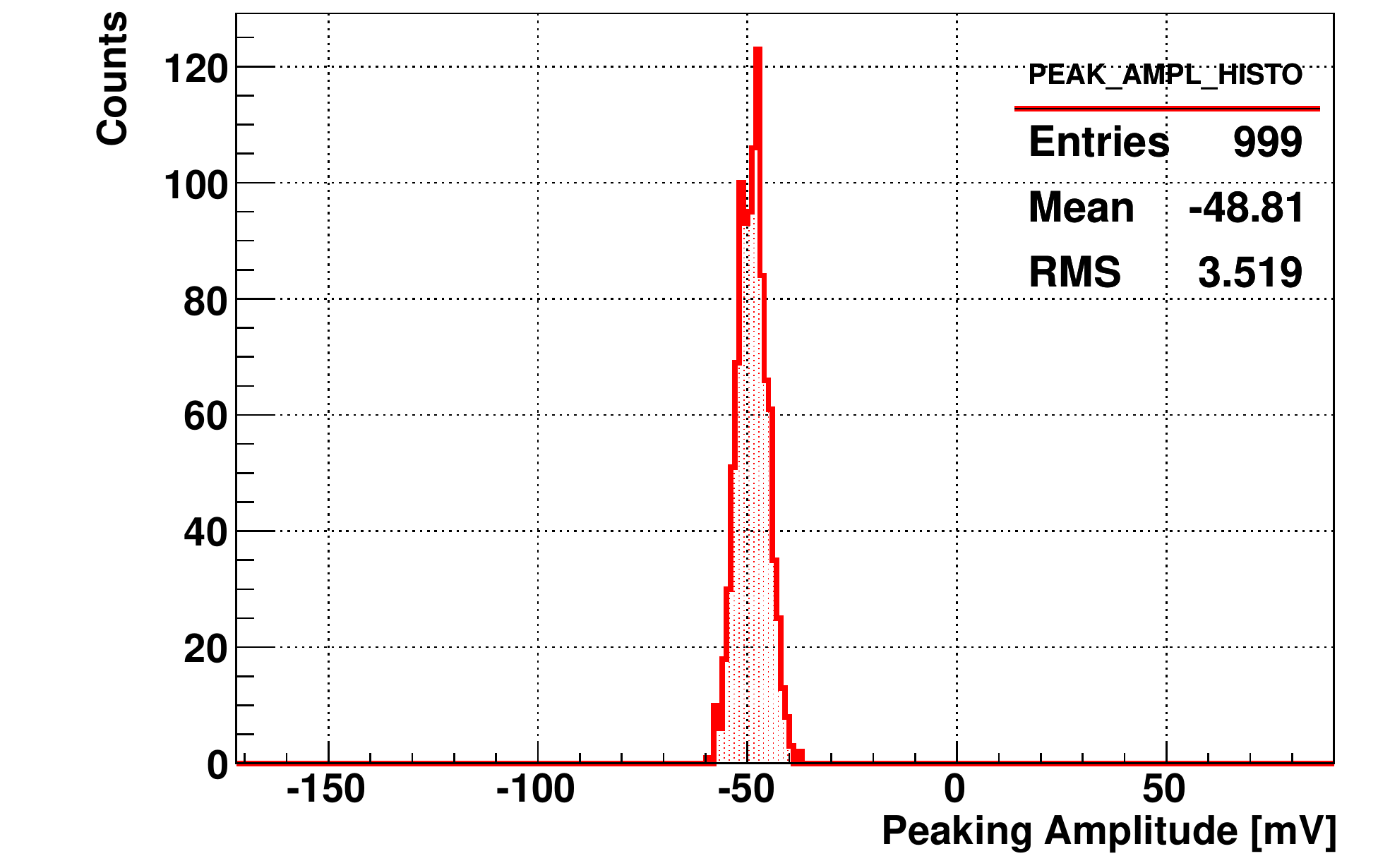}
  \vspace{-0.8cm}
  \caption{Waveform and peaking amplitude distribution measured with 
           \mbox{ACQIRIS} electronics for \underline{$7.2$ V} voltage 
           amplitude sent to the LED.}
  \label{fig:ACQIRIS_LED_7.2}
\end{figure*}

\begin{figure*}[h!]
  \hspace{-1.0cm}
  \includegraphics[height=5cm,width=7.cm]
           {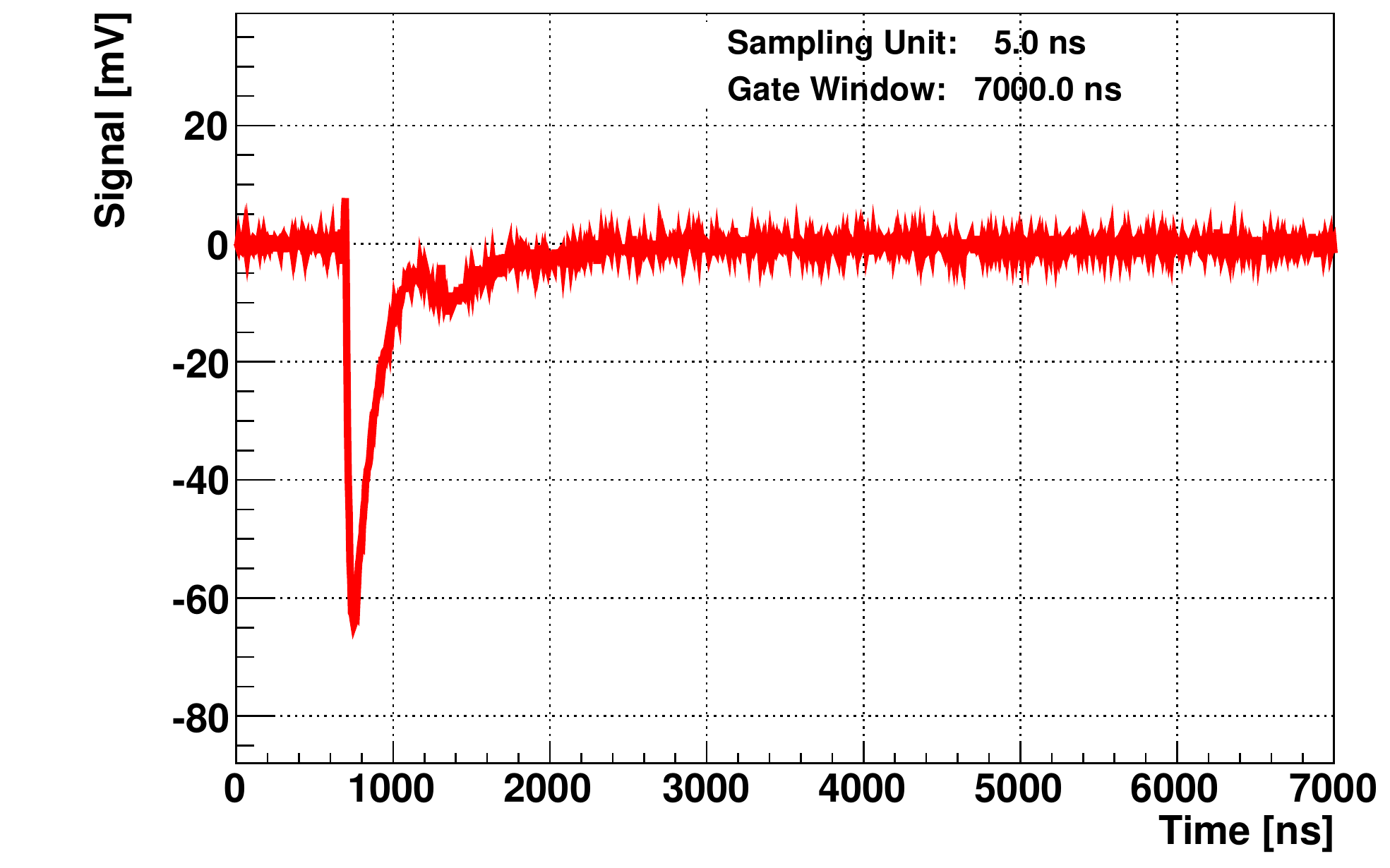}
  \includegraphics[height=5cm,width=7.cm]
           {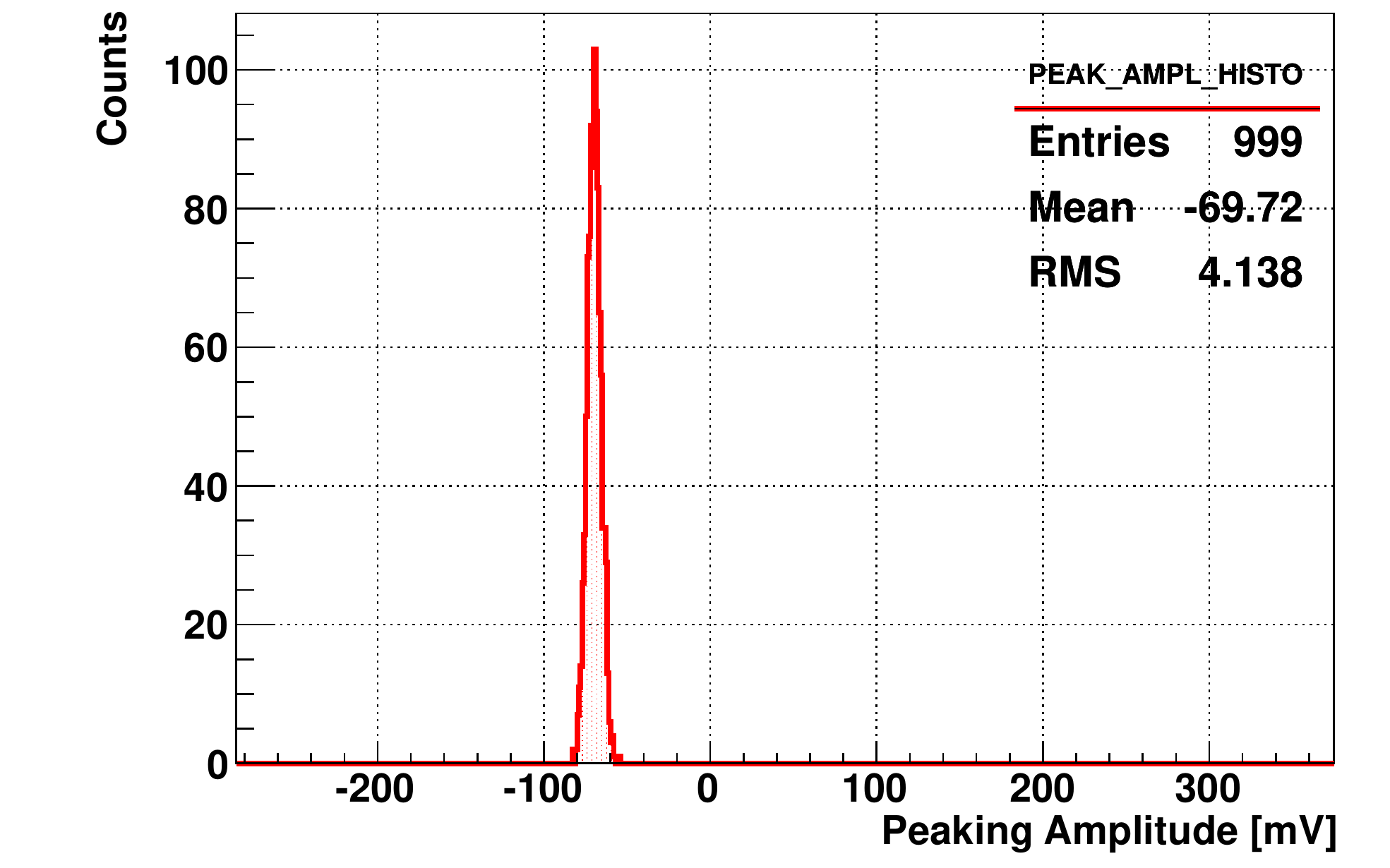}
  \vspace{-0.8cm}
  \caption{Waveform and peaking amplitude distribution measured with 
           \mbox{ACQIRIS} electronics for \underline{$8.1$ V} voltage 
           amplitude sent to the LED.}
  \label{fig:ACQIRIS_LED_8.1}
\end{figure*}

\begin{figure*}[h!]
  \hspace{-1.0cm}
  \includegraphics[height=5cm,width=7.cm]
           {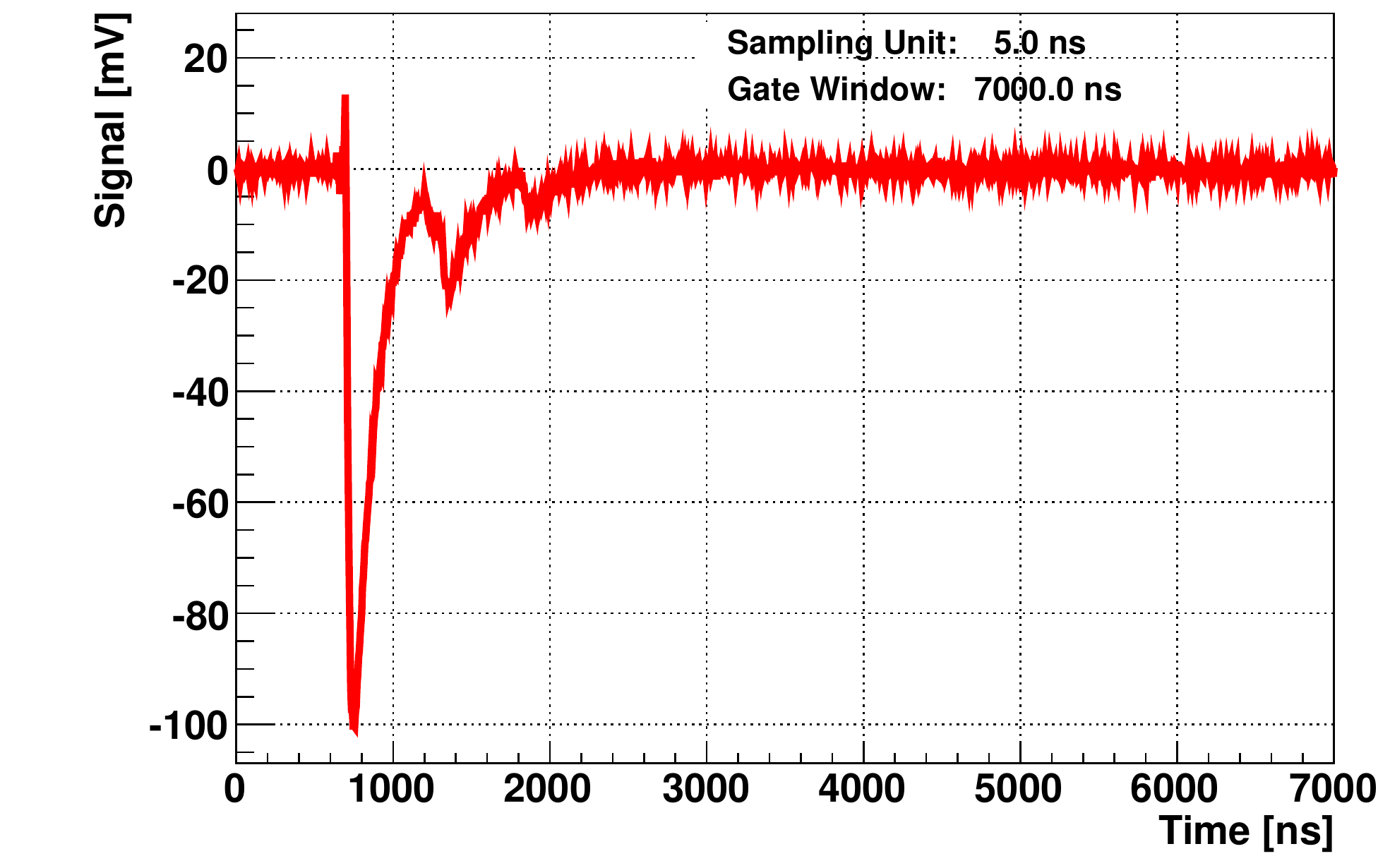}
  \includegraphics[height=5cm,width=7.cm]
           {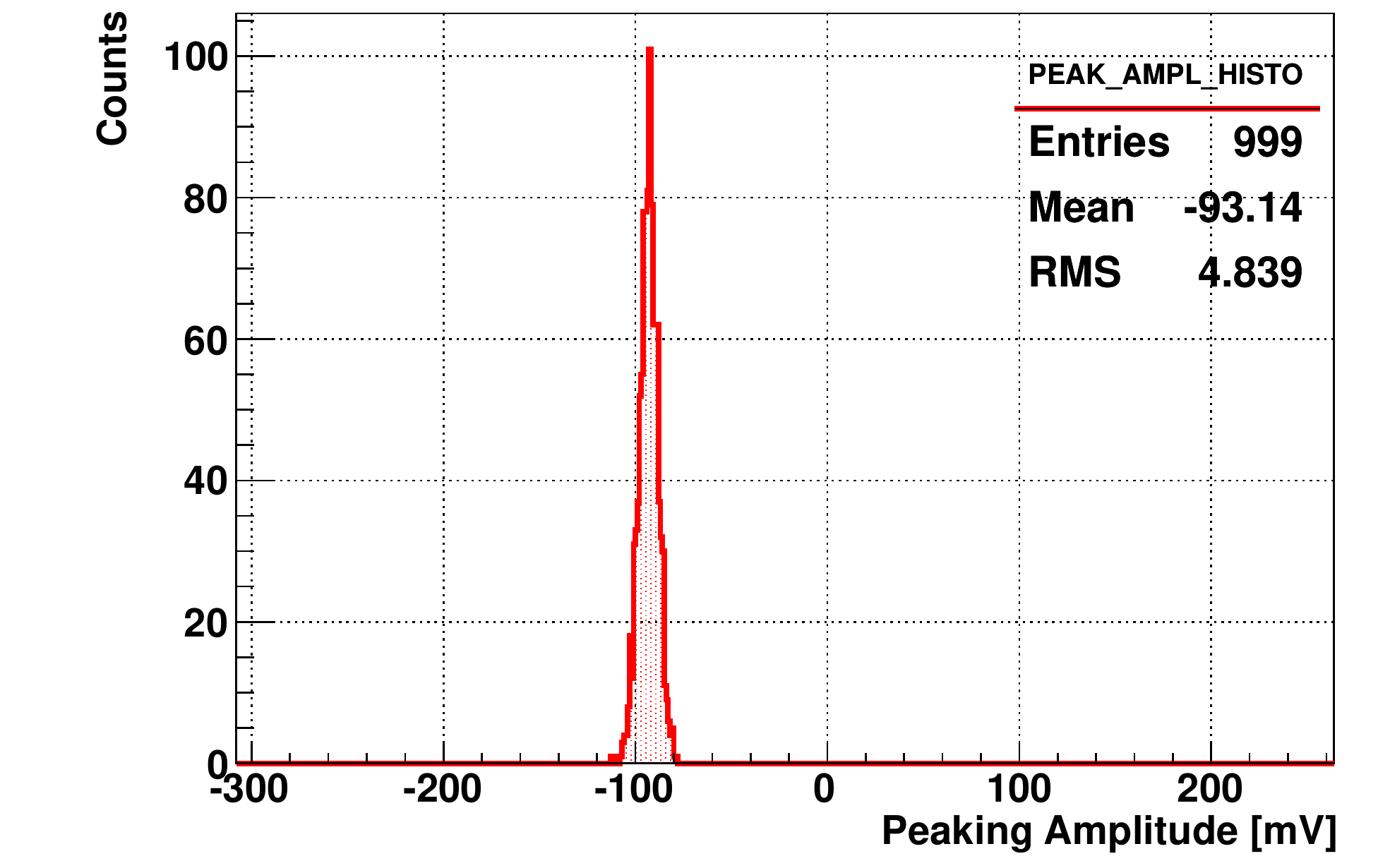}
  \vspace{-0.8cm}
  \caption{Waveform and peaking amplitude distribution measured with 
           \mbox{ACQIRIS} electronics for \underline{$9.0$ V} voltage 
           amplitude sent to the LED.}
  \label{fig:ACQIRIS_LED_9.0}
\end{figure*}

Within the similar experimental conditions, the same measurements 
were performed using the ACQIRIS readout electronics.
The peaking amplitude distributions and 
the typical waveforms (after pedestal subtraction) are presented
in the Fig.~\ref{fig:ACQIRIS_LED_5.4}~-~\ref{fig:ACQIRIS_LED_9.0}
for different values of the voltage input to the LED, and the results
are shown in Tab.~\ref{tab:ACQIRIS_SCAN_LED}.
\begin{table}[b!]
\hspace{-0.5cm}
\begin{tabular}{||c|c|c||}
\hline 
\hline
\multicolumn{2}{|c|}{ \textbf{ACQIRIS System: LED Intensity Scan} }\\
\hline
\hline
   \bf{Voltage Amplitude to LED [V]:} &\bf{Measured Peaking Amplitude [mV]:}\\
\hline 
\hline 
    5,4 &    -18 \\
    6.3 &    -31 \\
    7.2 &    -49 \\
    8.1 &    -70 \\
    9.0 &    -93 \\
\hline
\hline
\end{tabular}
\caption{Scan performed with the ACQIRIS readout system increasing 
         the LED intensity by changing the amplitude of the voltage 
         pulse sent to the LED.}
\label{tab:ACQIRIS_SCAN_LED}
\end{table}

\begin{figure}[t!]
 \begin{center}
 \includegraphics[height=7.5cm, width=9.5cm]{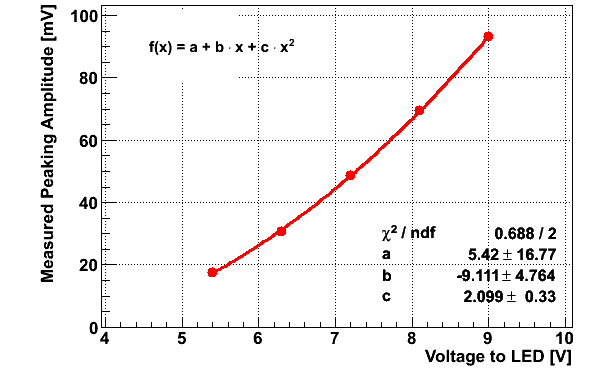}
  \vspace{-0.5cm}
  \caption{In the ACQIRIS system the peaking amplitude of the 
           incoming waveforms
           was measured for different values of the amplitude for the voltage
           pulse sent to the LED. It is evident the non-linearity 
           of the LED light output.}
  \label{fig:LED_NON_LINEARITY}
  \end{center}
\end{figure}

It is interesting to observe, as above mentioned, the non-linear 
LED intensity with respect to the amplitude of the input voltage pulse,
Fig.~\ref{fig:LED_NON_LINEARITY}. 
This is the reason why one cannot use directly the setting of the 
voltage amplitude to calibrate the output of the photomultiplier. 

\subsection{Linearity of the Experimental Setup System}
\label{sec:WAVEFORMS_LINEARITY}
\begin{figure}[t!]
  \vspace{-0.5cm}
  \begin{center}
 \includegraphics[height=7.5cm, width=9.5cm]{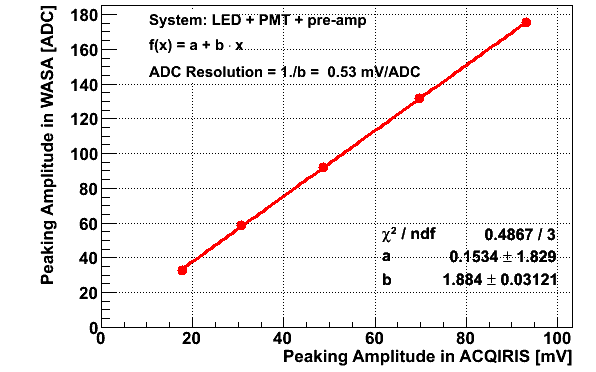}
  \vspace{-0.5cm}
  \caption{The linearity and the resolution of the WASA ADC were
           investigated by using the peaking amplitude values 
           obtained by both readout systems.}
  \label{fig:LED_PMT_SYSTEM_LINEARITY}
  \end{center}
\end{figure}

By comparing the results of the peaking amplitude obtained 
with both readout systems in the same experimental setup configuration
the linearity of the WASA system can be investigated
in the system modified with the frequency filtering stage activated. 
The results are presented in Fig.~\ref{fig:LED_PMT_SYSTEM_LINEARITY}, 
showing a very good linearity 
of the WASA ADC in the investigated region. Note that the WASA 
data have the pedestal (in this configuration at approximately 
$1880$ ADC units) subtracted.

Out of the linear fit the ADC resolution can be also 
again calculated, resulting in $0.53$ mV per ADC. 
This value is slightly different from what extracted 
in the previous measurement with the voltage pulse generator 
and the high-pass filter, possibly indicating a systematical 
uncertainty of few percent present in either the analysis
chain, or in the hardware configuration, or in both. 
%

\subsection{Waveform Analysis: WASA vs \mbox{ACQIRIS} Comparison}
\label{sec:WAVEFORMS_LED}
By using the ADC resolution found in this experimental configuration 
the charge injected in the WASA system can be integrated, converted in 
Coulomb units and finally compared with the results obtained using the 
ACQIRIS electronics for the five chosen values of the voltage pulse 
amplitude.

\begin{figure*}[b!]
  \hspace{-1cm}
  \includegraphics[height=5cm,width=7.cm] {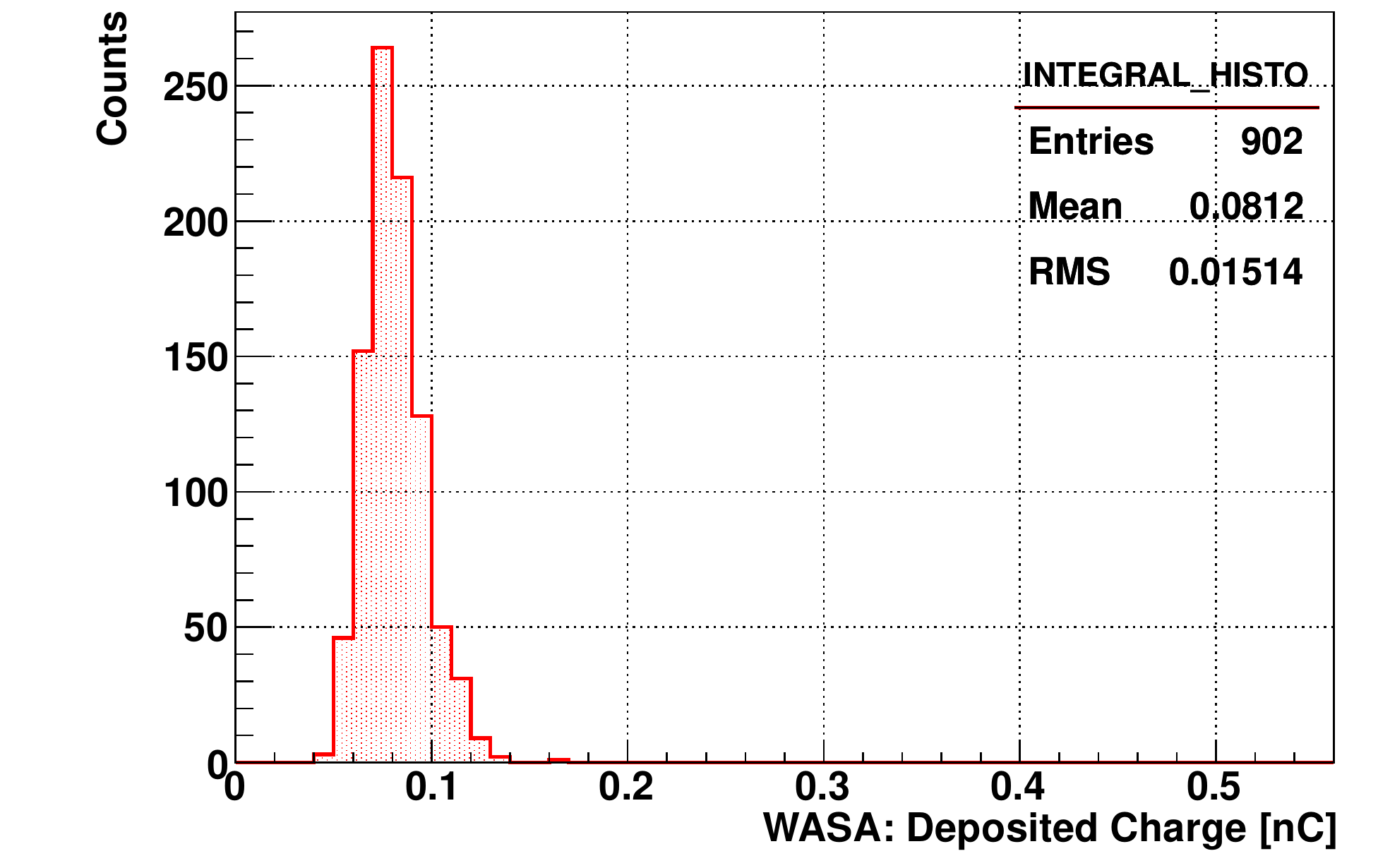}
  \includegraphics[height=5cm,width=7.cm] {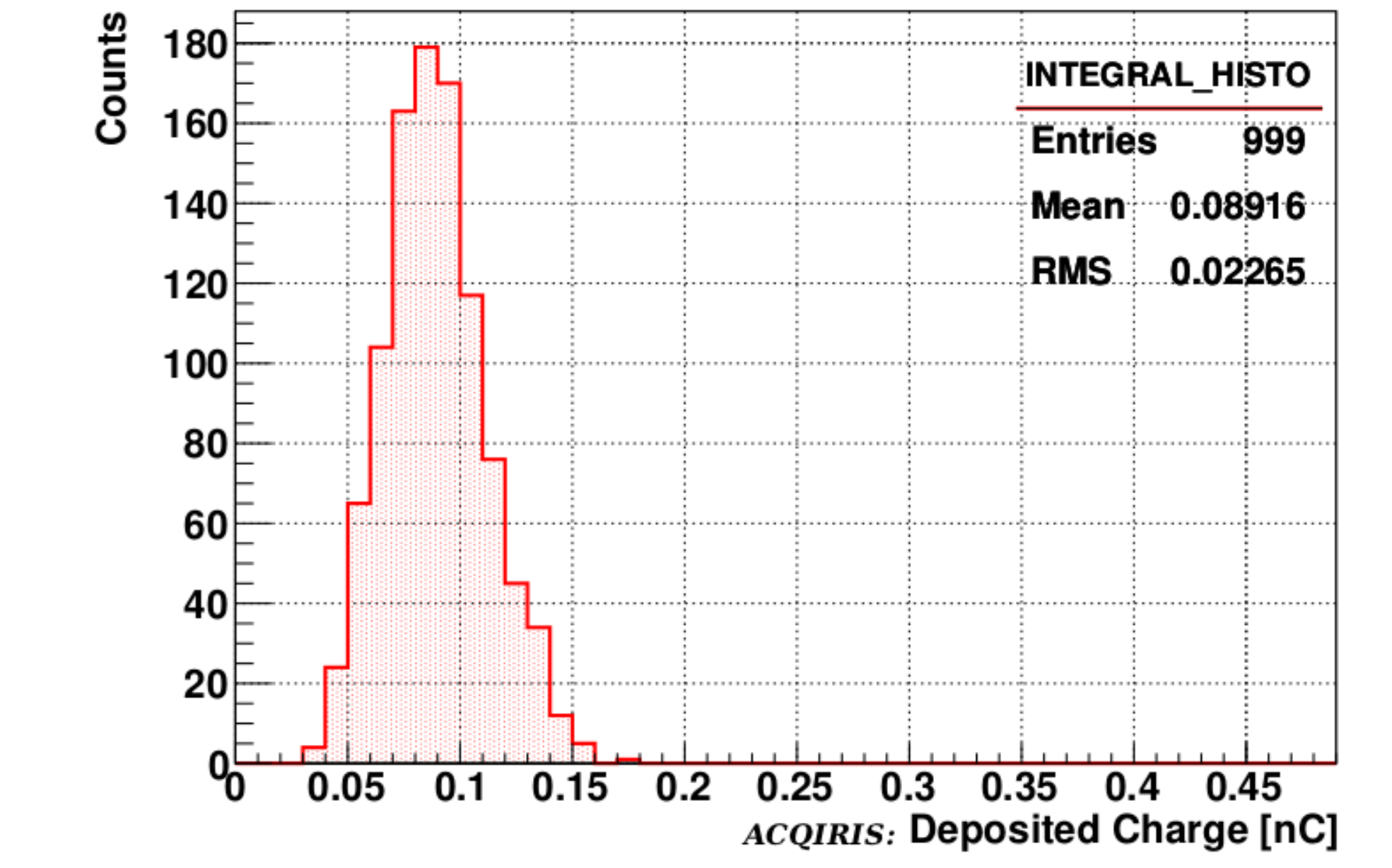}
  \vspace{-0.8cm}
  \caption{Integrated charge measured with both the 
           WASA and the \mbox{ACQIRIS} 
           for \underline{$5.4$ V} voltage 
           amplitude sent to the LED.}
  \label{fig:COMPARE_CHARGE_LED_5.4}
\end{figure*}

\begin{figure*}[h!]
  \hspace{-1cm}
  \includegraphics[height=5cm,width=7.cm]
           {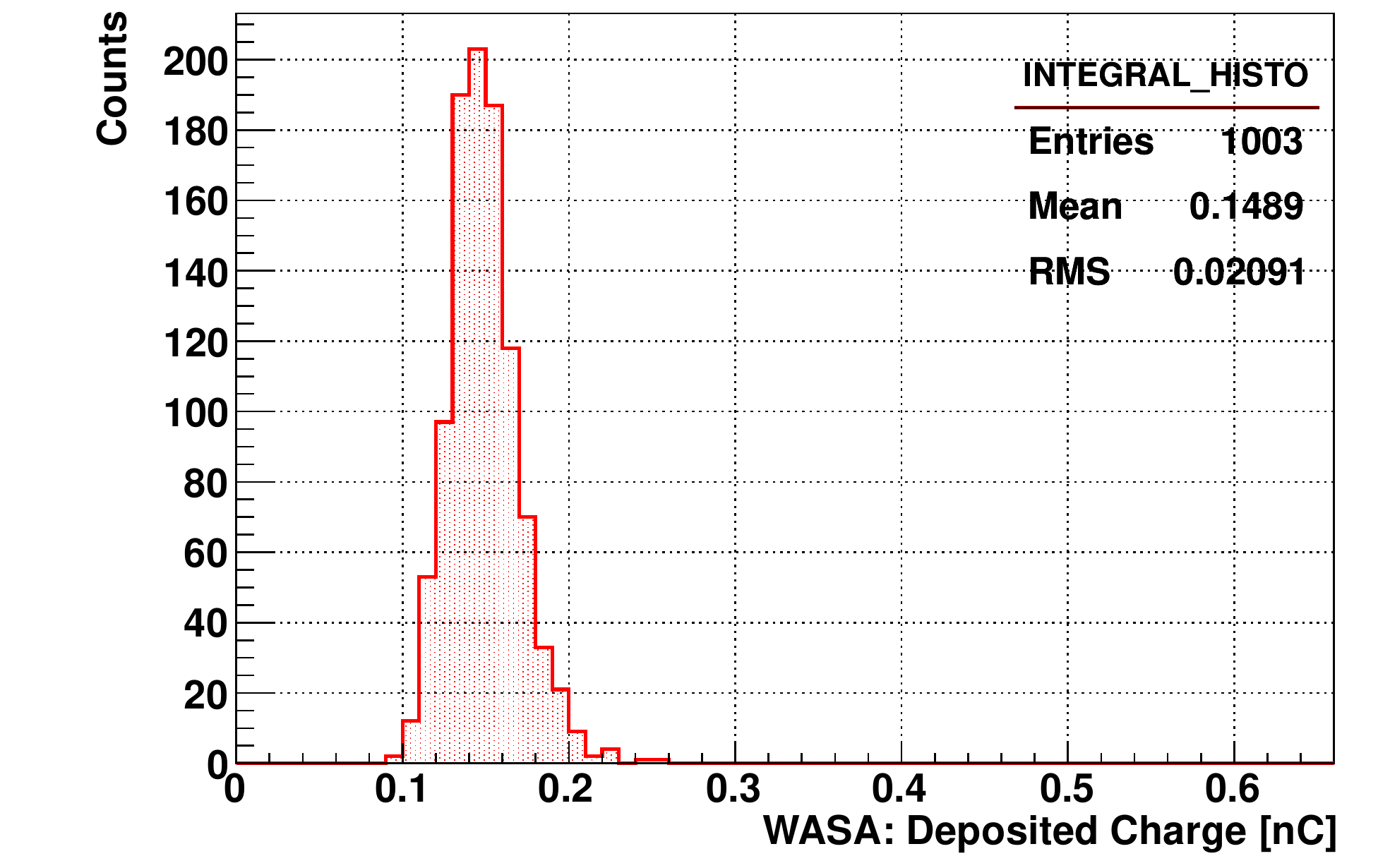}
  \includegraphics[height=5cm,width=7.cm]
           {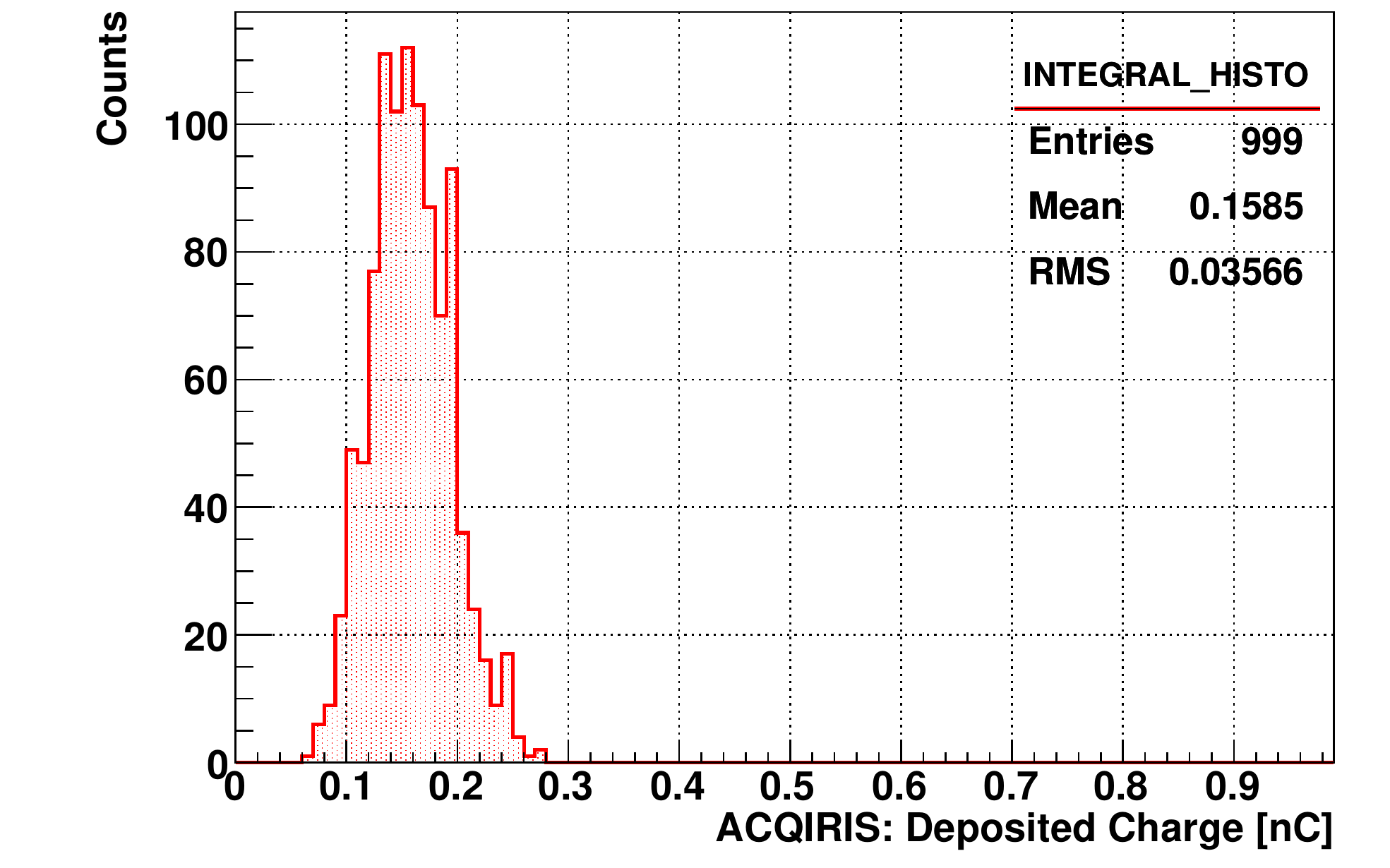}
  \vspace{-0.8cm}
  \caption{Integrated charge measured with both the 
           WASA and the \mbox{ACQIRIS} 
           for \underline{$6.3$ V} voltage 
           amplitude sent to the LED.}
  \label{fig:COMPARE_CHARGE_LED_6.3}
\end{figure*}

\begin{figure*}[h!]
  \hspace{-1.0cm}
  \includegraphics[height=5cm,width=7.cm]
           {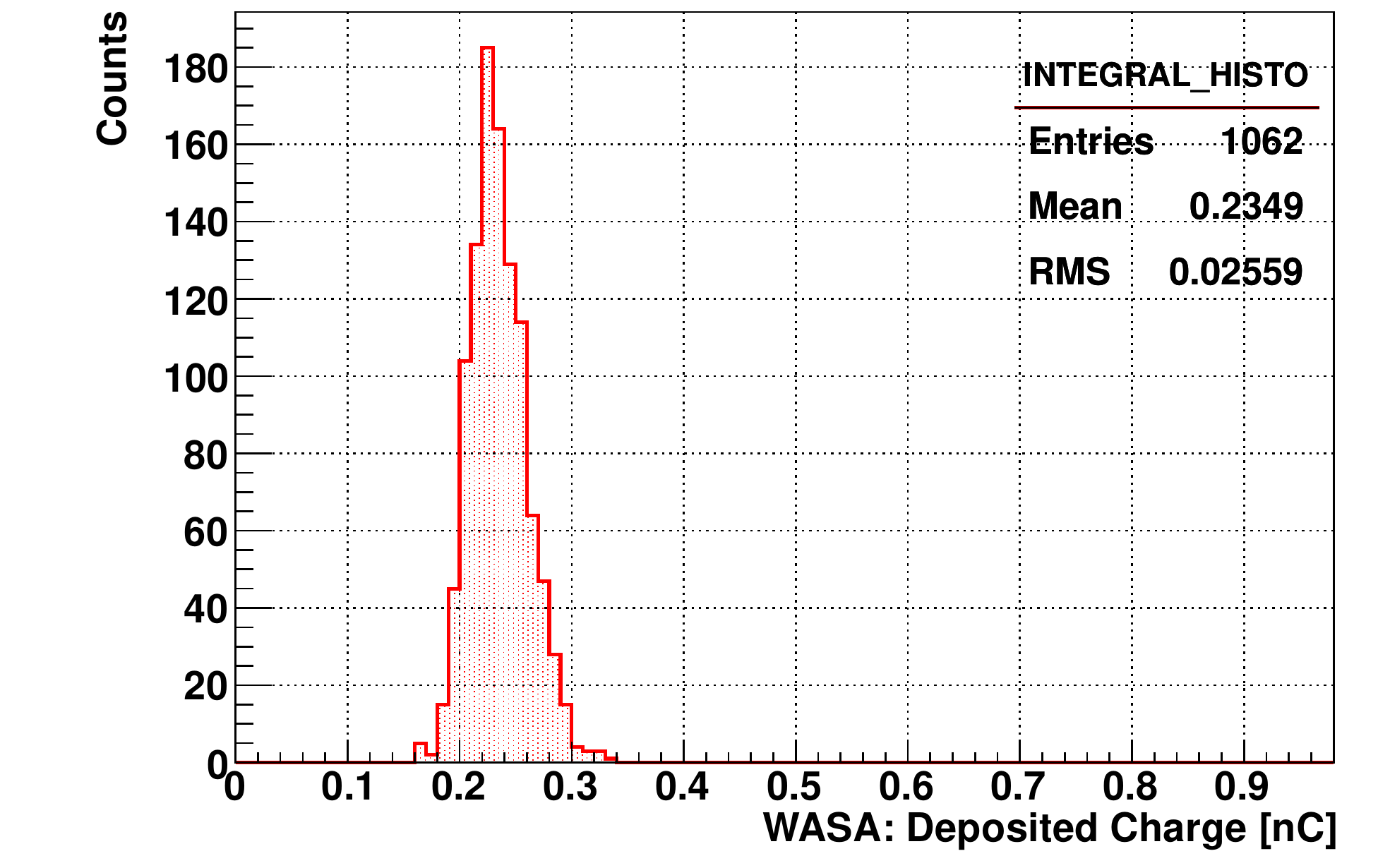}
   \includegraphics[height=5cm,width=7.cm]
           {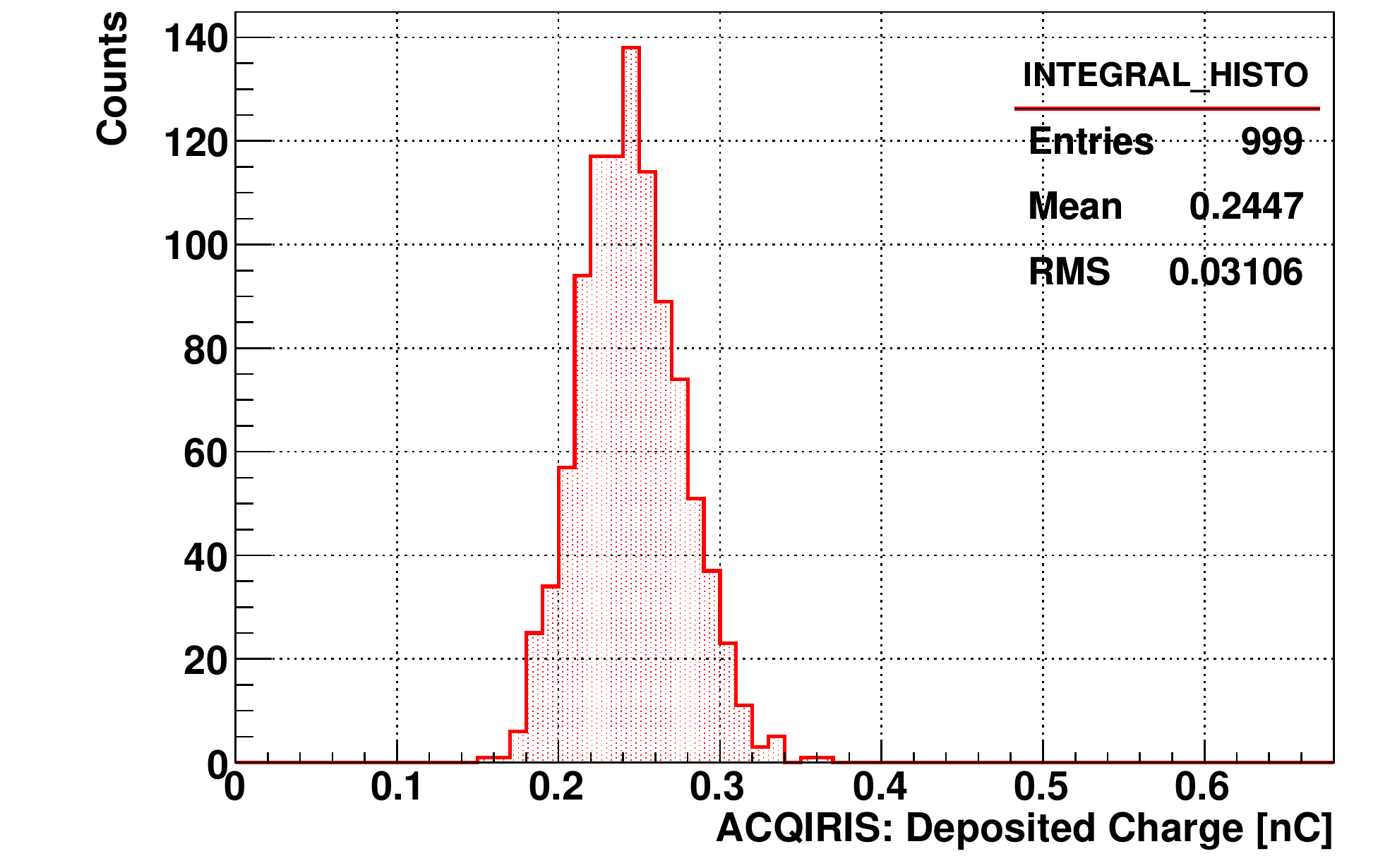}
   \vspace{-0.8cm}
  \caption{Integrated charge measured with both the 
           WASA and the \mbox{ACQIRIS} 
           for \underline{$7.2$ V} voltage 
           amplitude sent to the LED.}
  \label{fig:COMPARE_CHARGE_LED_7.2}
\end{figure*}

\begin{figure*}[h!]
  \hspace{-1.0cm}
  \includegraphics[height=5cm,width=7.cm]
           {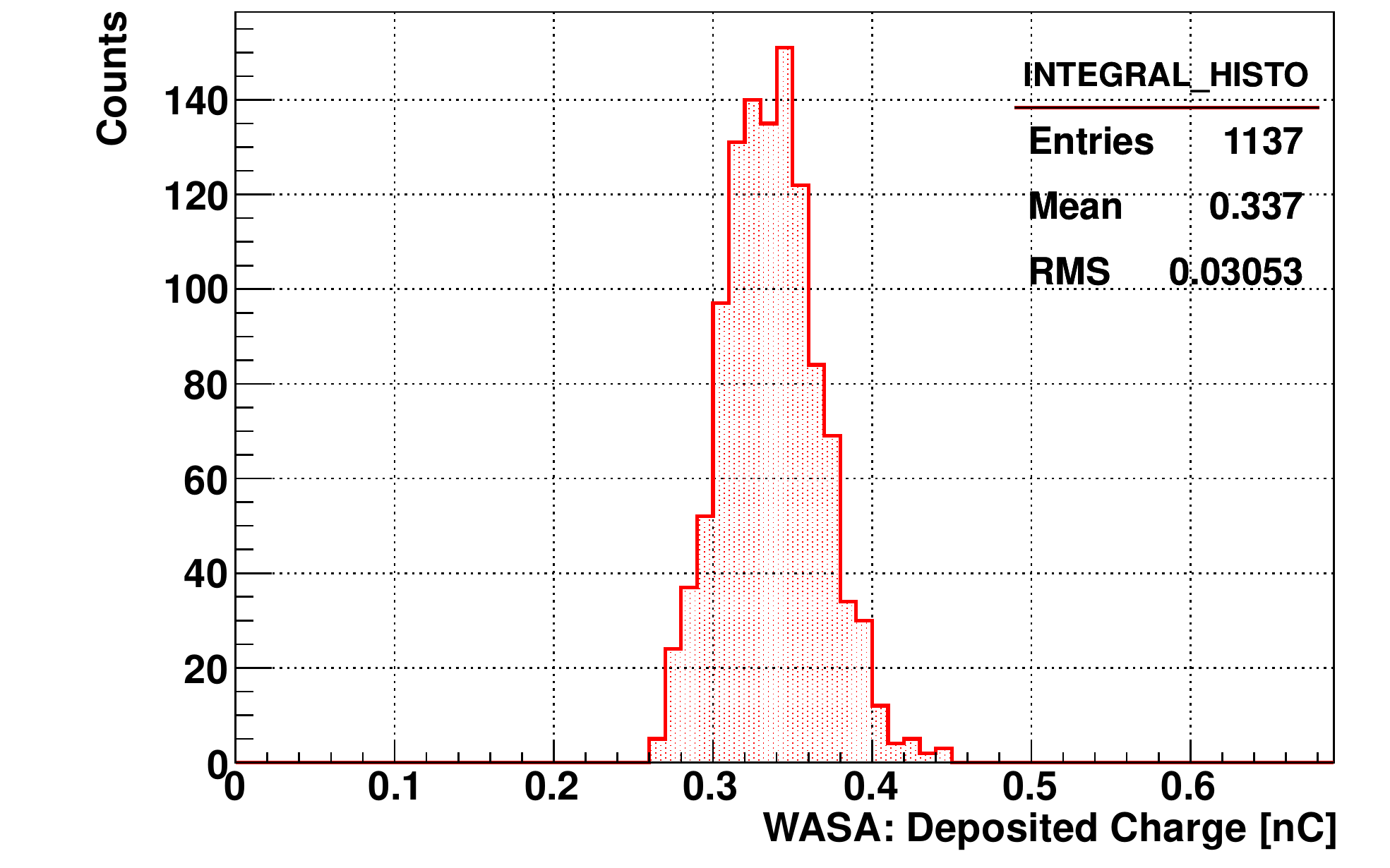}
  \includegraphics[height=5cm,width=7.cm]
           {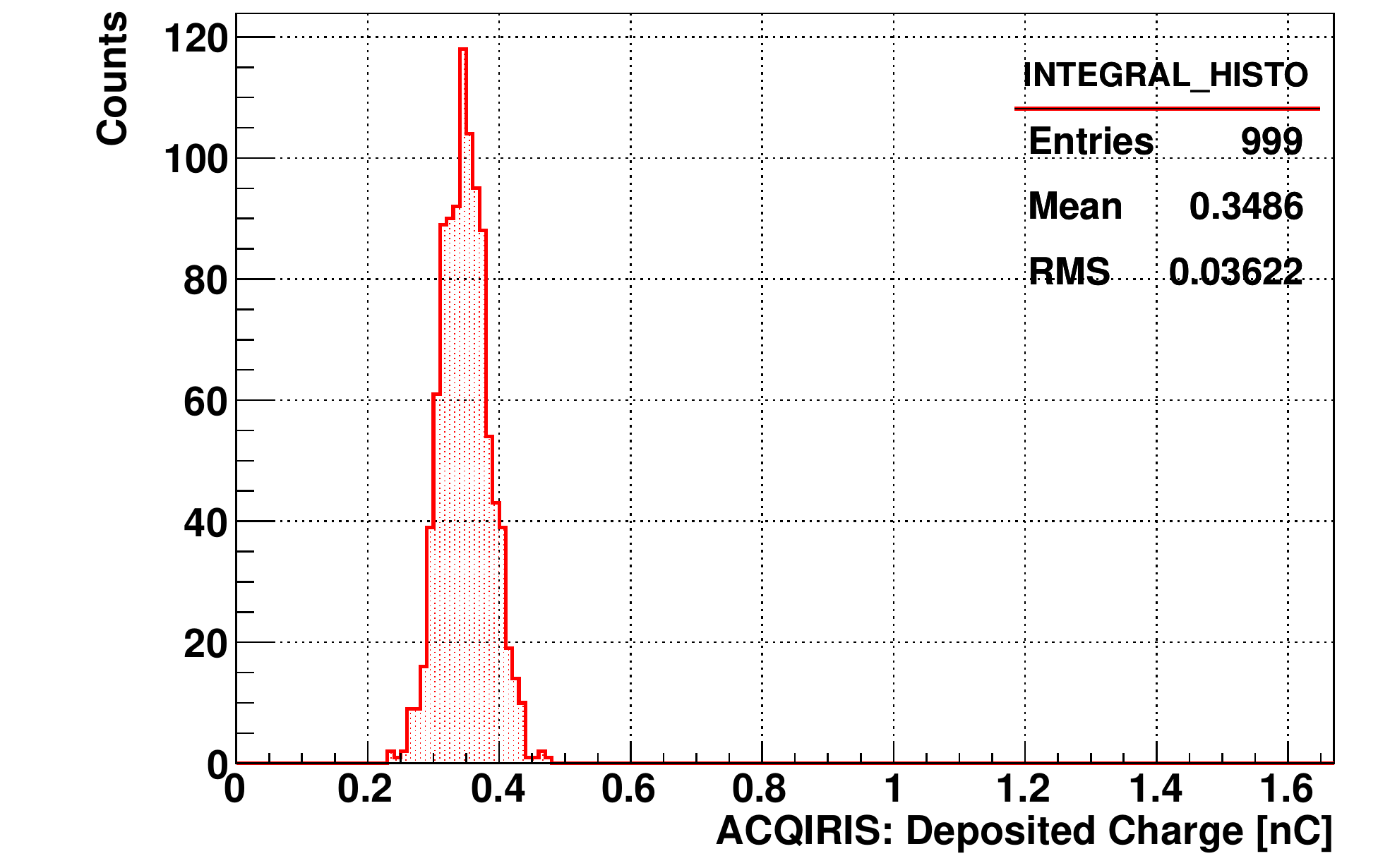}
  \vspace{-0.8cm}
  \caption{Integrated charge measured with both the 
           WASA and the \mbox{ACQIRIS}
           for \underline{$8.1$ V} voltage 
           amplitude sent to the LED.}
  \label{fig:COMPARE_CHARGE_LED_8.1}
\end{figure*}

\begin{figure*}[h!]
  \hspace{-1.0cm}
  \includegraphics[height=5cm,width=7.cm]
           {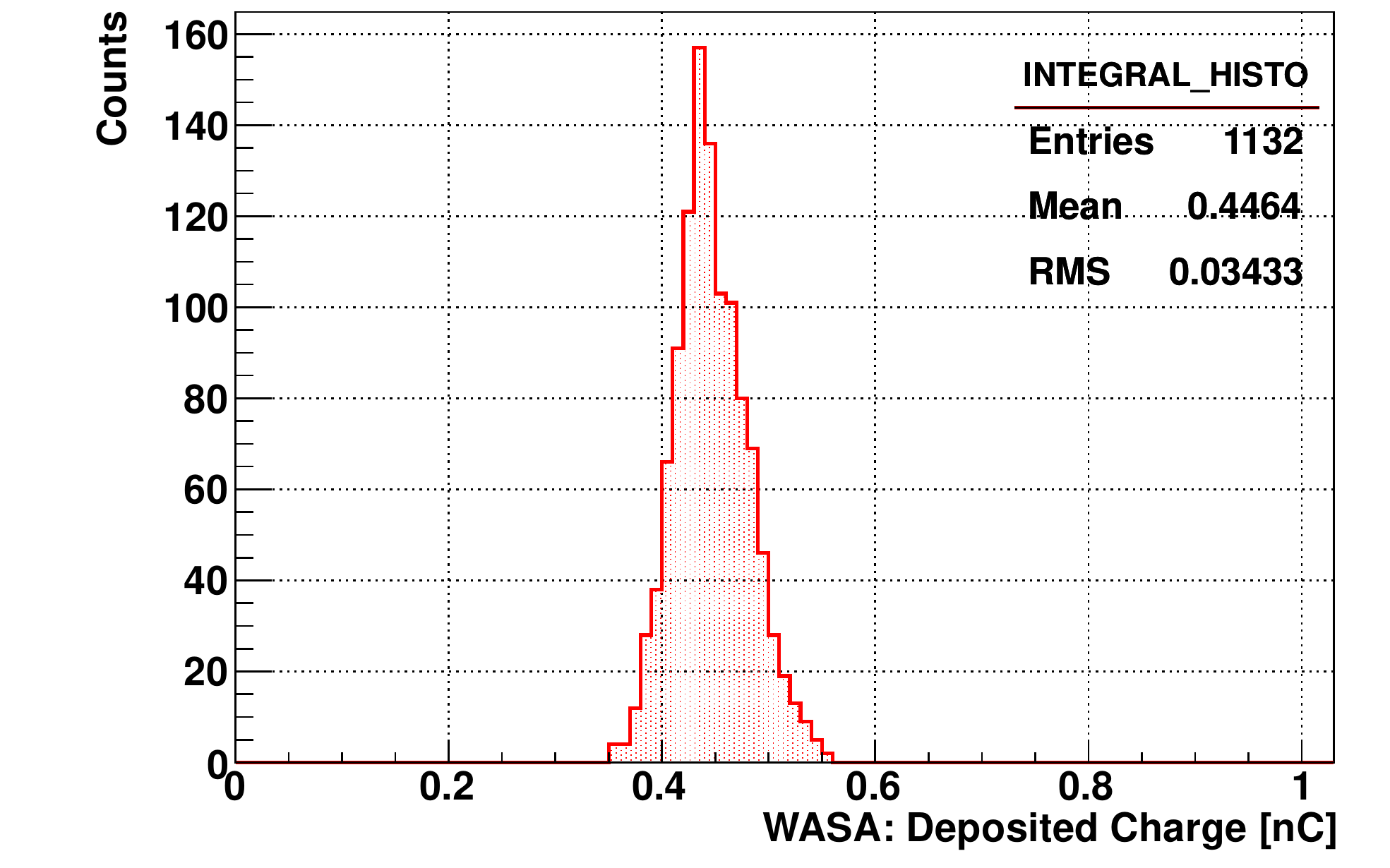}
  \includegraphics[height=5cm,width=7.cm]
           {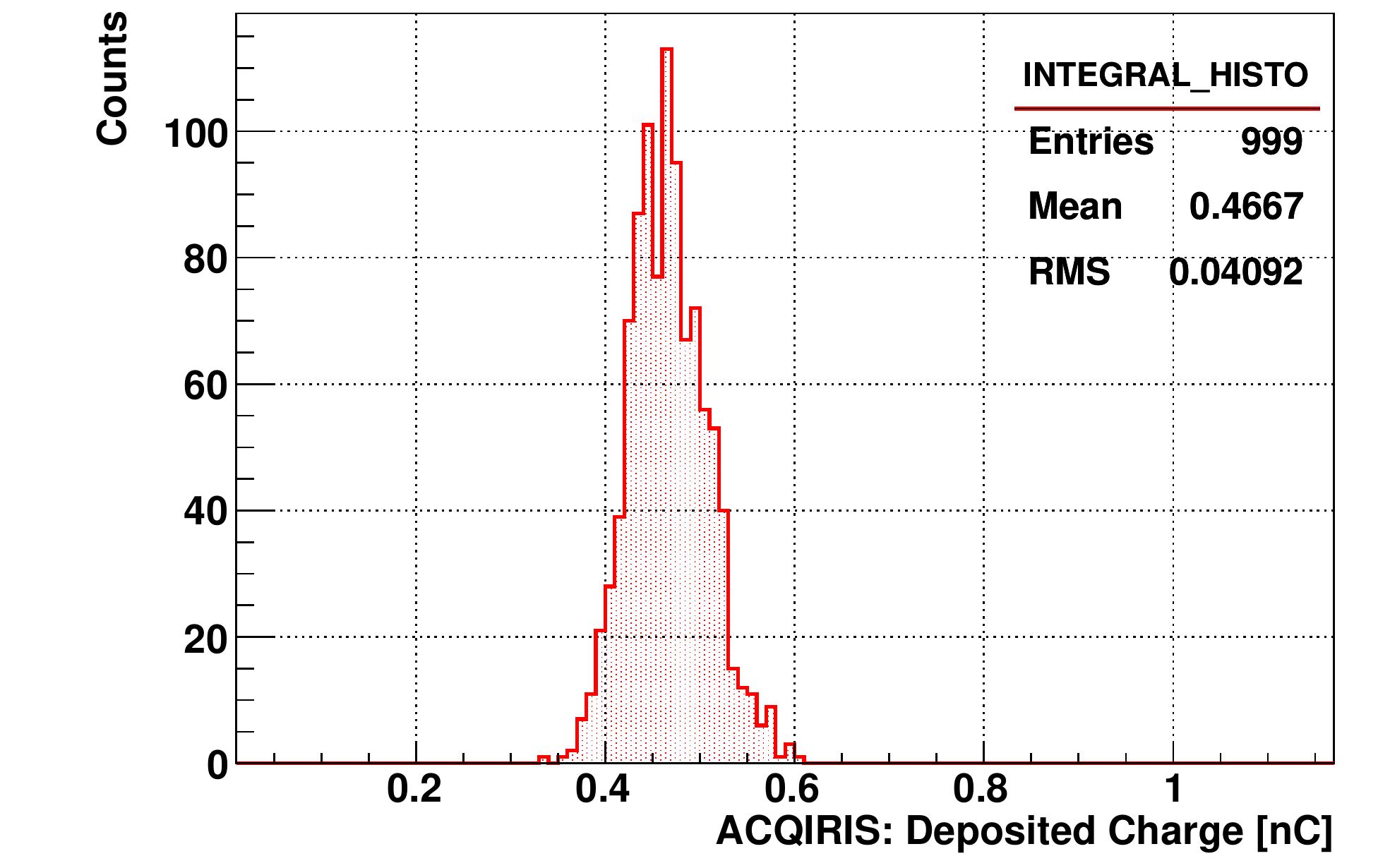}
  \vspace{-0.8cm}
  \caption{Integrated charge measured with both the 
           WASA and the \mbox{ACQIRIS} 
           for \underline{$9.0$ V} voltage 
           amplitude sent to the LED.}
  \label{fig:COMPARE_CHARGE_LED_9.0}
\end{figure*}

The comparison of the integrated charge measured by the two systems is 
presented in 
Fig.~\ref{fig:COMPARE_CHARGE_LED_5.4}~-~\ref{fig:COMPARE_CHARGE_LED_9.0} 
for the values 
$5.4$, $6.3$, $7.2$, $8.1$ and $9.0$ V, respectively, of the amplitude of the 
voltage pulse sent to the LED. 

For all experimental setups the integrated charge appears
to be systematically slightly larger for the ACQIRIS system
than what measured by the WASA system.
The charge is quite small within the used setup of the 
measurements, possibly biasing (in the calculation stage) 
the evaluation of the 
relative difference at very small charge values (up to $8\%$);
in principle this could be avoided running the PMT at higher 
gain. 
At large charge values instead the relative difference is 
approximately $4\%$ stable.
Note that the statistical uncertainty of the extracted 
means is at the per-mill level, in principle small enough  
not to bias the comparison of the measurements by the 
two systems. 

\section{Design of a new Transimpedence Amplifier:\\
Analysis with the Voltage Pulse Generator}

\subsection{Transimpedence Amplifier Description}
%
In order to make the data processing faster (e.g. with respect
to the WASA differential ADC driver) a new transimpedence amplifier 
has been designed and build at ZEA-2 by J. Heggen. 
The test-bench configuration is similar to the setup presented 
in Fig.~\ref{fig:TestBench_Pulser} with the introduction of the 
amplifier in between the SUCOBOX and the readout WASA/ACQIRIS module. 

\begin{figure*}[h!]
   \hspace{-1cm}
      \includegraphics[height=5.cm,width=14.cm]{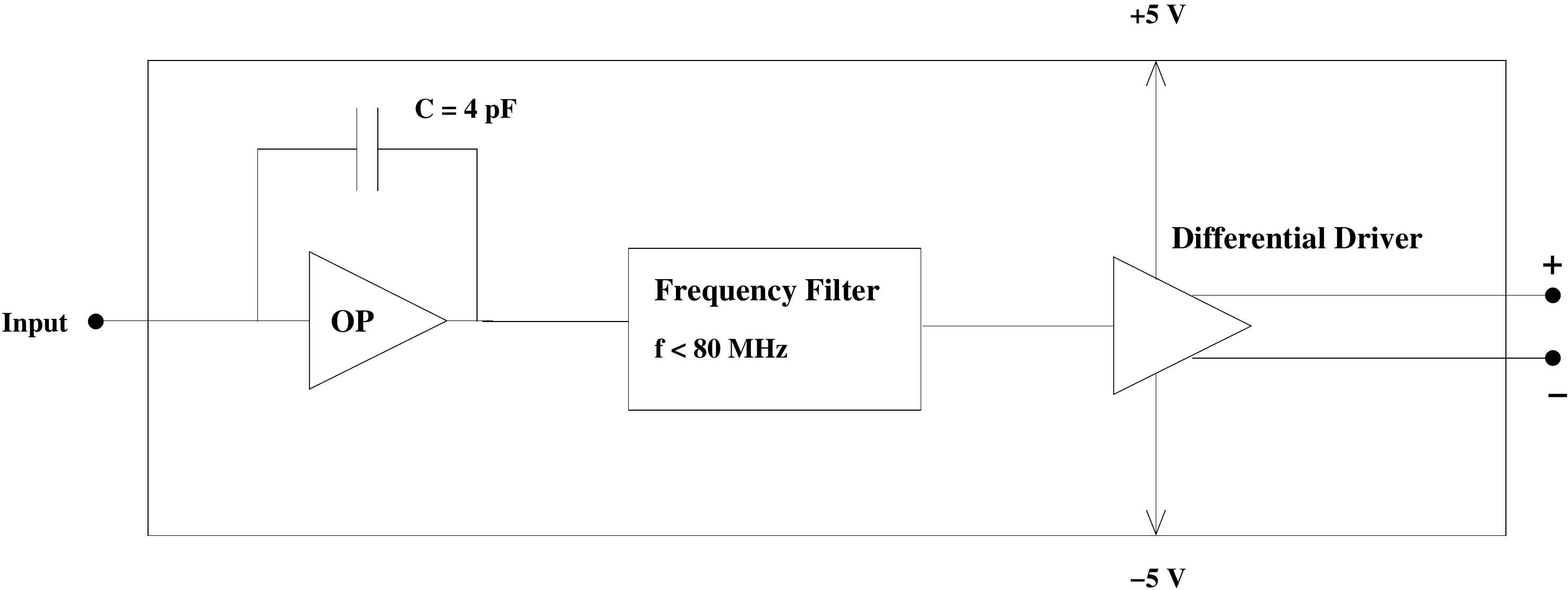}\\
  \vspace{-0.8cm}
  \caption{Schematic of the developed amplifier made of, in sequence,
           the transimpedence in parallel with a very low capacitance,
           a frequency filter and a differential driver.}
  \label{fig:TRANSIMPEDENCE}
\end{figure*}

The schematic of the transimpedance amplifier is presented in 
Fig.~\ref{fig:TRANSIMPEDENCE}. 
The system is made of three stages, an operational amplifier (OP) 
in parallel with a very small
capacitance such to minimize the charge integration and its characteristic 
time, thus fastening the signal processing. A frequency filters up to 
\mbox{$80$ MHz} follows, and feeds a differential driver which performs the
single-ended-to-differential conversion. 

The bipolar output should be directly connected to the ADC converter.
The use of a differential signal cable could help in reducing the signal 
distortion due to external noise sources, expecially when the device
is located close to the current source, as the photomultiplier.
%

\subsection{Measurements with WASA System:\\ No Frequency Filter}
\begin{figure}[h!]
\hspace{-1.5cm} 
      \includegraphics[height=5cm,width=7.cm]{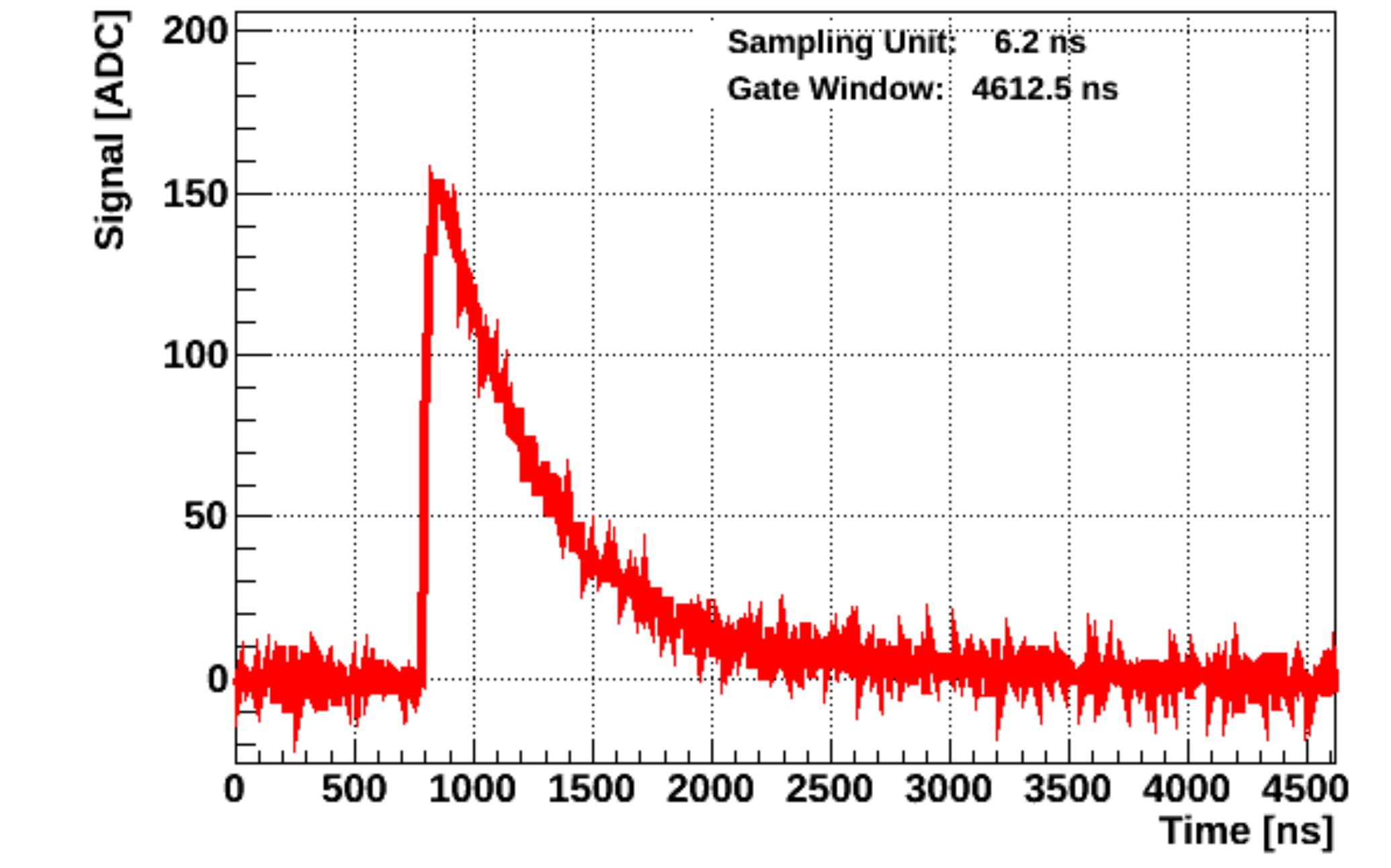}
      \includegraphics[height=5cm,width=7.cm]{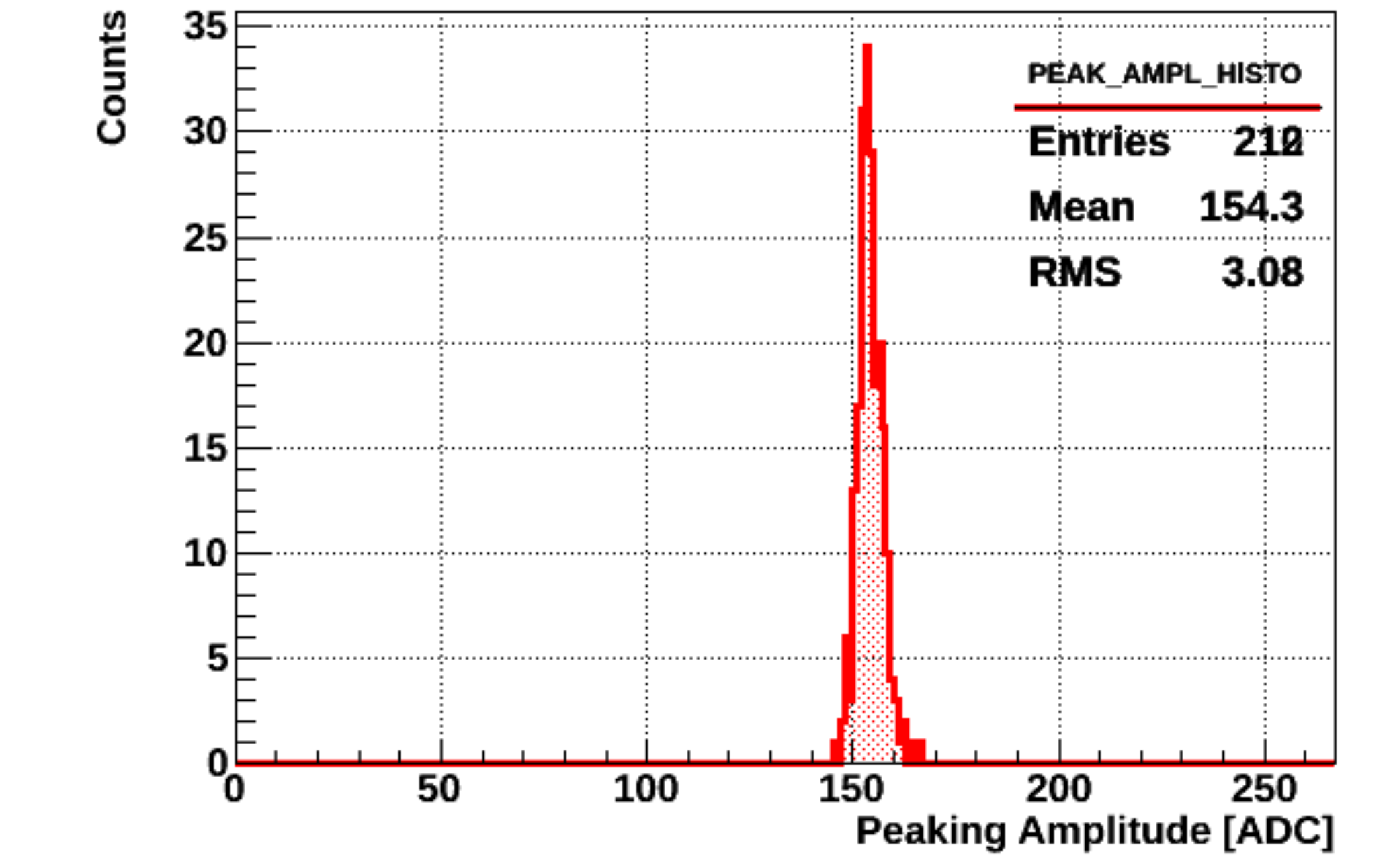}\\

\hspace{-1.5cm} 
      \includegraphics[height=5cm,width=7.cm]{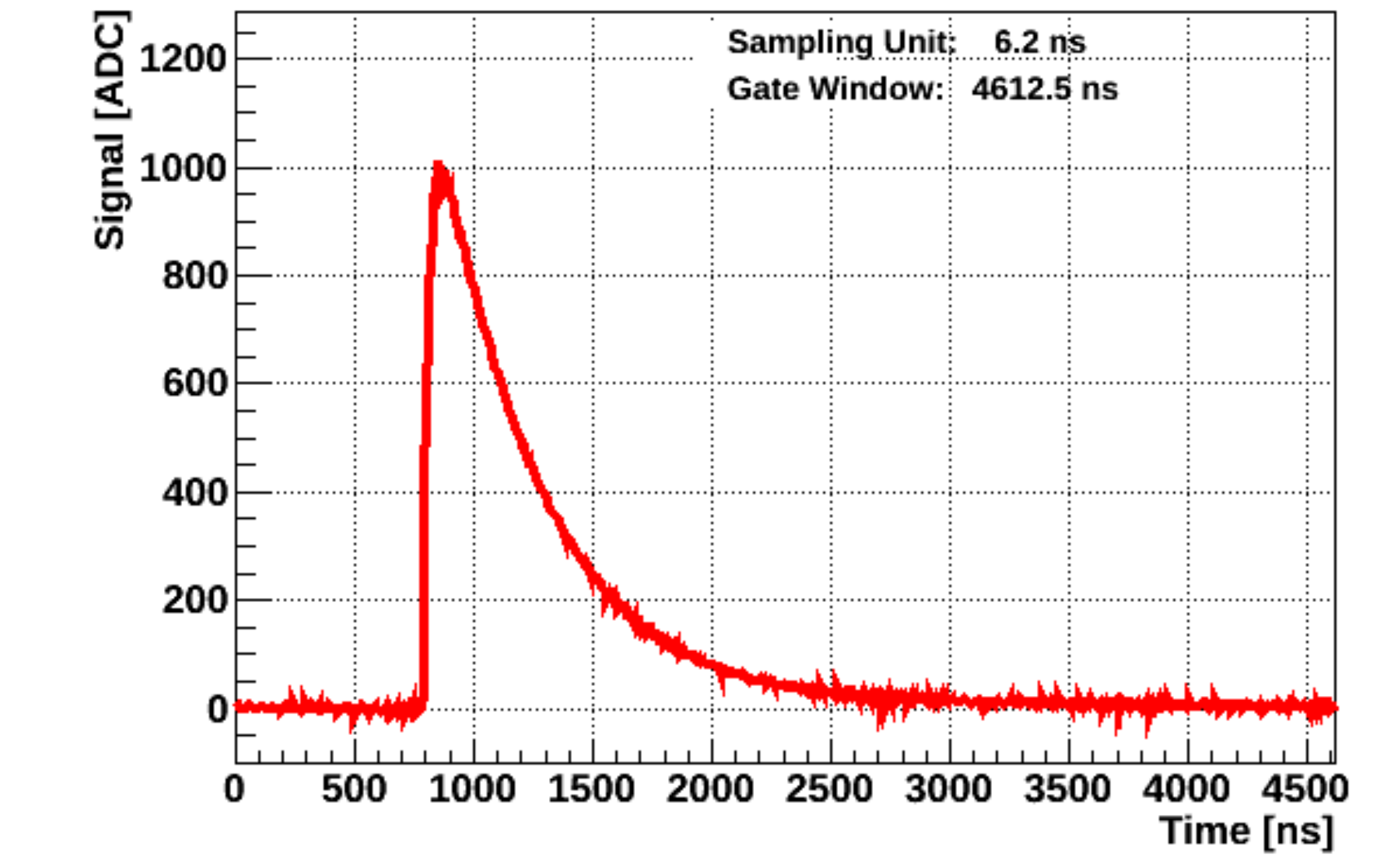}
      \includegraphics[height=5cm,width=7.cm]{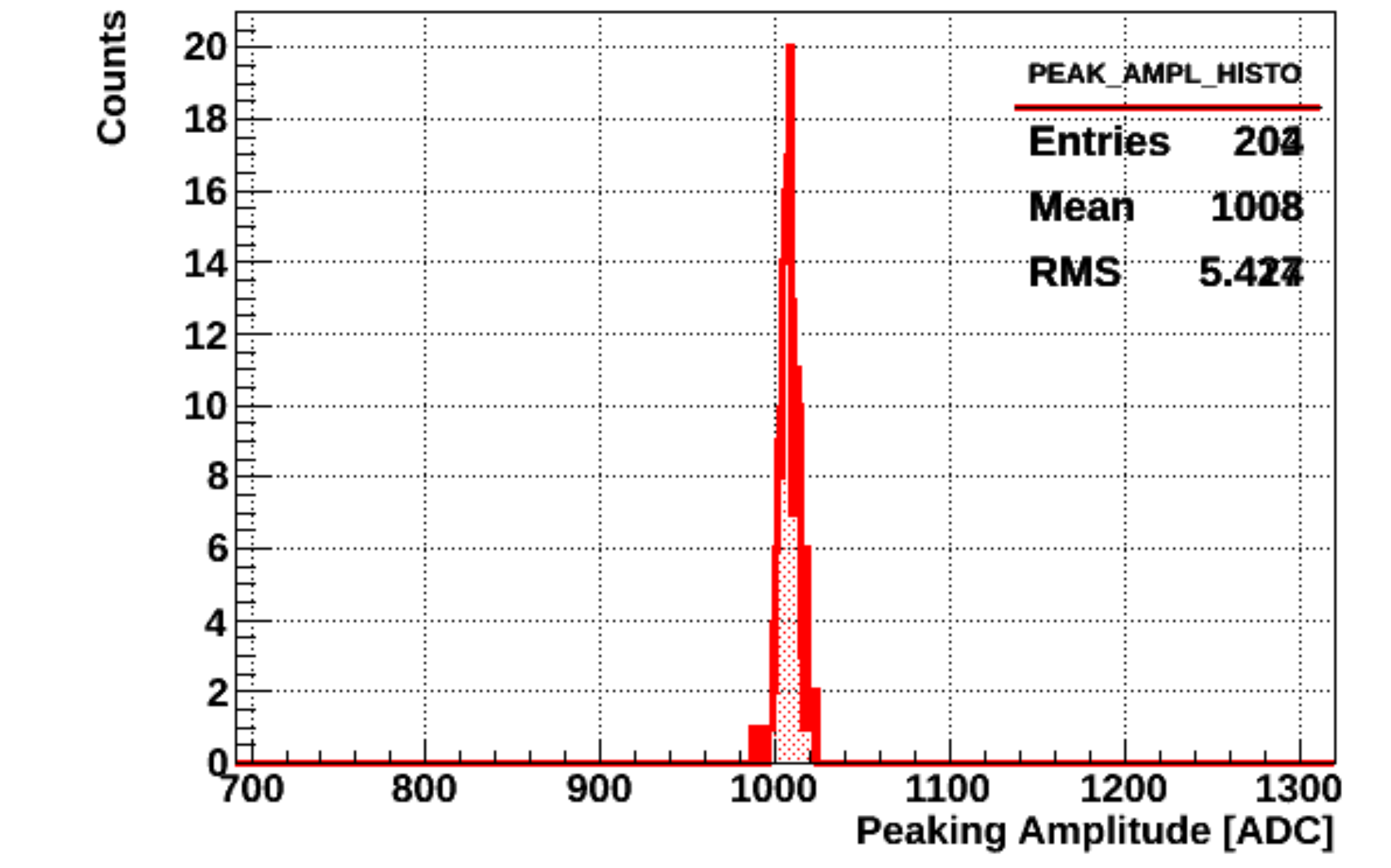}\\

  \vspace{-0.3cm}
  \caption{Waveform 
           analysis with the WASA electronics using the developed 
           transimpedence amplifier above described. 
           Measurement done {\bf without frequency filtering}.
           Examples  obtained 
           with $3$ mV (upper panels) and \mbox{$20$ mV}
           (bottom panels) of amplitude for the voltage pulse  
           provided by the pulse generator.}
  \label{fig:NEWAMPL_WASA_PULSER_EXAMPLE}
\end{figure}

%
To test the performance of the new amplifier the signal filtering has 
been first removed in both the amplifier and in the
differential amplifier in the QDC WASA board (additionally here the 
capacitor at the signal input has been bridged).  
This procedure in the QDC module helped avoiding signal undershooting.  
In addition, 
the negative output of the amplifier was kept to the ground level.
Typical results of the measurements are presented in the upper (bottom) 
panel of Fig.~\ref{fig:NEWAMPL_WASA_PULSER_EXAMPLE} 
for the input voltage amplitude of $3$ ($20$) mV.

At the oscilloscope for the amplitude values of $1$, $2$ and $3$ mV at 
the pulser generator the 
peaking amplitude values of $22.5$, $45.0$ and $67.5$ mV are measured, 
resulting 
in a gain factor of 22.5 for the whole system (RC plus amplifier plus 
cable attenuation).
As already mentioned, an unknown systematical uncertainty can
potentially be present when extracting those values directly from the 
oscilloscope. 
As described in the previous sections, for each event the calculated 
pedestal is subtracted to the waveform sampling data bringing, as expected, 
the baseline to zero.
%

\subsubsection{Linearity of the Experimental Setup System}
%
A scan is performed, injecting in the WASA  system 
waveforms with different values of the input voltage pulse amplitude, 
as shown in Tab.~\ref{tab:NEWAMPL_WASA_SCAN_PULSER}.
\begin{table}[t!]
\hspace{-1.5cm}
\begin{tabular}{|c|c|c|c|c|c|c|c|c|c|c|c|c|c|c|}
\hline 
\hline
\multicolumn{15}{|c|}{ \textbf{WASA System: Linear Scan} }\\
\hline
\hline
\multicolumn{15}{|c|}{ \textbf{Output Pulser [mV]:} }\\
\hline
                 \small   3 &  
                 \small   5 &  
                 \small   8 &  
                 \small  11 &  
                 \small  14 &  
                 \small  17 &  
                 \small  20 &  
                 \small  23 &  
                 \small  26 &  
                 \small  29 &  
                 \small  32 &  
                 \small  35 &  
                 \small  38 &  
                 \small  41 &  
                 \small  42  \\ 
\hline
\multicolumn{15}{|c|}{ \textbf{Measured Peaking Amplitude [ADC]:}}\\
\hline
   \small   154 
   &  \small   255 
   &  \small   405 
   &  \small   555 
   &  \small   707 
   &  \small   858 
   &  \small  1008 
   &  \small  1160 
   &  \small  1310 
   &  \small  1459 
   &  \small  1608 
   &  \small  1757 
   &  \small  1907 
   &  \small  2057 
   &  \small  2106 \\ 
\hline
\end{tabular}\\

\caption{Scan performed with the WASA readout system
         using the new transimpedence amplifier and 
         {\bf without frequency filtering}. 
         The total error in the measured peaking amplitude is 
         conservatively considered as being unity.}
\label{tab:NEWAMPL_WASA_SCAN_PULSER}

\end{table}

\begin{figure}[t!]
  \hspace{-2cm} \vspace{-0.5cm}
  \includegraphics[height=5.cm,width=8cm]
     {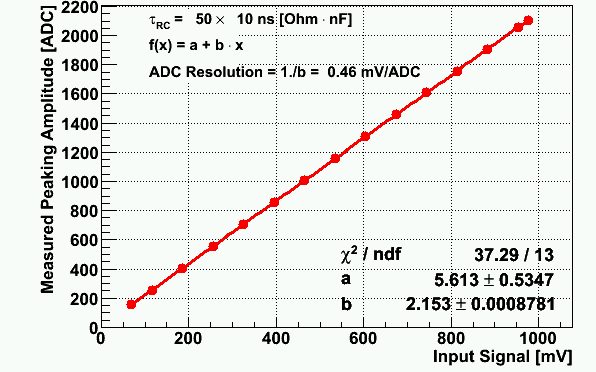}
  \includegraphics[height=5.cm,width=8cm]
     {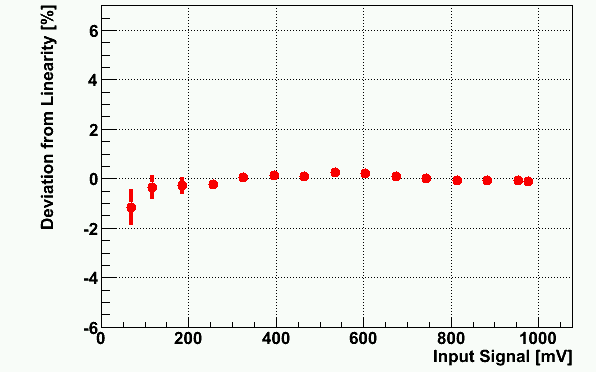}\\

  \hspace{-2cm} 
  \includegraphics[height=5.cm,width=8cm]
     {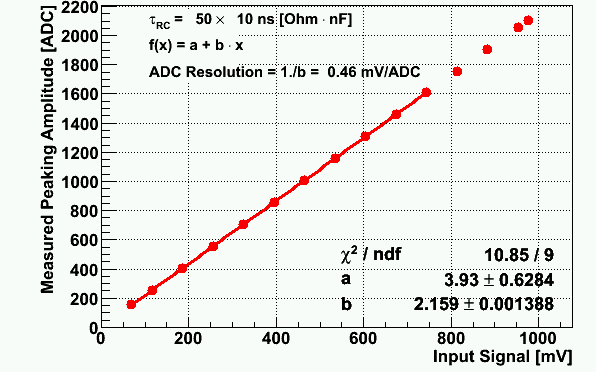}
  \includegraphics[height=5.cm,width=8cm]
     {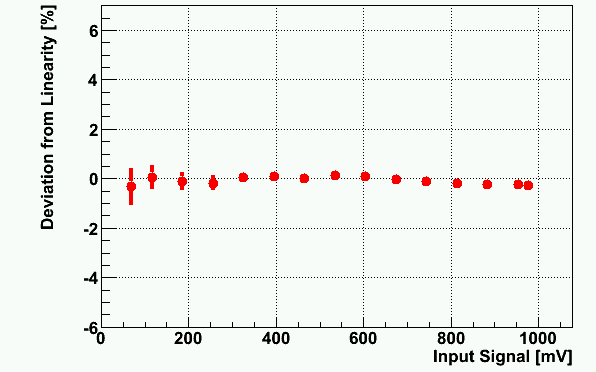}\\

  \vspace{-0.5cm}
  \caption{The ADC resolution is here extracted via a linear fit 
           to the data of the scan in the amplitude of the 
           input signal to the WASA board. The fit is performed 
           in the whole scan region (upper panels) and in its lower part
           (lower panels), without affecting the extracted ADC resolution.
           Measurement done \bf{without frequency filtering}.}
  \label{fig:NEWAMPL_WASA_LINEARITY_PULSER}
\end{figure}

The above data are presented in Fig.~\ref{fig:NEWAMPL_WASA_LINEARITY_PULSER}. 
A linear function is fitted to the data in two different measurement 
regions, resulting in an ADC resolution of approximately $0.46$ mV per ADC.
Note that the input signal is obtained using the values  
measured at the oscilloscope for $1$, $2$ and $3$ mV (above described)
and rescaling them for the larger amplitude values of the generated 
pulse. 

From the fit results, the deviation to linearity is calculated in 
percent, and is presented in the left panels of the picture. 
The deviation is typically well below $1\%$ in the investigated region.
This extracted ADC resolution will not be used to calculate the 
total injected charge for each waveform due to the
possible systematical uncertainty affecting the precise measurement 
of the peaking amplitude at the oscilloscope.

\subsubsection{Waveform Analysis: WASA vs \mbox{ACQIRIS} Comparison}
\label{sec:NEWAMPL_WAVEFORM_PULSER}
\begin{figure}[t!]
\hspace{-1cm} 
      \includegraphics[height=5cm,width=7.cm]
               {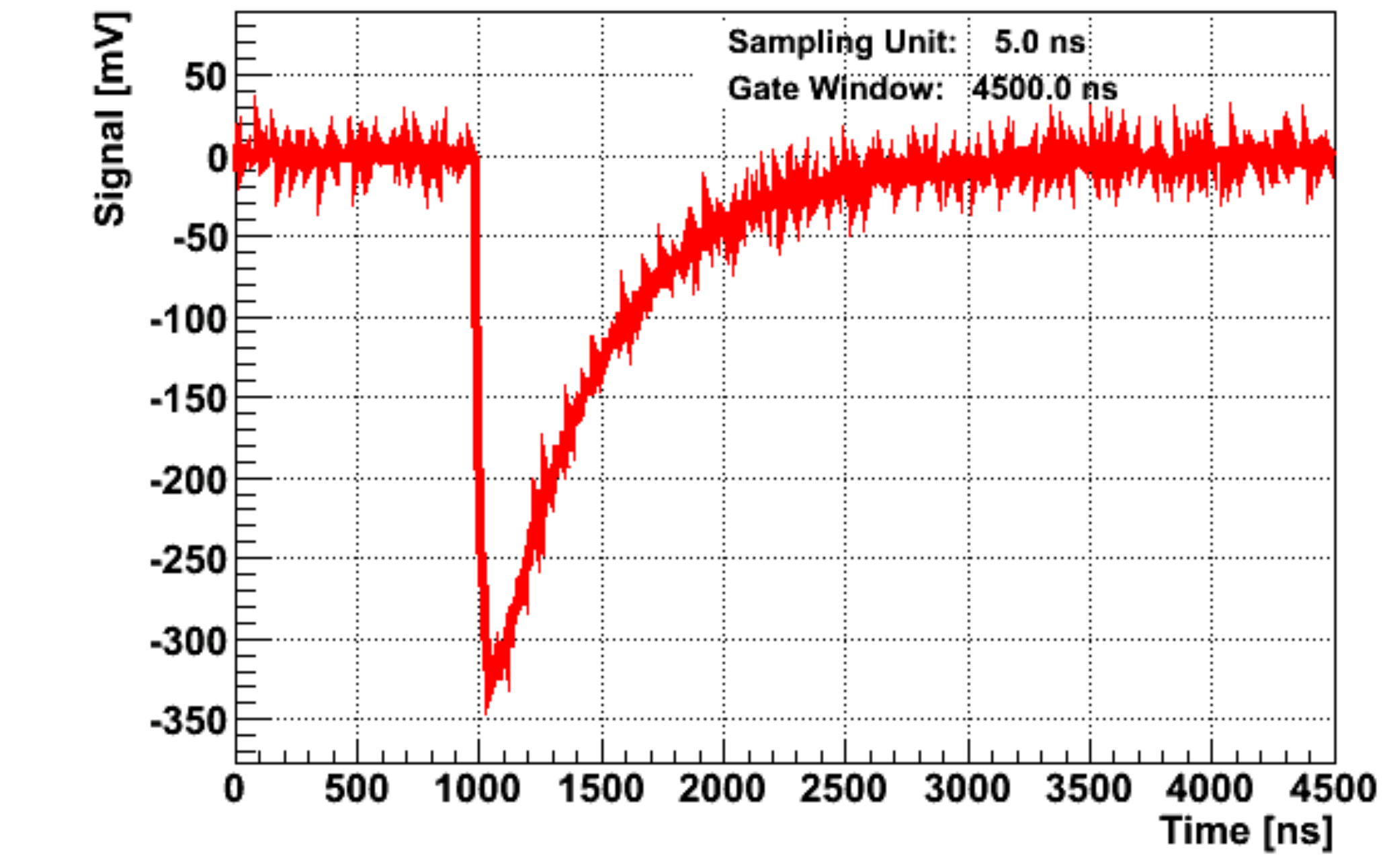}
      \includegraphics[height=5cm,width=7.cm]
               {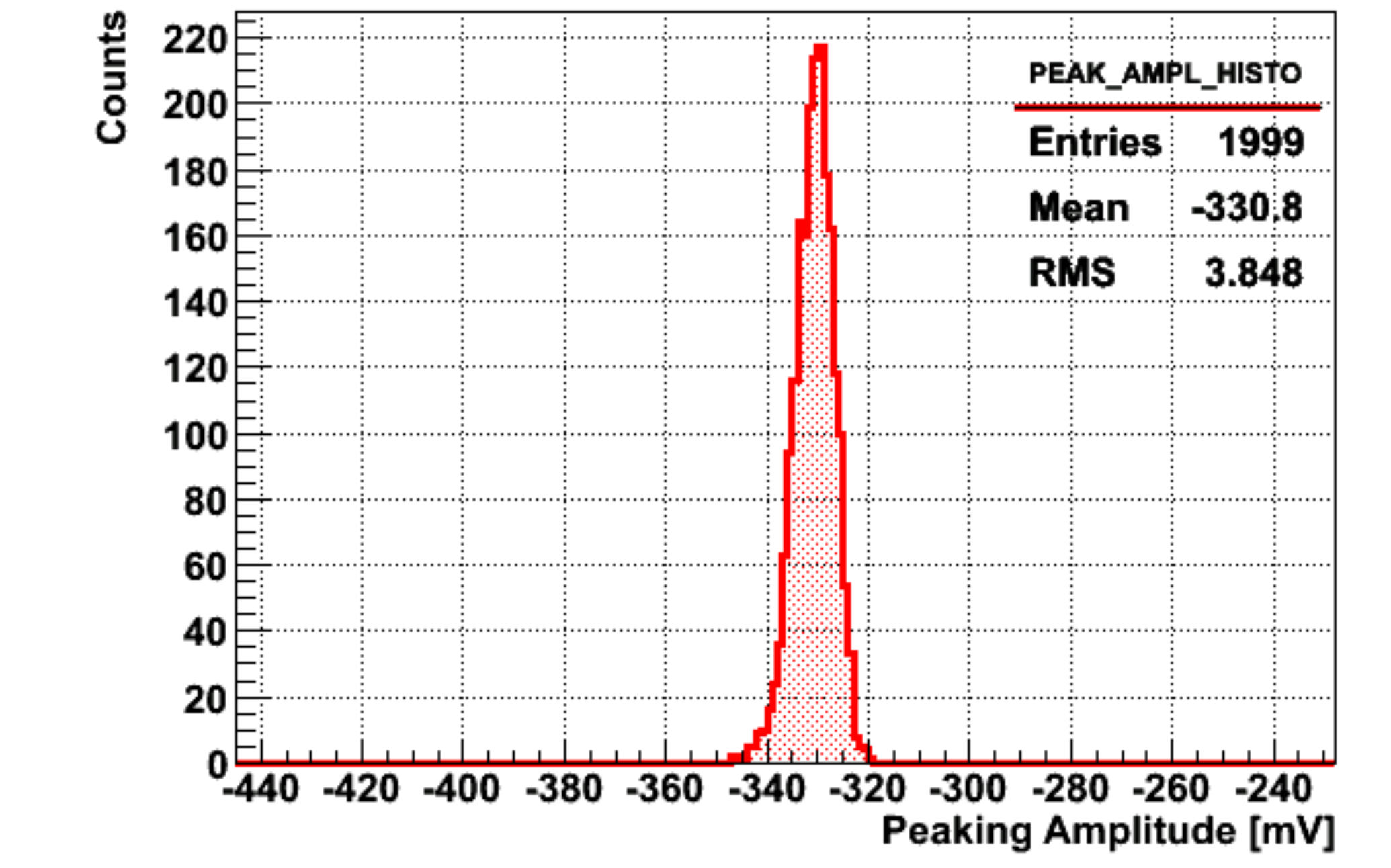}\\

\hspace{-1cm} 
      \includegraphics[height=5cm,width=7.cm]
               {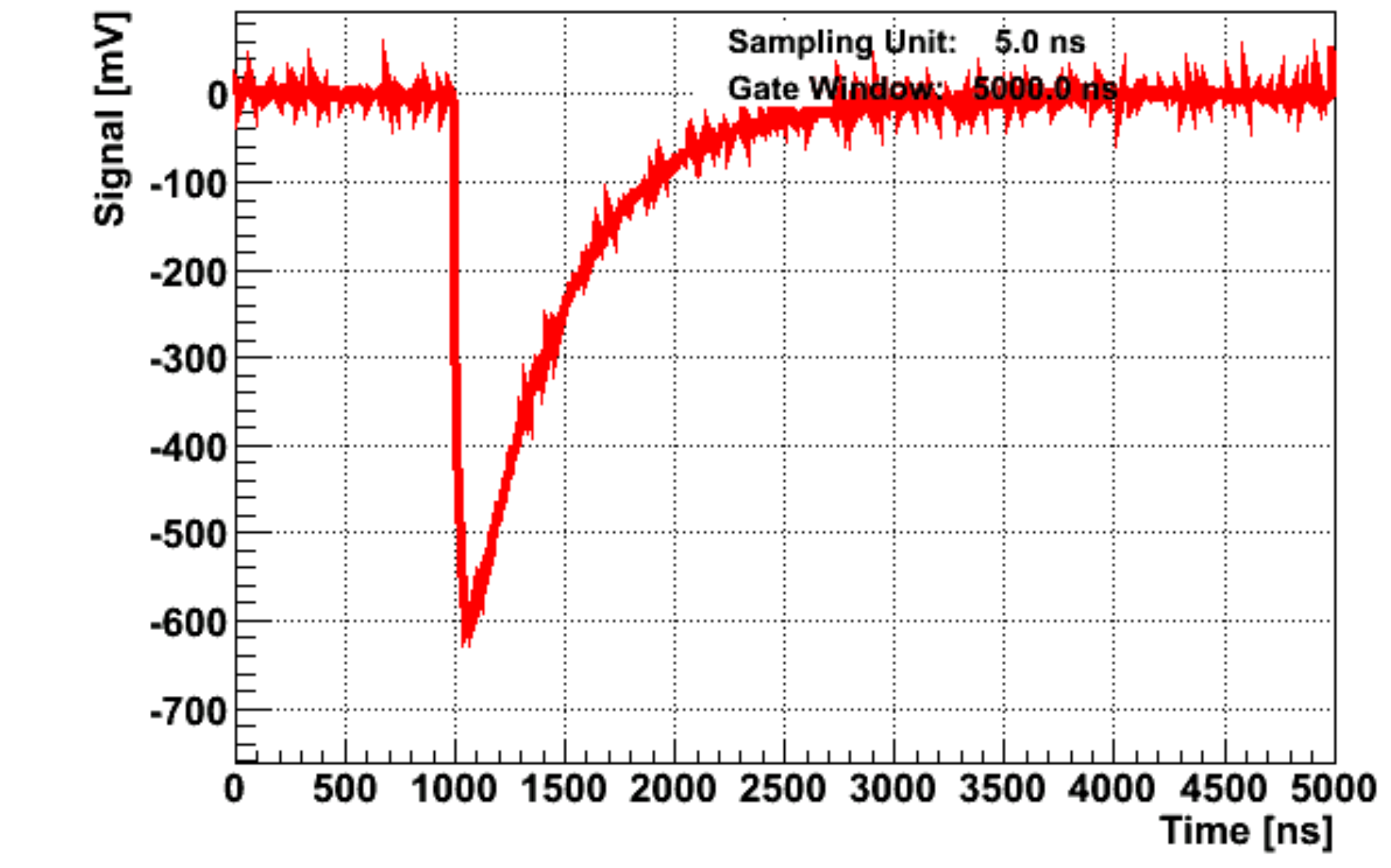}
      \includegraphics[height=5cm,width=7.cm]
               {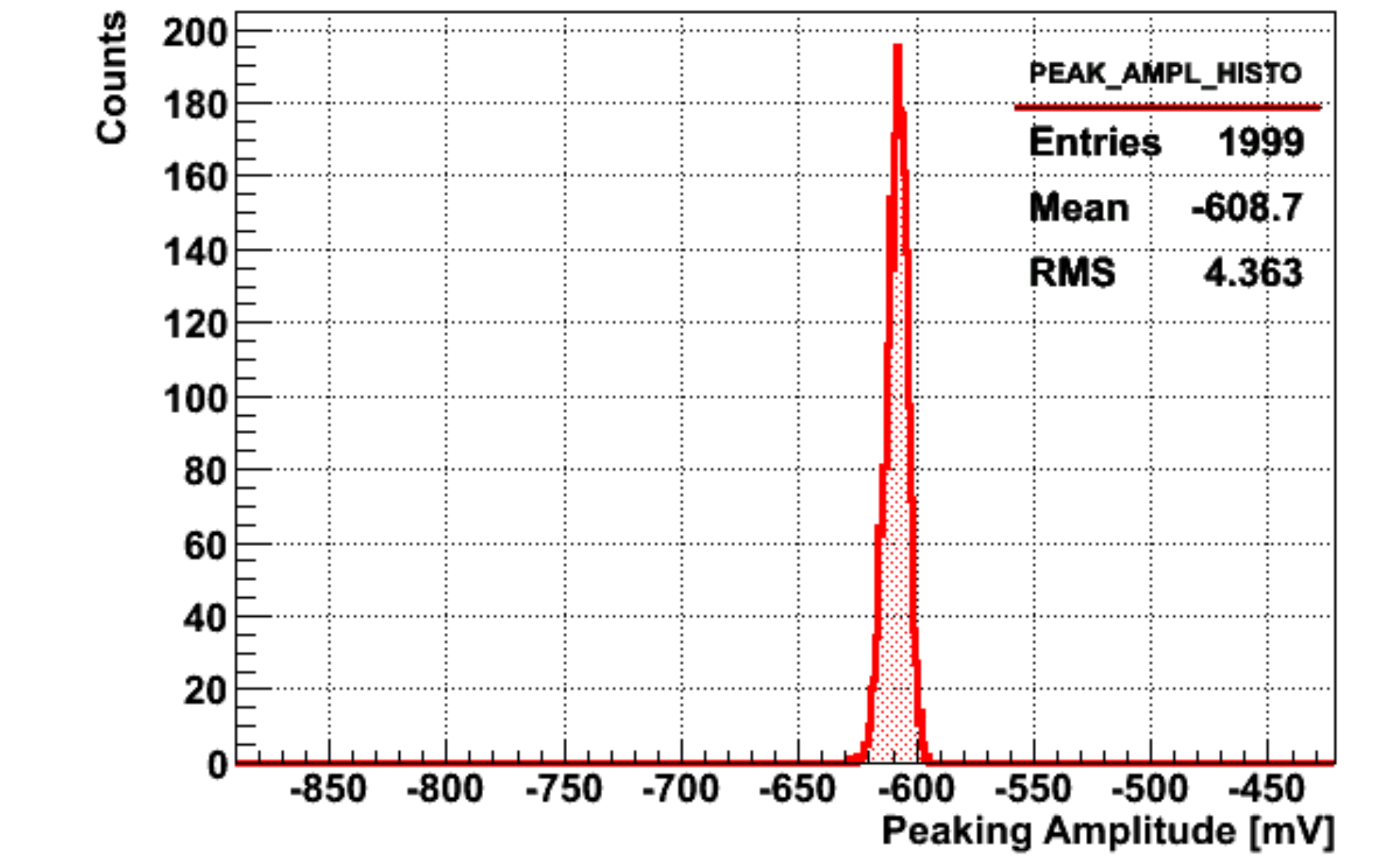}\\

\hspace{-1cm}
      \includegraphics[height=5cm,width=7.cm]
               {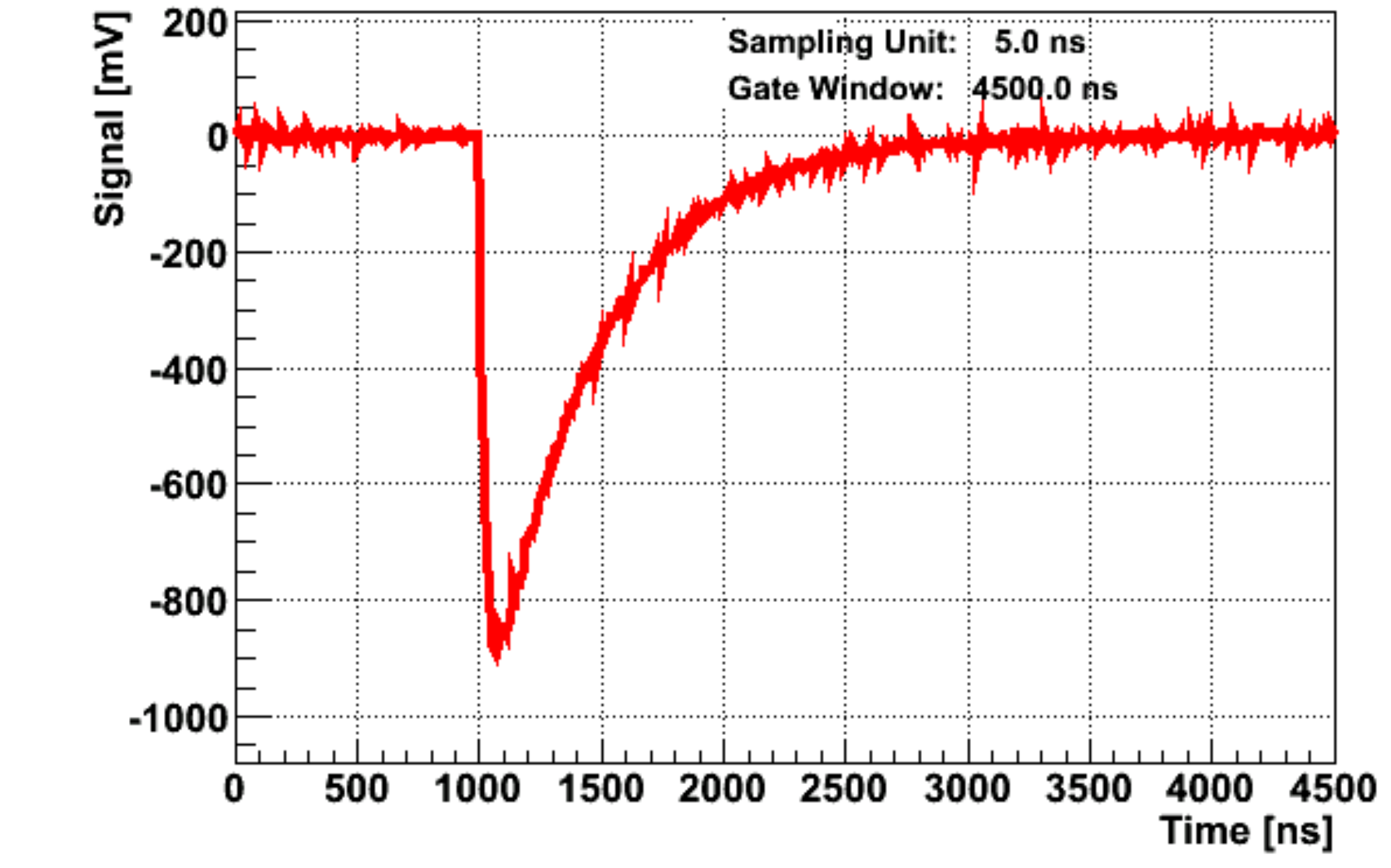}
      \includegraphics[height=5cm,width=7.cm]
               {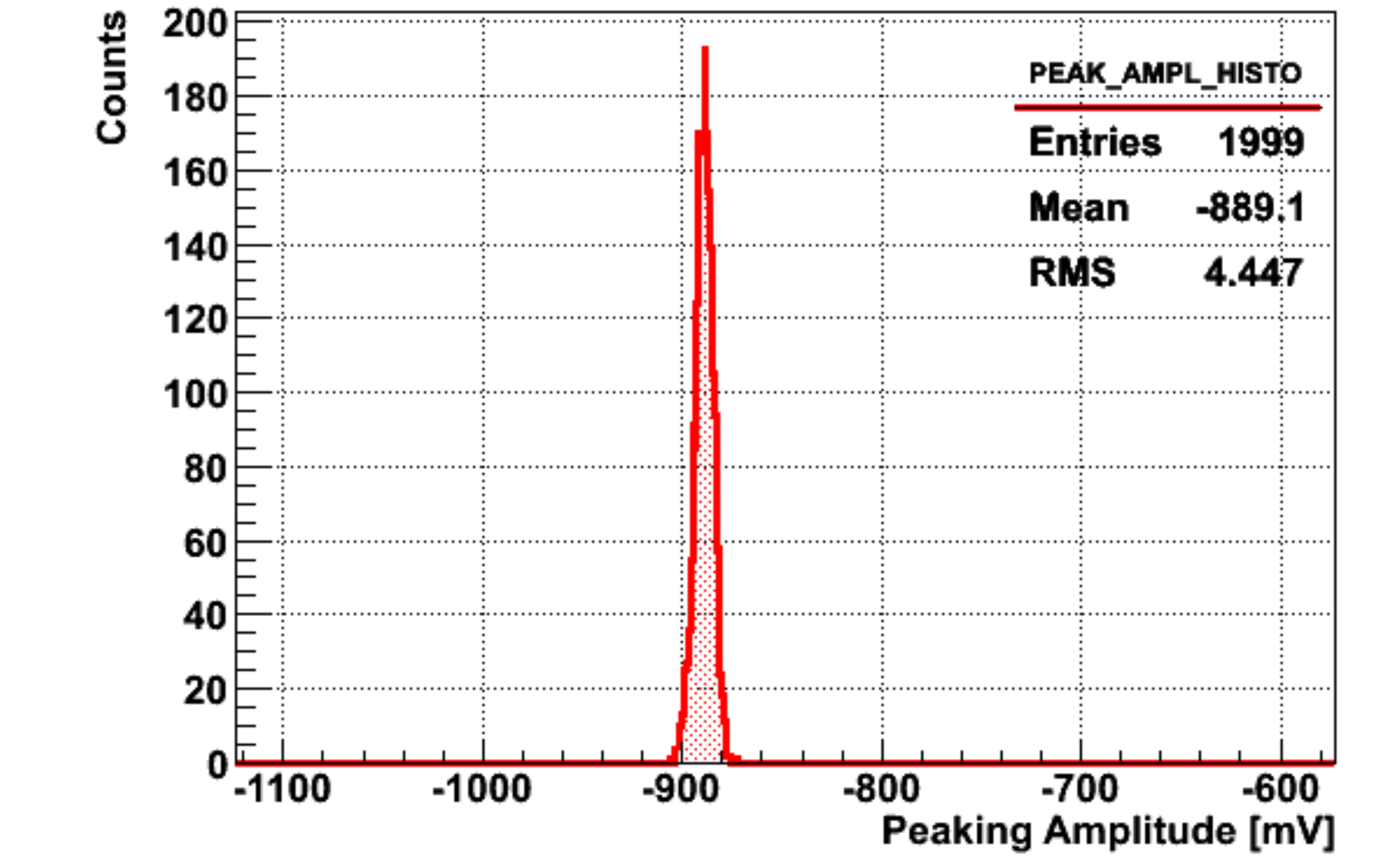}\\

  \caption{Waveform analysis performed with the ACQIRIS electronics 
           using the new transimpedence 
           amplifier and  {\bf without frequency filtering}.
           Results obtained 
           with $14$, $26$ and $38$ mV (upper, middle and bottom panels, 
           respectively)
           voltage amplitude provided by the pulse generator.}
  \label{fig:NEWAMPL_ACQIRIS_PULSER_EXAMPLE}
\end{figure}

With the same experimental setup some measurements have 
been performed using also ACQIRIS readout electronics.
The values of $14$, $26$ and $38$ mV for the amplitude of 
the input voltage pulse  were used, and the 
corresponding relevant 
results are presented in Fig.~\ref{fig:NEWAMPL_ACQIRIS_PULSER_EXAMPLE}. 

\begin{figure}[h!]
 \hspace{-2.cm}  \vspace{-0.25cm}
 \includegraphics[height=5.3cm,width=7.5cm]
               {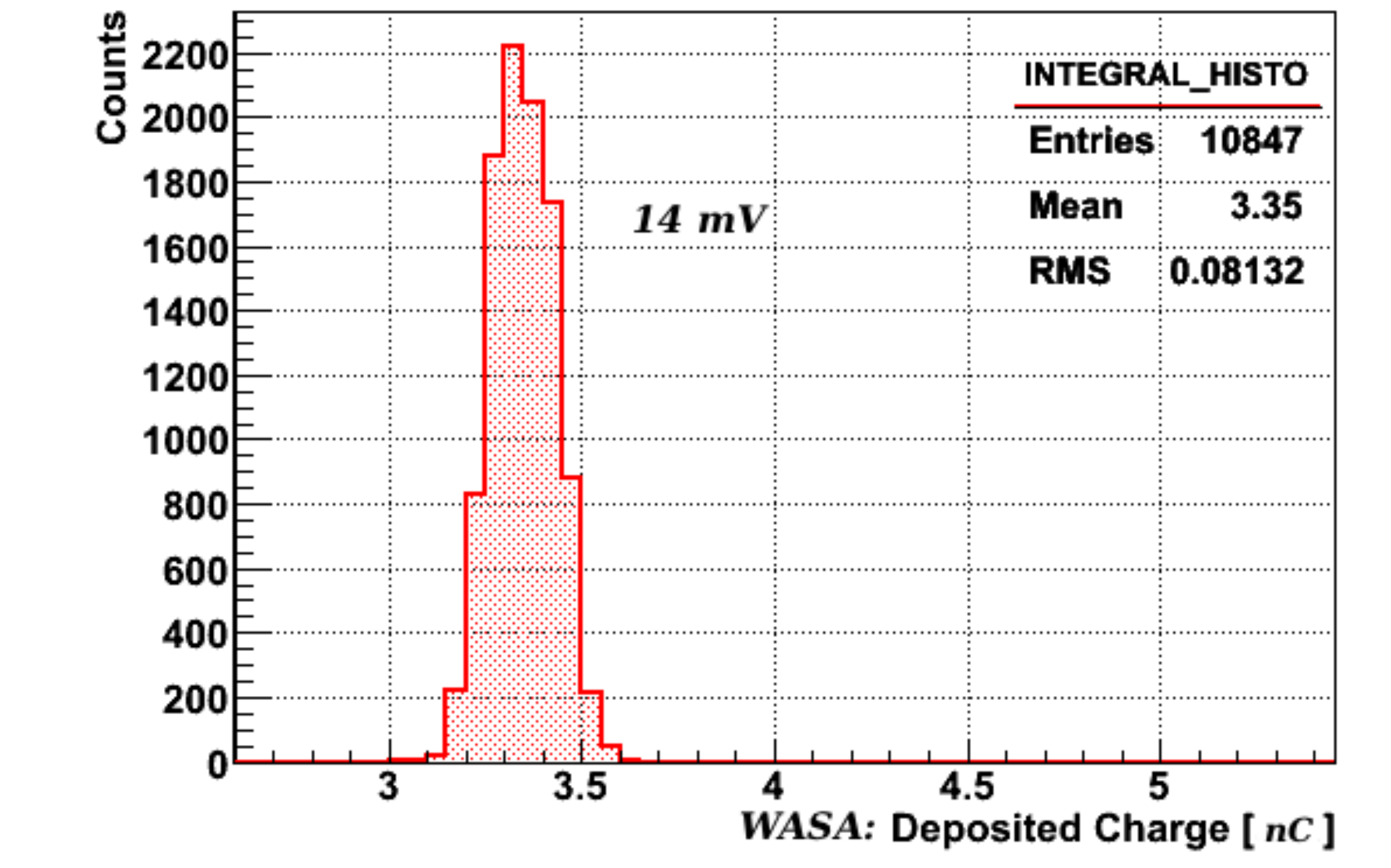}
 \includegraphics[height=5.3cm,width=7.5cm]
               {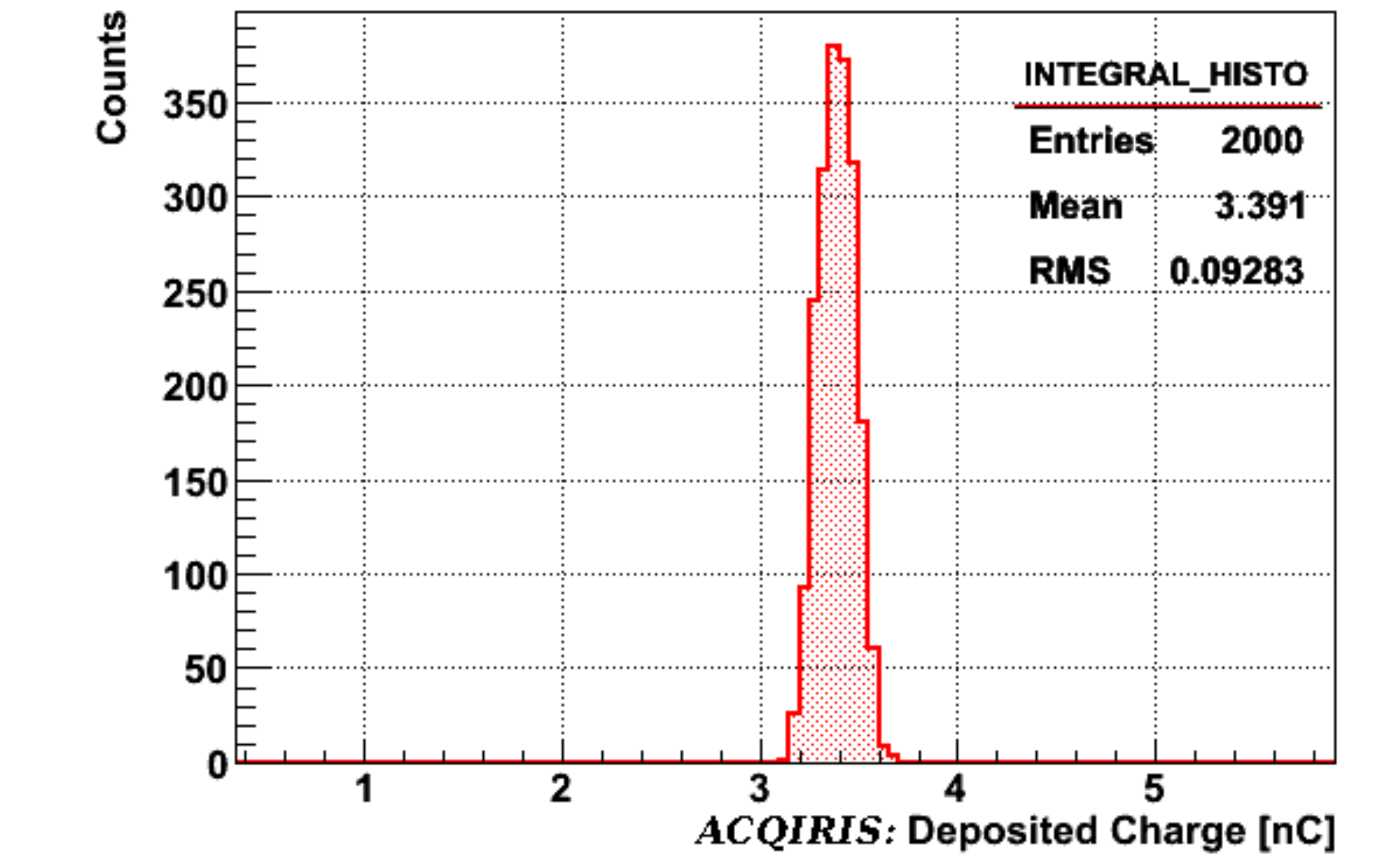}\\

 \hspace{-2.cm} \vspace{-0.25cm}
 \includegraphics[height=5.3cm,width=7.5cm]
                 {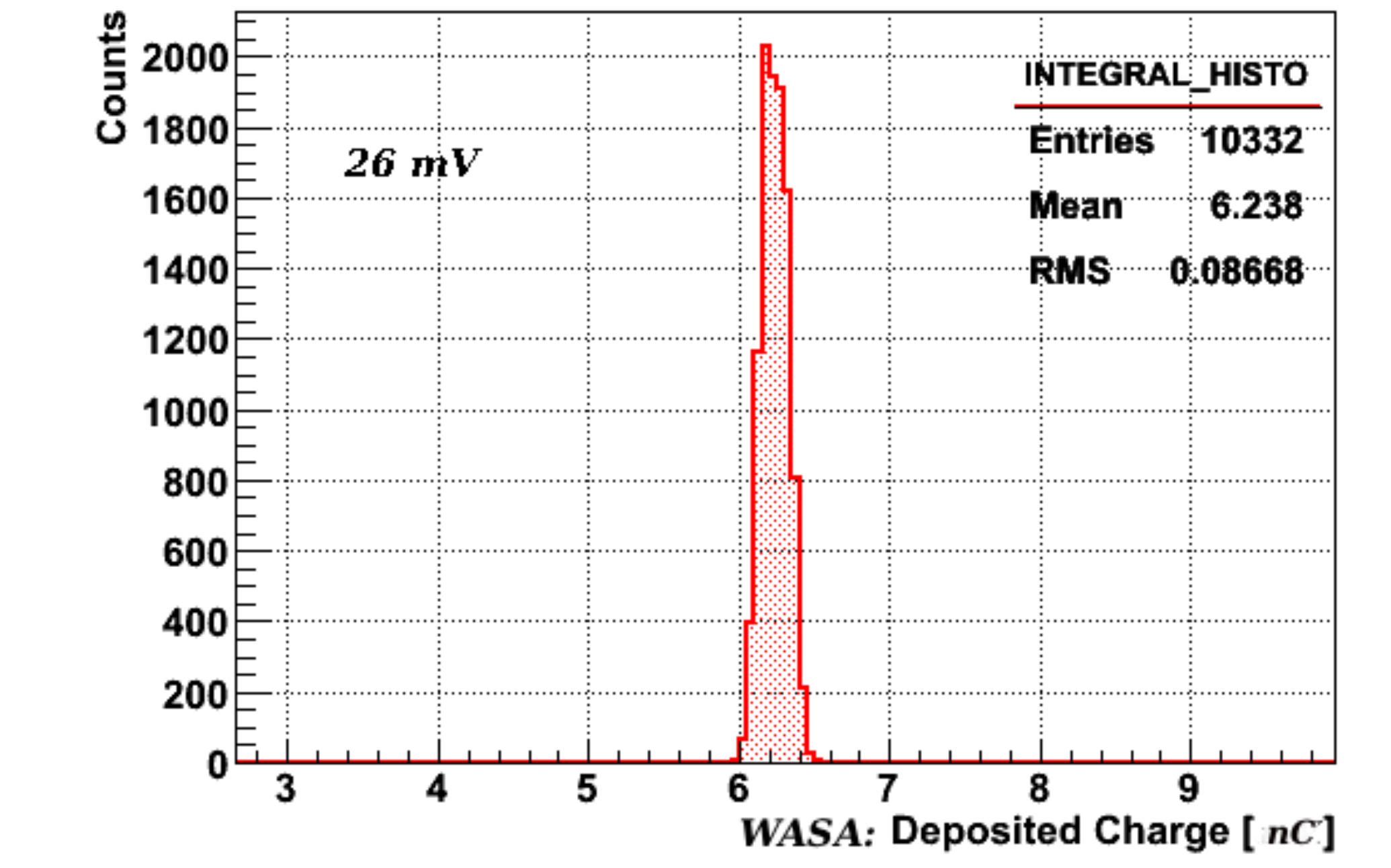}
 \includegraphics[height=5.3cm,width=7.5cm]
                 {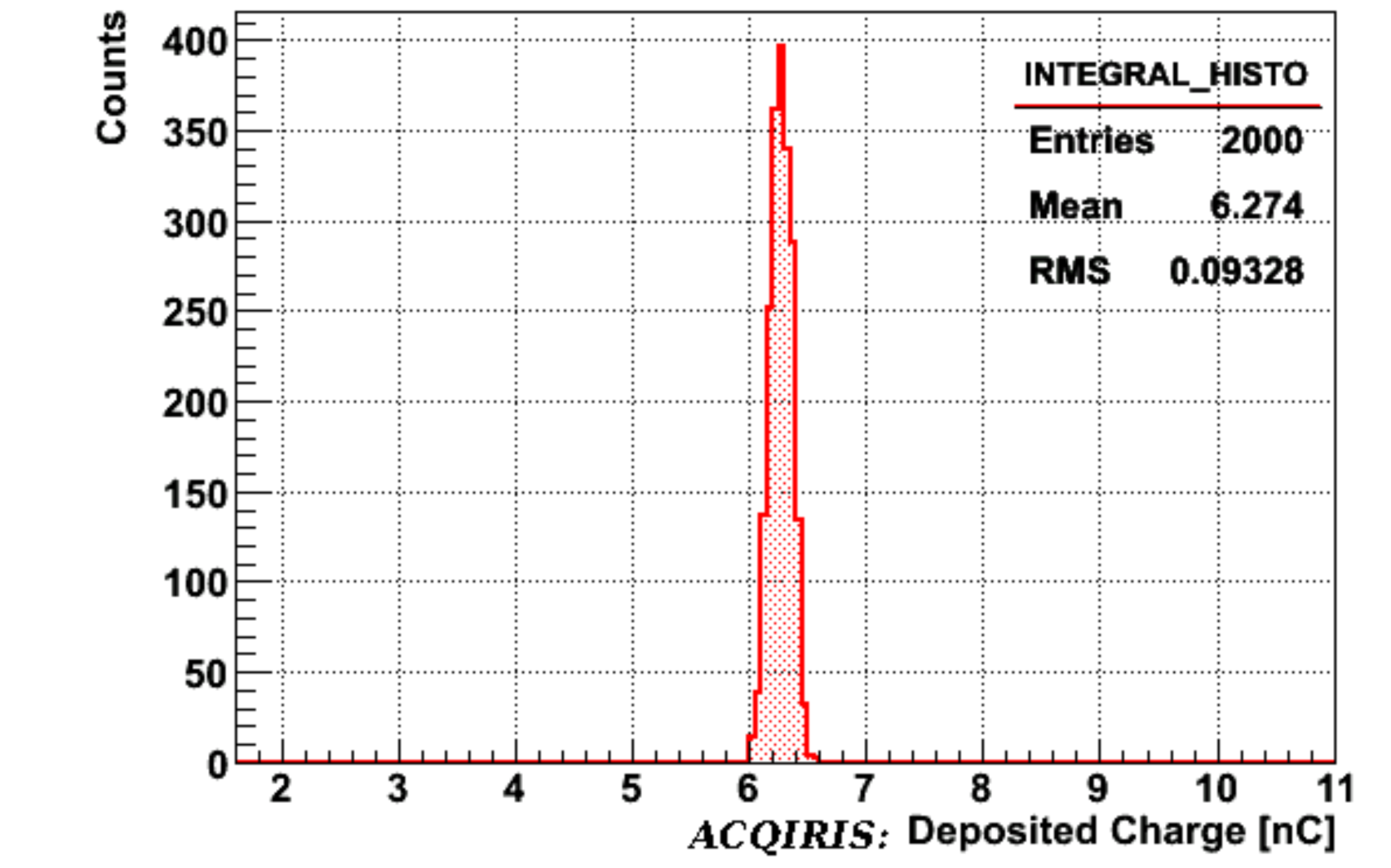}\\

 \hspace{-2.cm}
 \includegraphics[height=5.3cm,width=7.5cm]
               {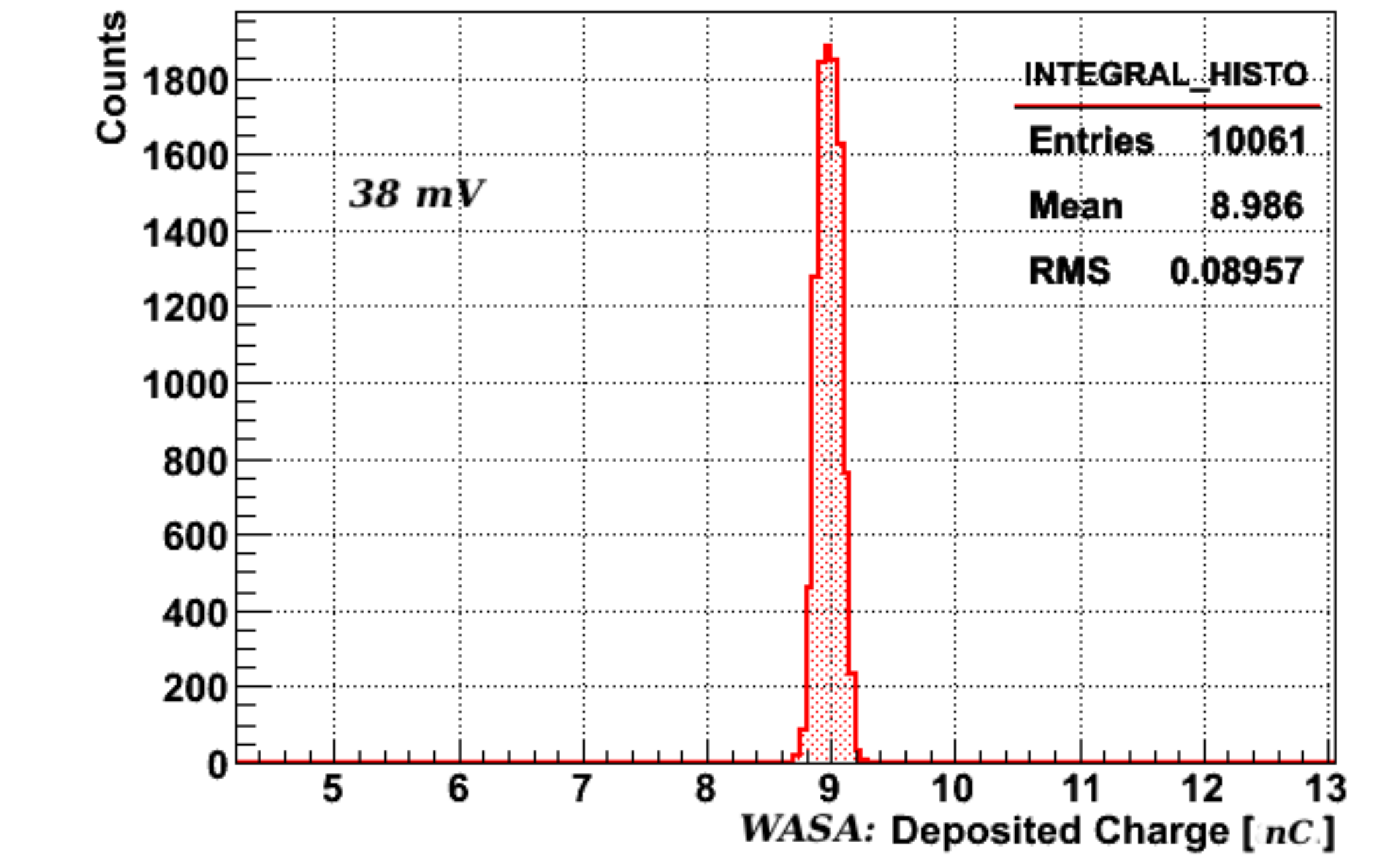}
 \includegraphics[height=5.3cm,width=7.5cm]
               {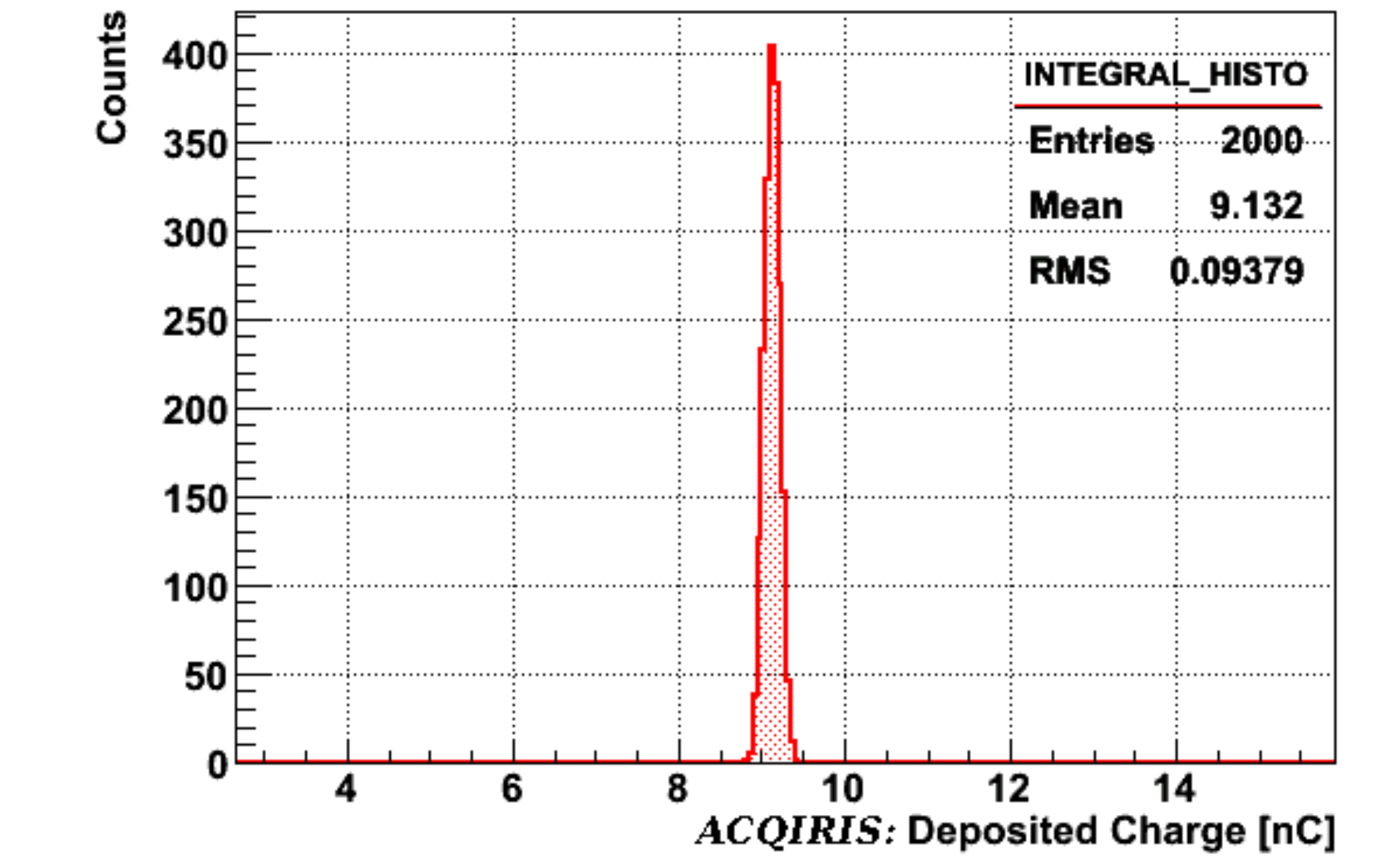}\\

  \vspace{-0.7cm}
  \caption{Integrated charge injected in the WASA and in the ACQIRIS 
           readout systems with the new transimpedance amplifier  
           {\bf without frequency filtering} 
           for the amplitude values of $14$, $26$ and $38$ mV 
           (upper, middle and bottom panels, 
           respectively) for the voltage pulse.}
  \label{fig:NEWAMPL_COMPARE_CHARGE_PULSER}
\end{figure}

Using the data obtained with the ACQIRIS system (in mV) the WASA 
data (in ADC) can be calibrated and converted in mV ($0.46$ mV per ADC unit).
The comparison of the integrated charge as measured by the two systems is 
presented in Fig.~\ref{fig:NEWAMPL_COMPARE_CHARGE_PULSER} for the above 
three amplitude values of the voltage pulse. 

For all three experimental setups a relative $1-2\%$ difference appears, 
better as in the previous setup.
Preliminary, for the moment this value can be considered as the systematical
uncertainty of the measurement. 

\subsection{Measurements with WASA System:\\ With Frequency Filter}
\begin{figure}[b!]
\hspace{-1.35cm} 
      \includegraphics[height=5cm,width=7.cm]{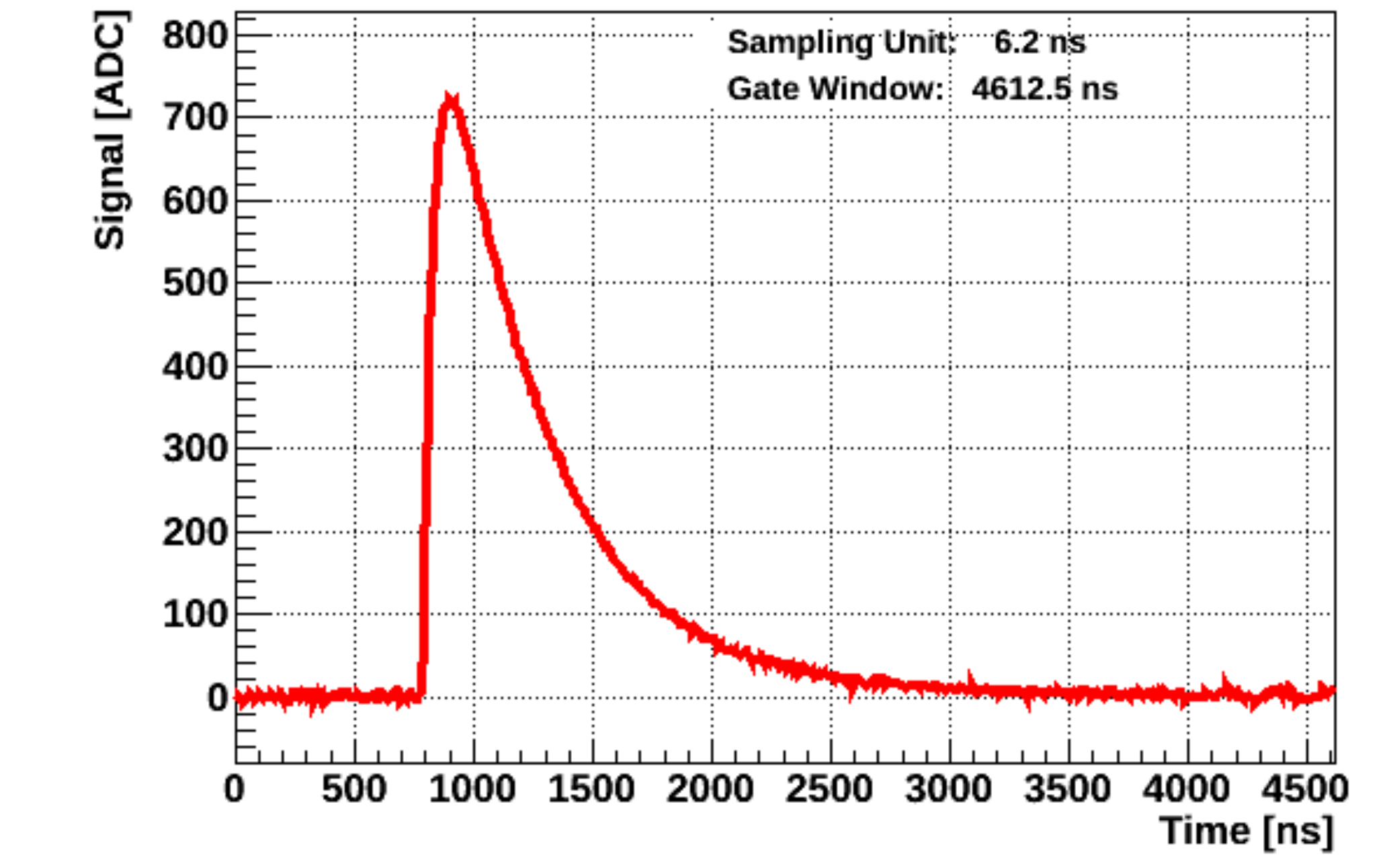}
      \includegraphics[height=5cm,width=7.cm]{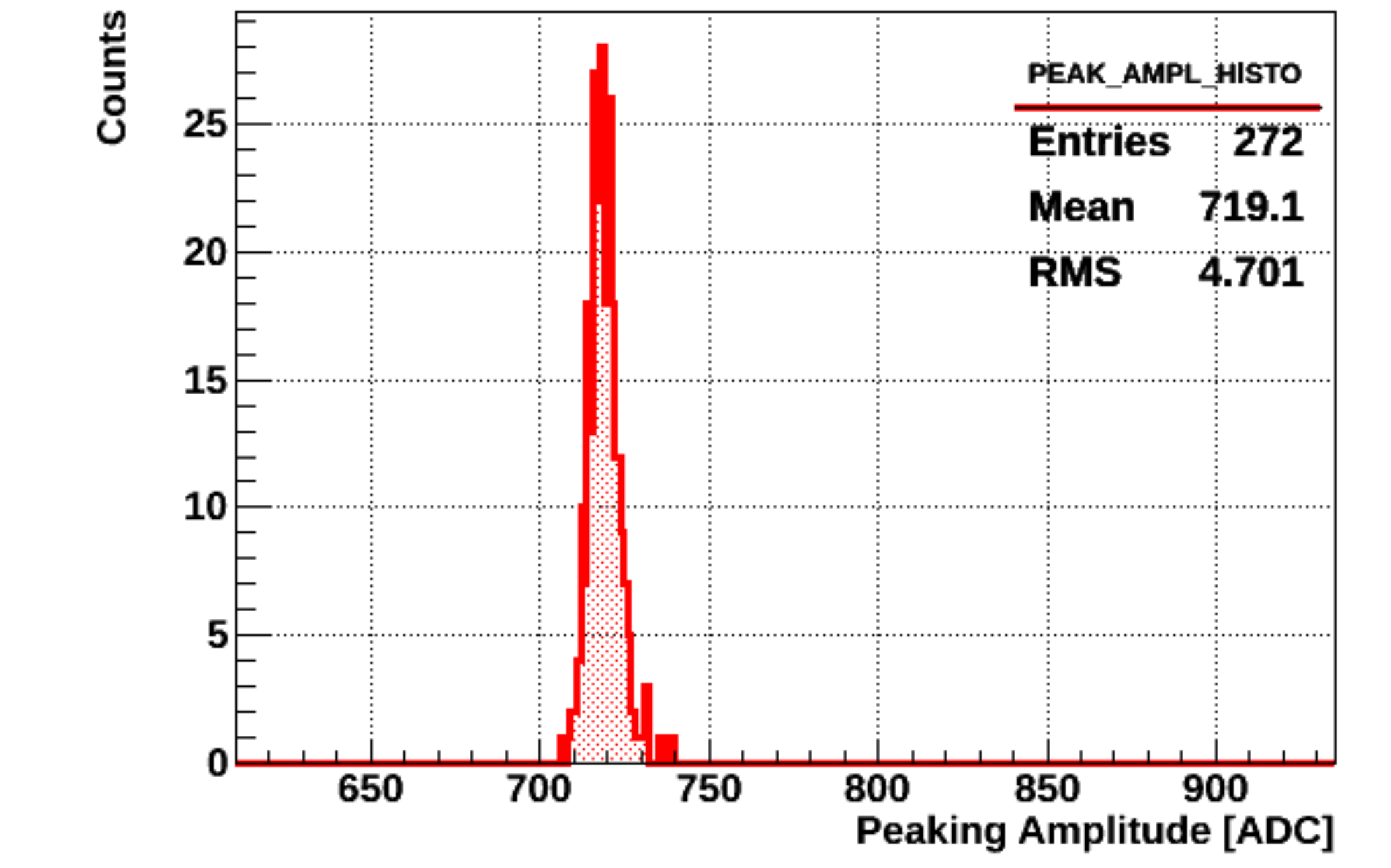}\\

\hspace{-1.35cm} 
      \includegraphics[height=5cm,width=7.cm]{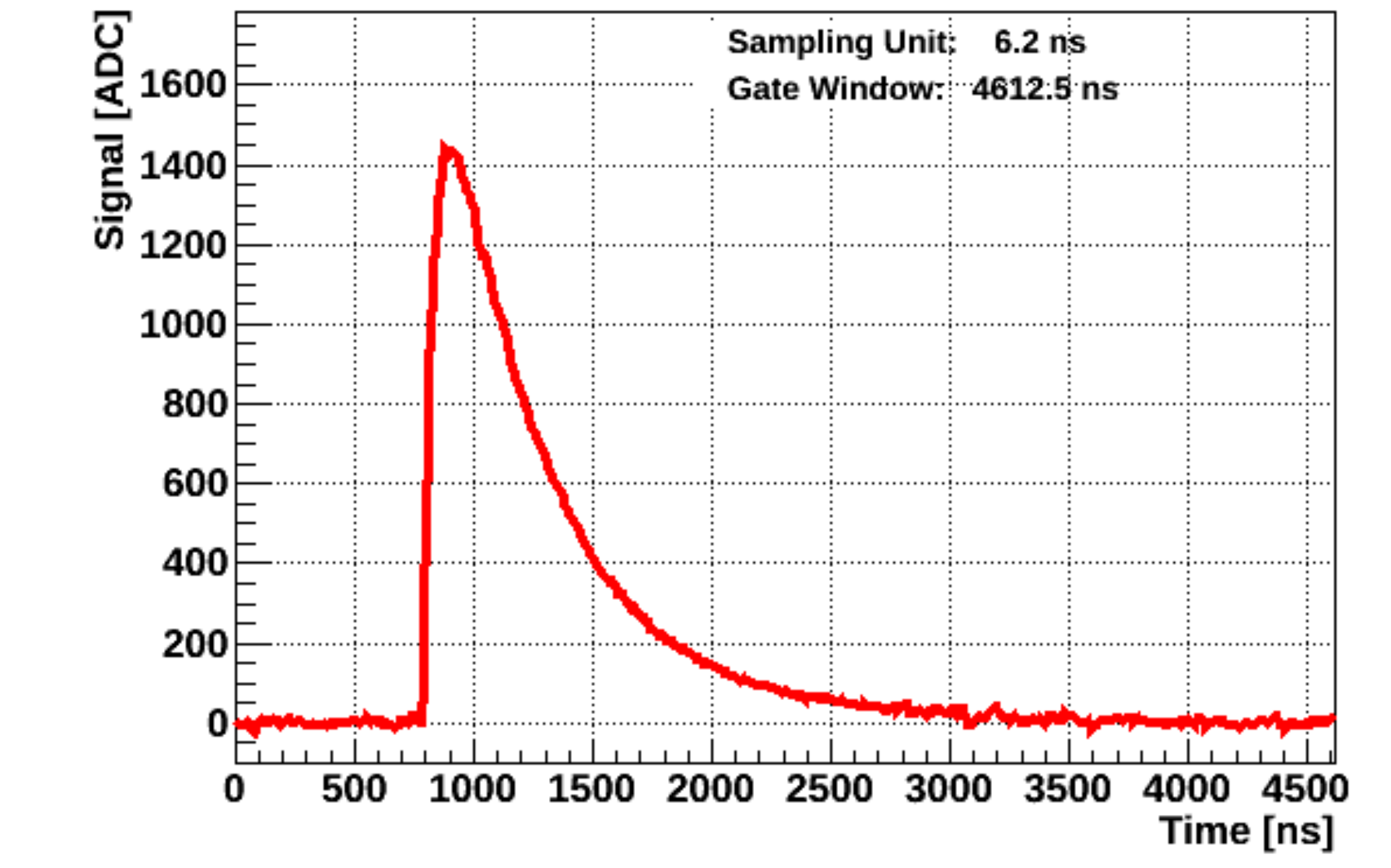}
      \includegraphics[height=5cm,width=7.cm]{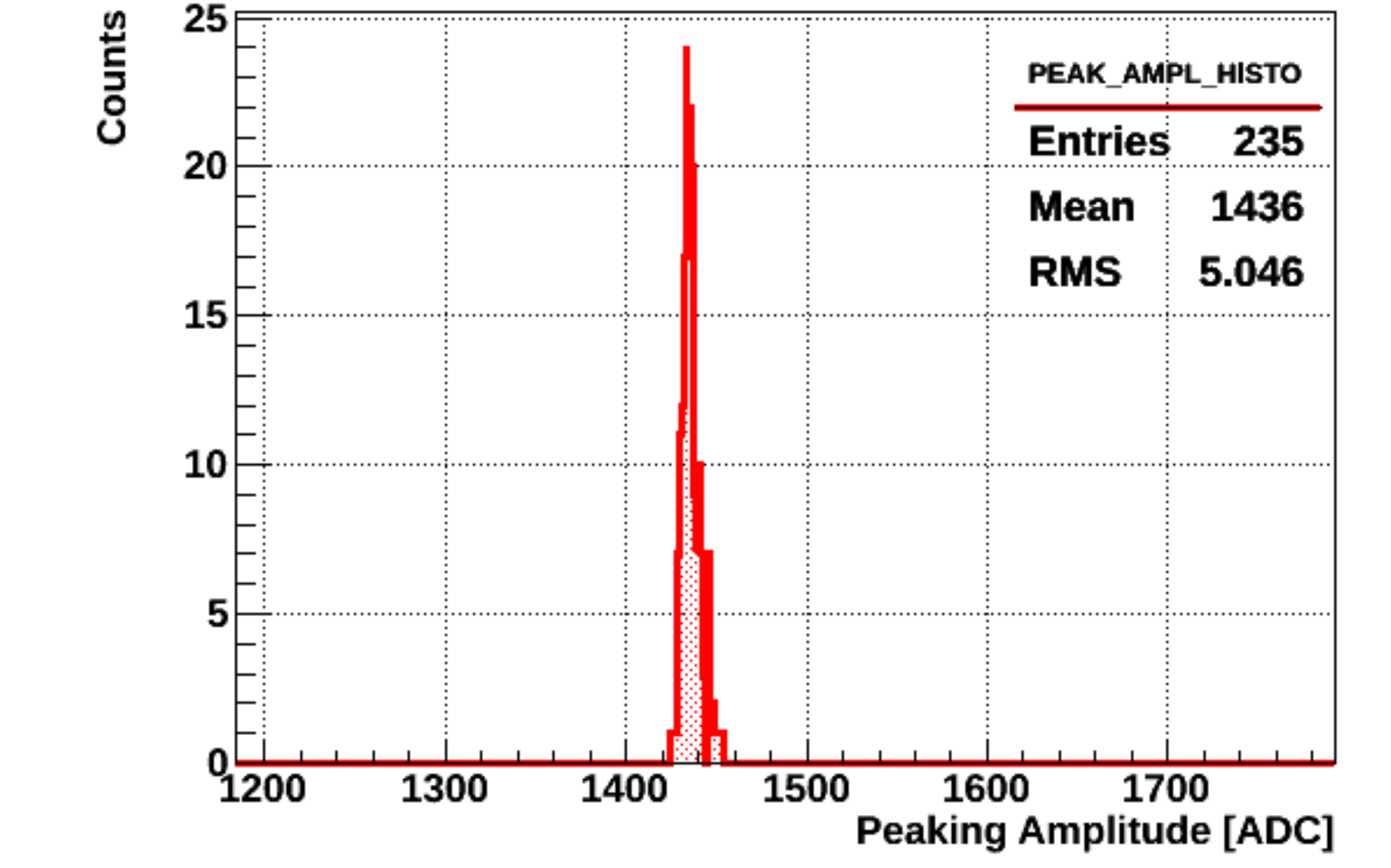}\\

  \caption{Waveform 
           analysis with the WASA electronics using the developed 
           transimpedence amplifier above described. 
           Measurements done {\bf with frequency filtering}.
           These examples were obtained 
           with $10$ mV (upper panels) and \mbox{$20$ mV}
            (bottom panels) of voltage 
           amplitude provided by the pulse generator.}
  \label{fig:NEWAMPL_FILTER_WASA_PULSER_EXAMPLE}
\end{figure}

%
The next test was performed with
the signal filtering stage ($f<80$ MHz) active in the the new amplifier.
Note that the modifications done in the QDC WASA board were 
kept to avoid signal undershooting. Again, 
the negative output from the amplifier was kept to the ground level.
The results of the typical measurements is presented in the upper (bottom) 
panels of Fig.~\ref{fig:NEWAMPL_FILTER_WASA_PULSER_EXAMPLE} 
for the input voltage 
amplitude of $10$ ($20$) mV.

It is visible the higher gain of the amplifier with the 
frequency filter stage active, due to the presence of 
capacitors in that section of the device.
%

\subsubsection{Linearity of the Experimental Setup System}
%
To investigate the linearity of the system with the 
frequency filtering stage active a scan is performed, 
injecting into the WASA  system waveforms with different 
amplitude values for the input voltage 
pulse, as shown in Tab.~\ref{tab:NEWAMPL_FILTER_WASA_SCAN_PULSER}.
\begin{table}[b!]
  \begin{center}
\begin{tabular}{|c|c|c|c|c|c|c|c|c|}
\hline 
\hline
\multicolumn{9}{|c|}{ \textbf{WASA System: Linear Scan} }\\
\hline
\hline
   \multicolumn{9}{|c|}{ \textbf{Output Pulser [mV]:} }\\
\hline 
    \small  3 &
    \small  5 &
    \small 10 &
    \small 15 &
    \small 20 &
    \small 23 &
    \small 25 &
    \small 26 &
    \small 28 \\
\hline 
   \multicolumn{9}{|c|}{ \textbf{Measured Peaking Amplitude [ADC]:} }\\
\hline 
        \small  219 
    &   \small  362 
    &   \small  719 
    &   \small 1079 
    &   \small 1436 
    &   \small 1650 
    &   \small 1793 
    &   \small 1865 
    &   \small 2007 \\
\hline
\end{tabular}\\

\vspace{0.5cm}
\caption{Scan performed with the WASA readout system
         using the new transimpedence amplifier and 
         {\bf with frequency filtering}. 
         The total error in the measured peaking amplitude is 
         conservatively considered as being unity.}
\label{tab:NEWAMPL_FILTER_WASA_SCAN_PULSER}
  \end{center}
\end{table}

\begin{figure}[t!]
\hspace{-1cm} 
      \includegraphics[height=5cm,width=7.cm]
               {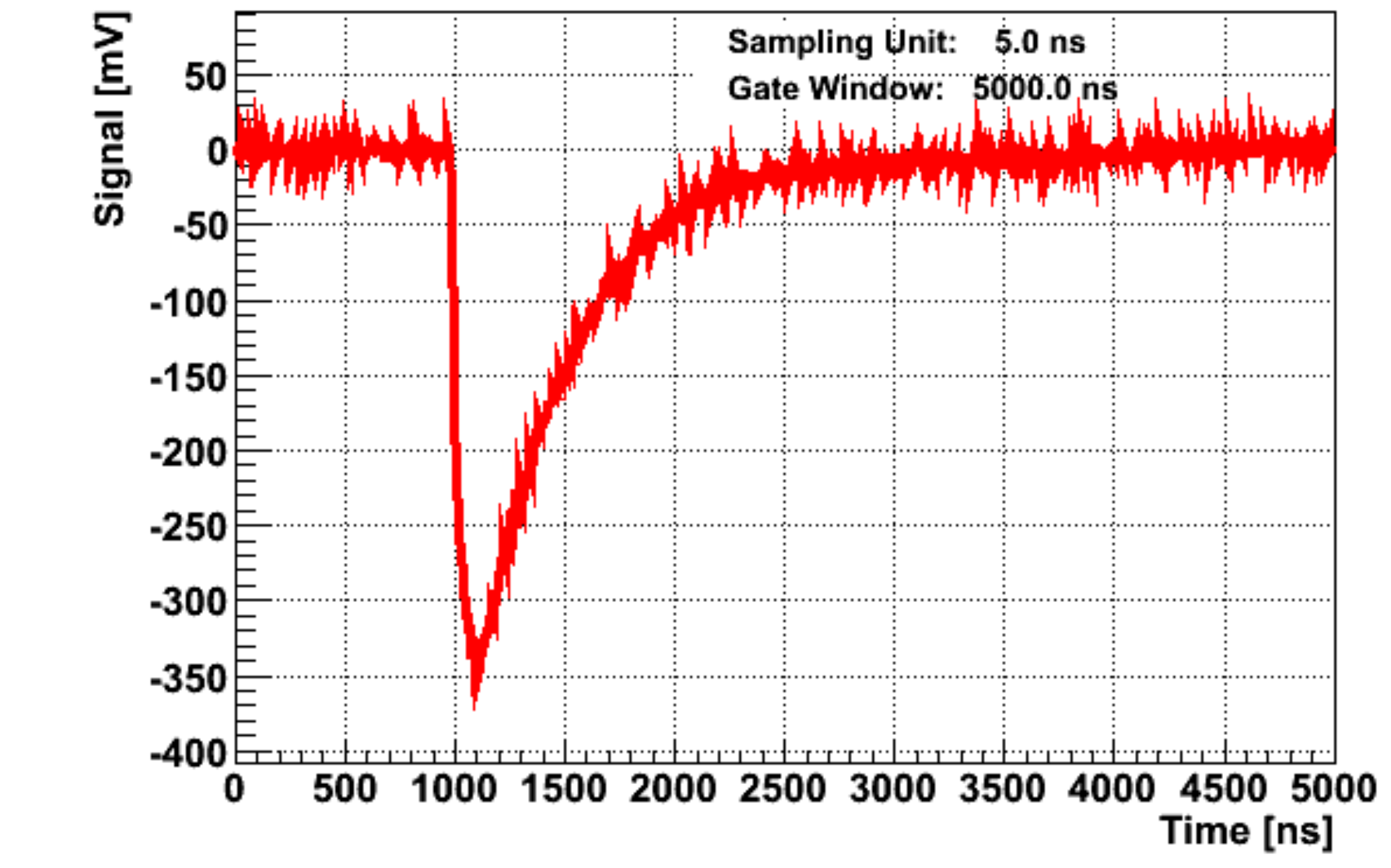}
      \includegraphics[height=5cm,width=7.cm]
               {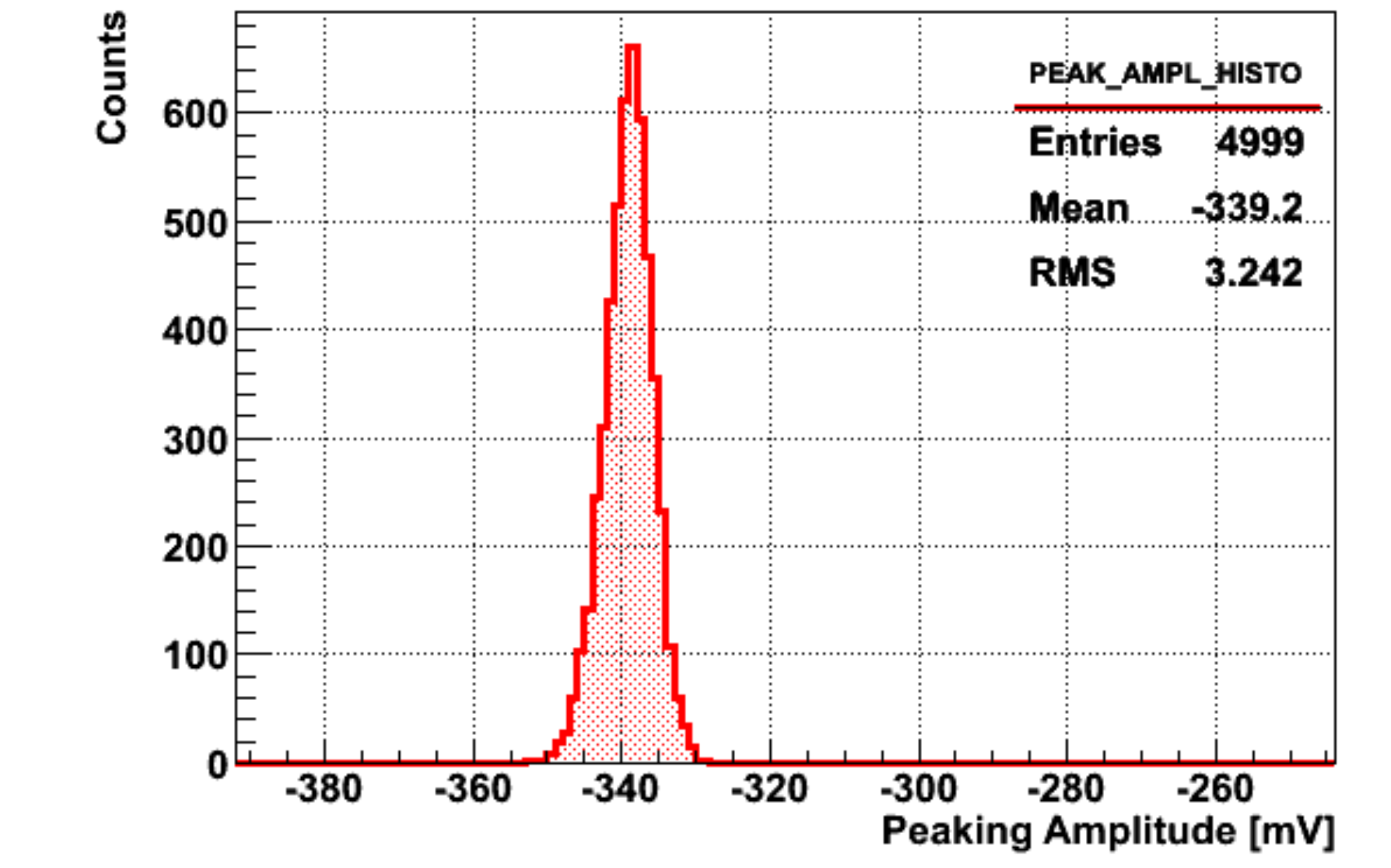}\\

\hspace{-1cm} 
      \includegraphics[height=5cm,width=7.cm]
               {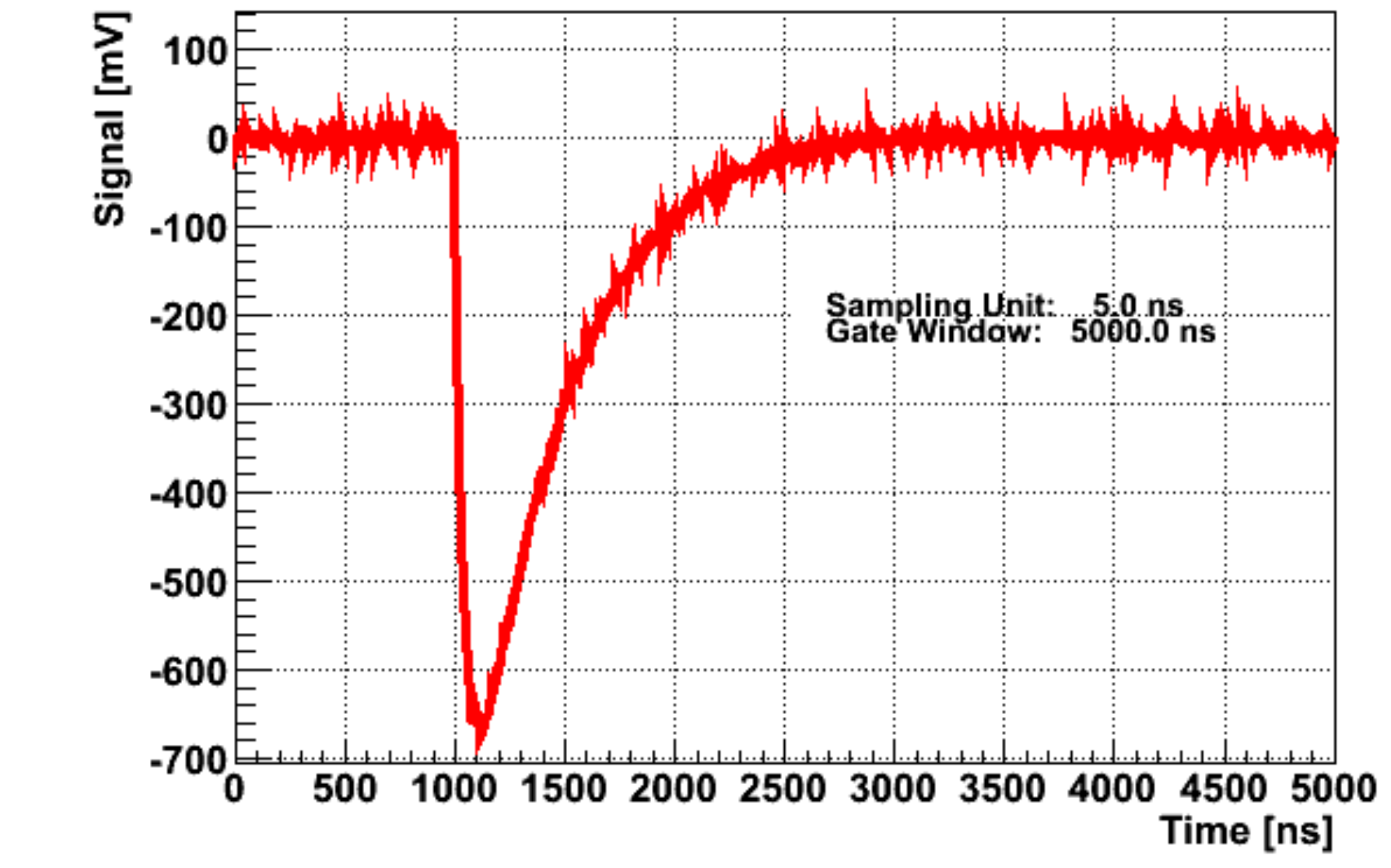}
      \includegraphics[height=5cm,width=7.cm]
               {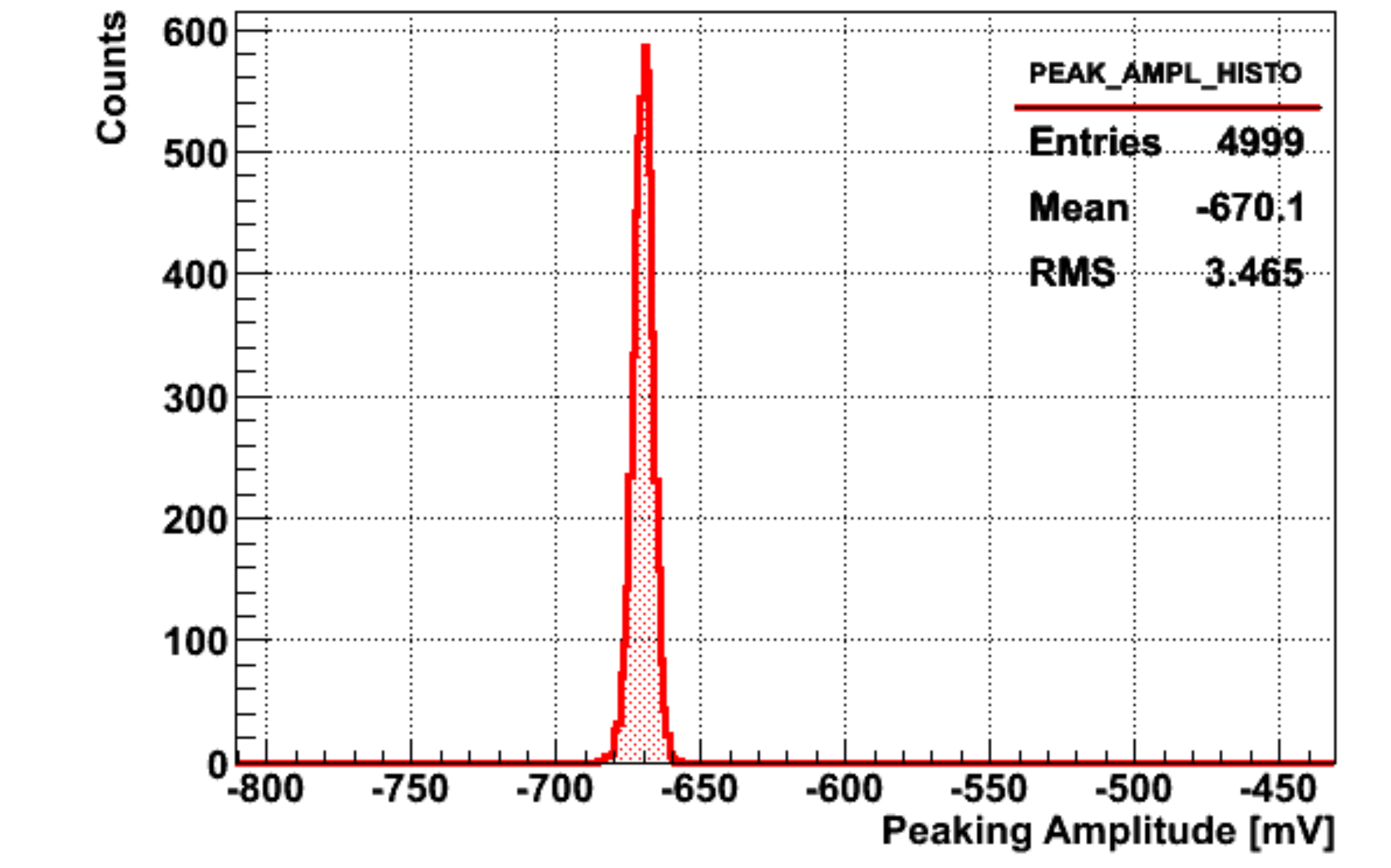}\\

\hspace{-1cm} 
      \includegraphics[height=5cm,width=7.cm]
               {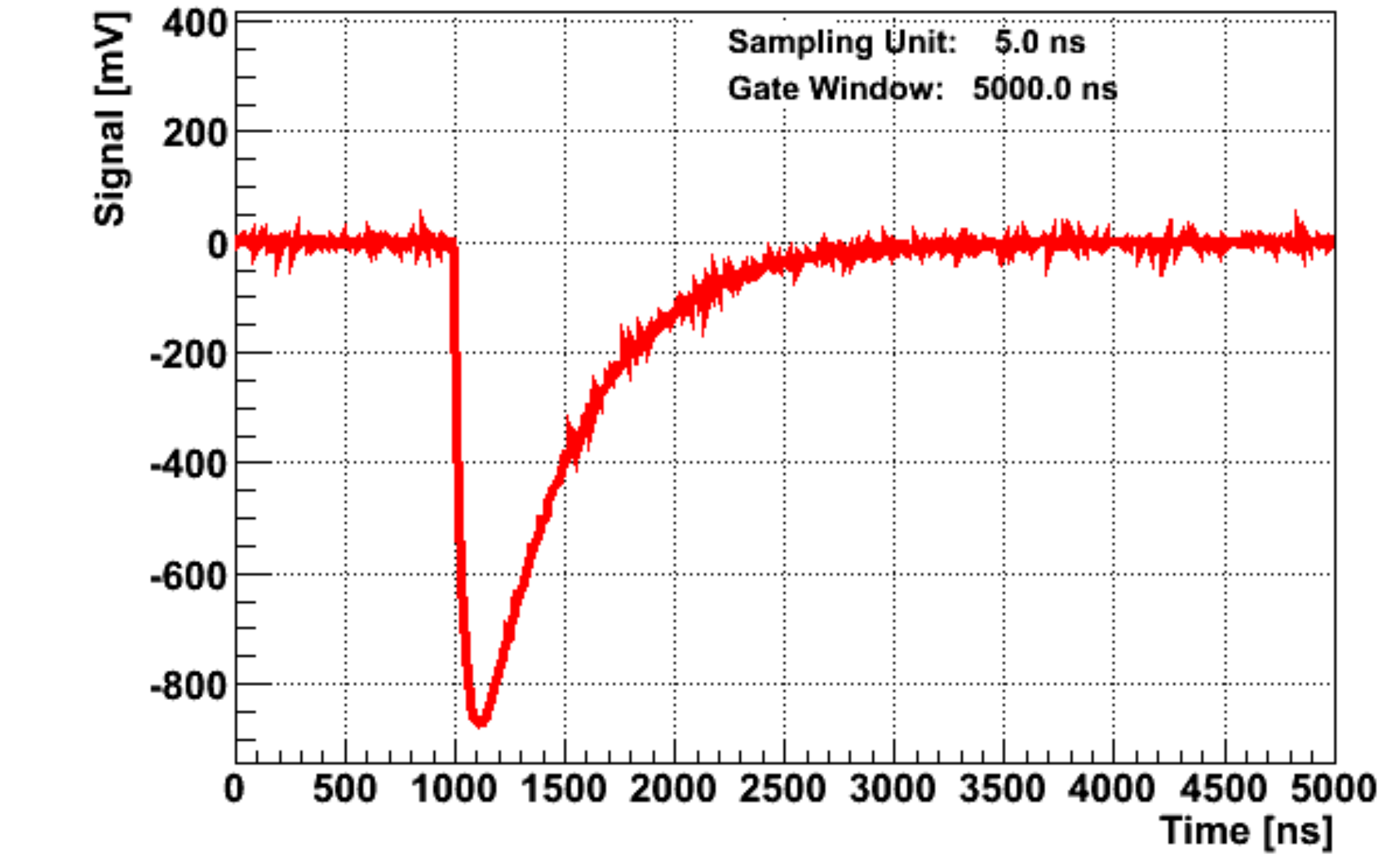}
      \includegraphics[height=5cm,width=7.cm]
               {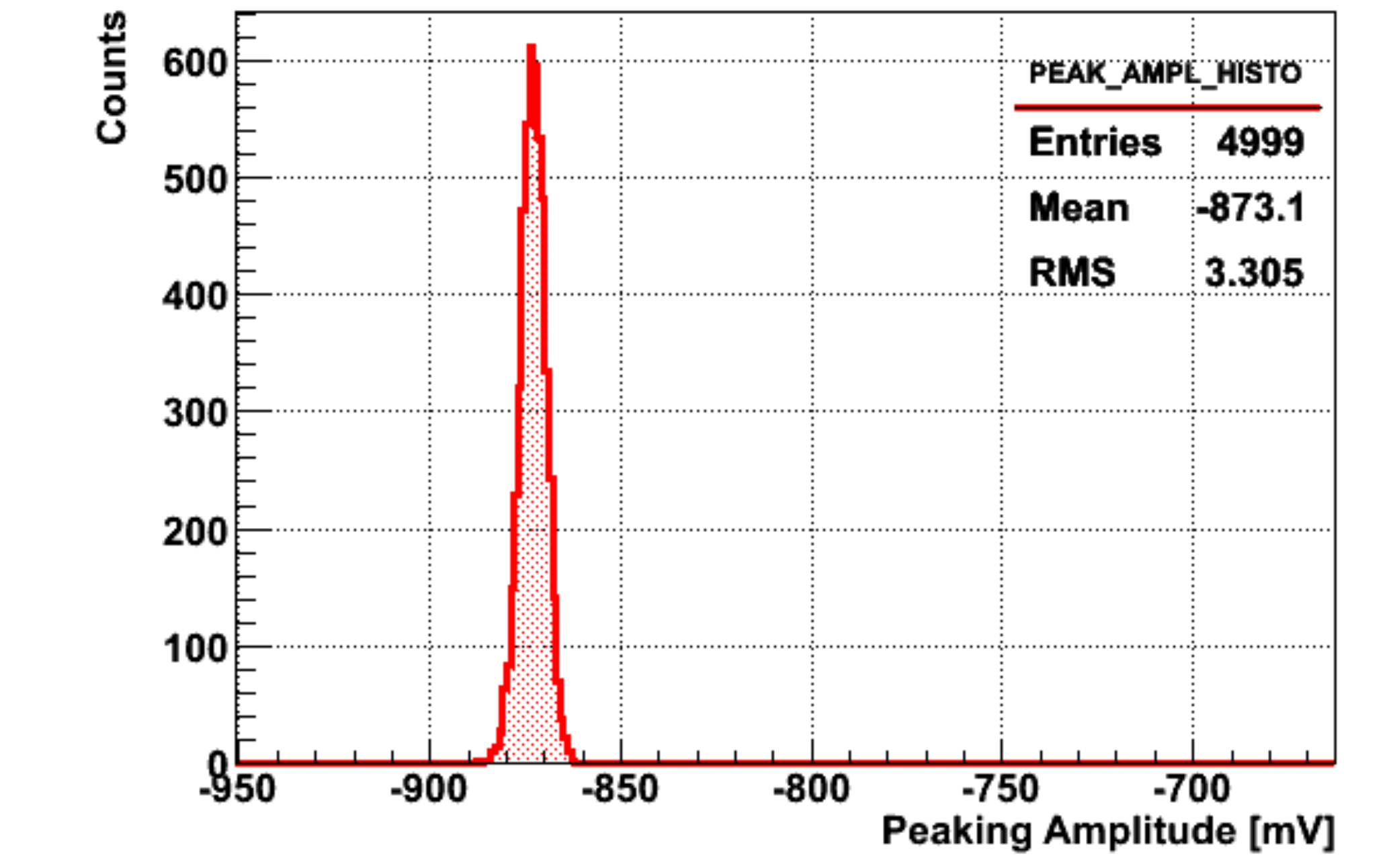}\\

  \caption{Waveform analysis with the ACQIRIS electronics with the new 
           transimpedence amplifier and {\bf with frequency filtering}. 
           Measurements performed 
           with $10$, $20$ and $26$ mV (upper, middle and bottom panels, 
           respectively) voltage amplitude provided by the pulse generator.}
  \label{fig:NEWAMPL_FILTER_ACQIRIS_PULSER_EXAMPLE}
\end{figure}

With the same experimental setup some measurements were
performed using also the ACQIRIS readout electronics.
The values of $10$, $20$ and $26$ mV for the amplitude of 
the input voltage pulse generator were used, and the 
corresponding relevant 
results are presented in Fig.~\ref{fig:NEWAMPL_FILTER_ACQIRIS_PULSER_EXAMPLE}. 

The three ACQIRIS measurements of the peaking amplitude 
show the gain values $33.9$, $33.5$ and $33.6$ 
for the used increasing values of the input voltage amplitude, with 
an absolute deviation of $0.4$ and of a relative deviation up to
$1\%$ (which could be considered as the systematical uncertainty 
of this measurement). 

\begin{figure}[t!]
  \hspace{-2.5cm}
  \includegraphics[height=6.5cm,width=8cm]
          {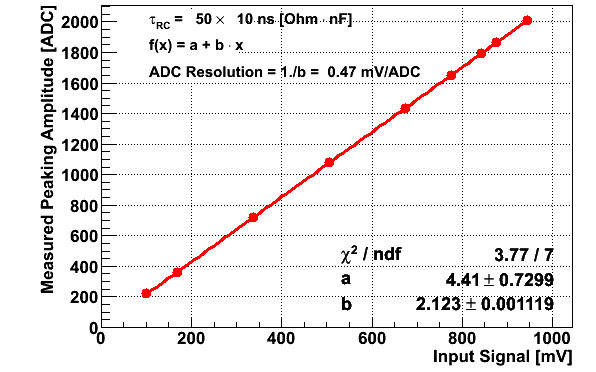}
  \hspace{-0.4cm}
  \includegraphics[height=6.5cm,width=8cm]
          {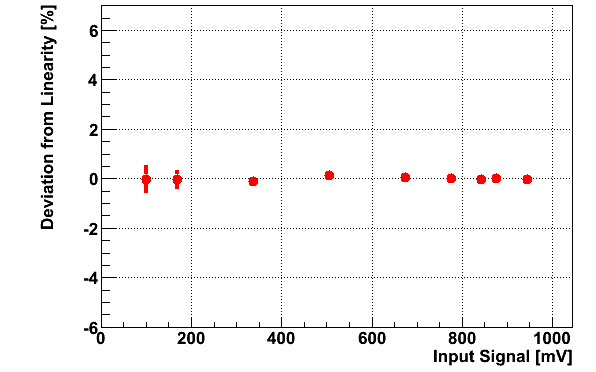}\\

  \caption{The ADC resolution is here extracted via a linear fit 
           to the data of the linear scan in the amplitude of the 
           input signal to the WASA board (left panel). 
           The deviation to linearity shows a system linear within 
           $1\%$ within all available ADC range (right panel).
           The measurement was performed with the new 
           transimpedence amplifier and {\bf with frequency filtering}.}
  \label{fig:NEWAMPL_FILTER_WASA_LINEARITY_PULSER}
  \vspace{0.7cm}
\end{figure}

Considering the mean gain value $33.7$ allows to evaluate the
peaking amplitude in mV of the input signal injected in the WASA 
readout electronics during the scan.
The WASA scan data are then presented, after the above mentioned rescaling
in Fig.~\ref{fig:NEWAMPL_FILTER_WASA_LINEARITY_PULSER}.

A 2-parameter linear function is fitted to the data, resulting in 
an ADC resolution of approximately $0.47$ mV per ADC unit.
From the fit results, the deviation to linearity is calculated in 
percent, and found to be well below $1\%$. It is presented in the 
right panel of the picture.

\subsubsection{Waveform Analysis: WASA vs \mbox{ACQIRIS} Comparison}
\label{sec:NEWAMPL_FILTER_WAVEFORM_PULSER}
The extracted ADC resolution will be used to 
calculate the total injected charge for each waveform in the WASA
readout system.

\begin{figure}[h!]
 \hspace{-2.cm} 
 \includegraphics[height=5.4cm,width=7.5cm]
         {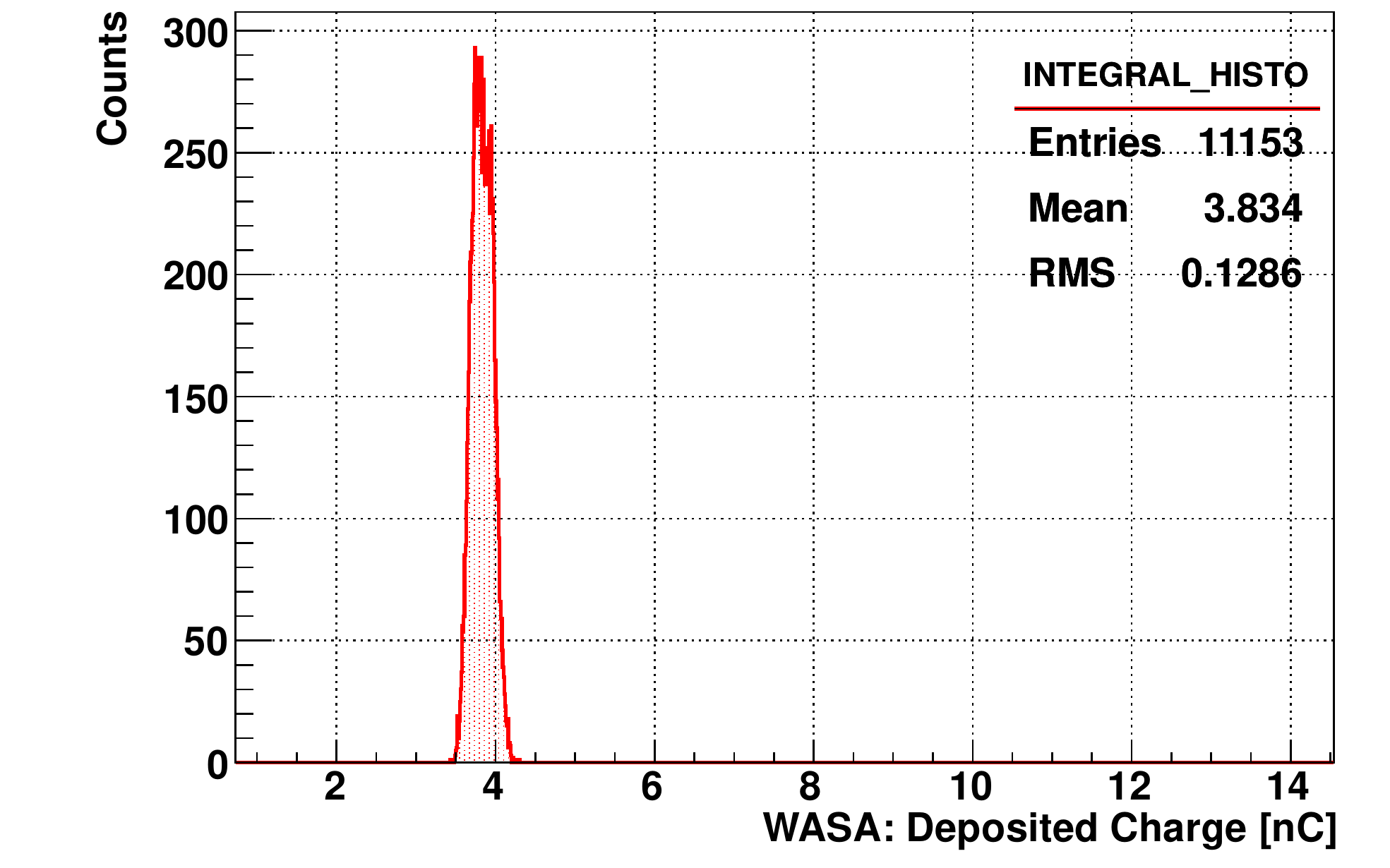}
 \includegraphics[height=5.4cm,width=7.5cm]
         {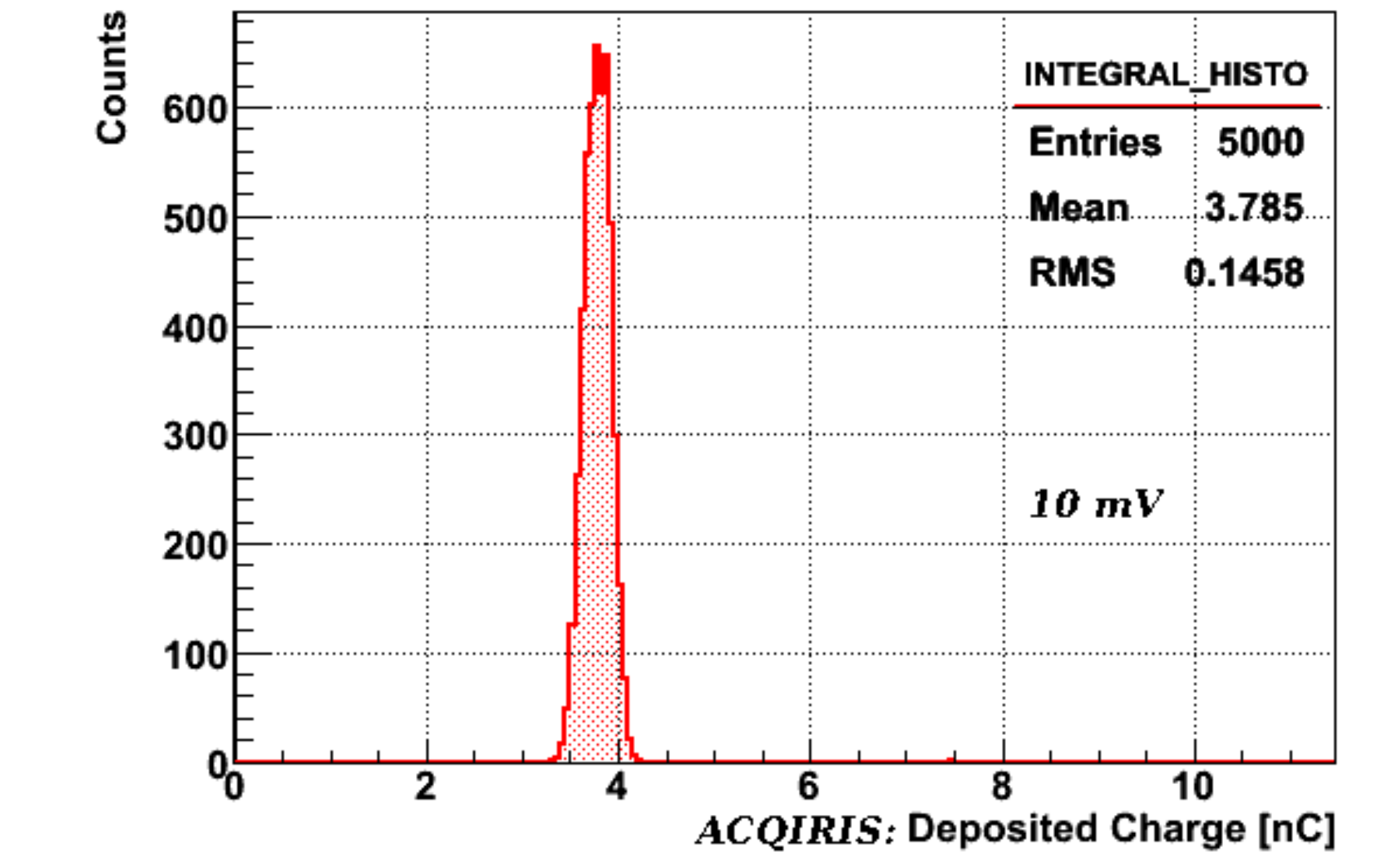}\\

  \caption{Integrated charge injected in the WASA and in the ACQIRIS 
           readout systems with the new transimpedance amplifier  
           {\bf with frequency filtering} 
           for the value $10$ mV of the voltage pulse amplitude.}
  \label{fig:NEWAMPL_FILTER_COMPARE_CHARGE_PULSER_10mV}
\end{figure}

\begin{figure}[h!]
 \hspace{-2.cm}
 \includegraphics[height=5.4cm,width=7.5cm]
         {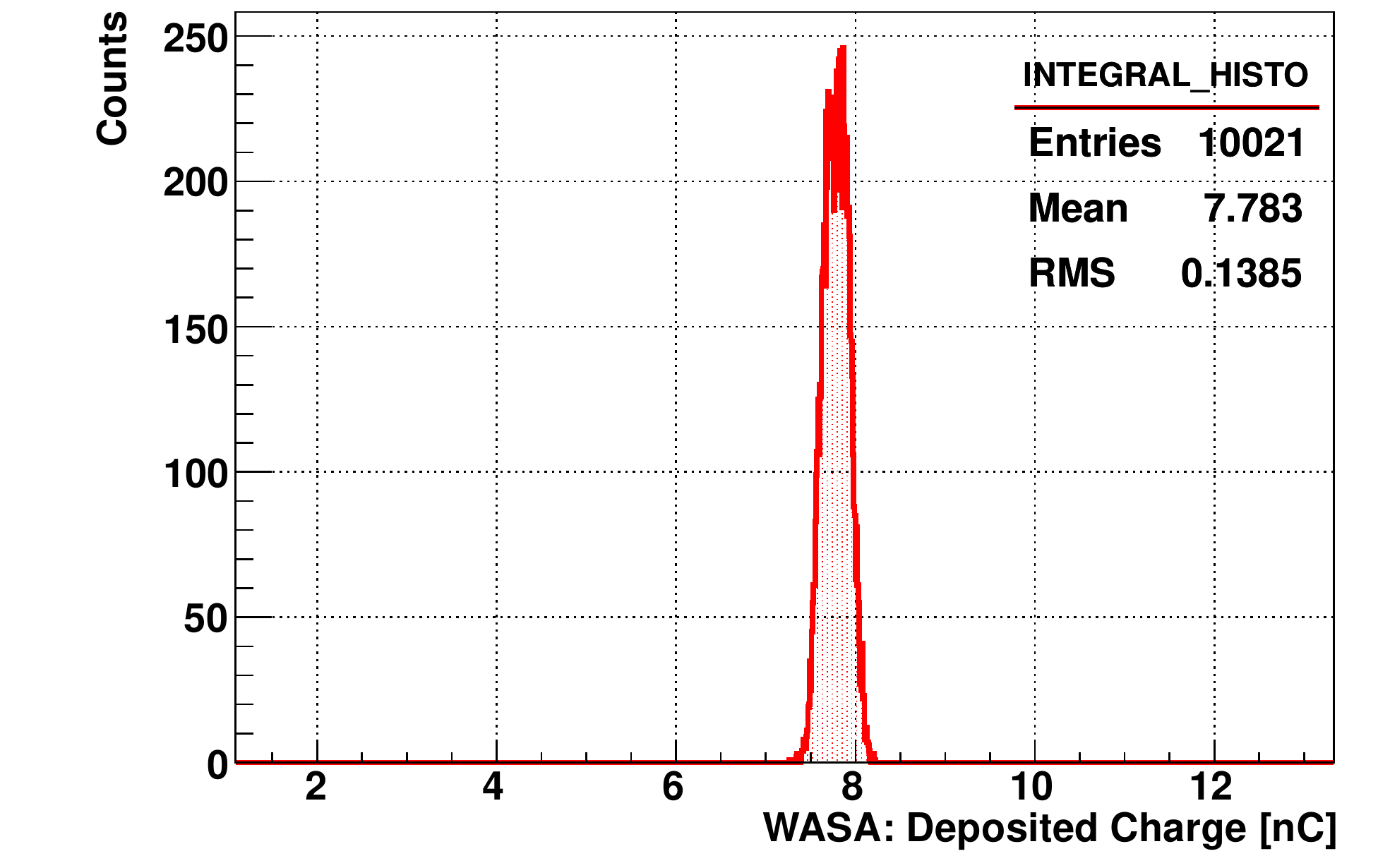}
 \includegraphics[height=5.4cm,width=7.5cm]
         {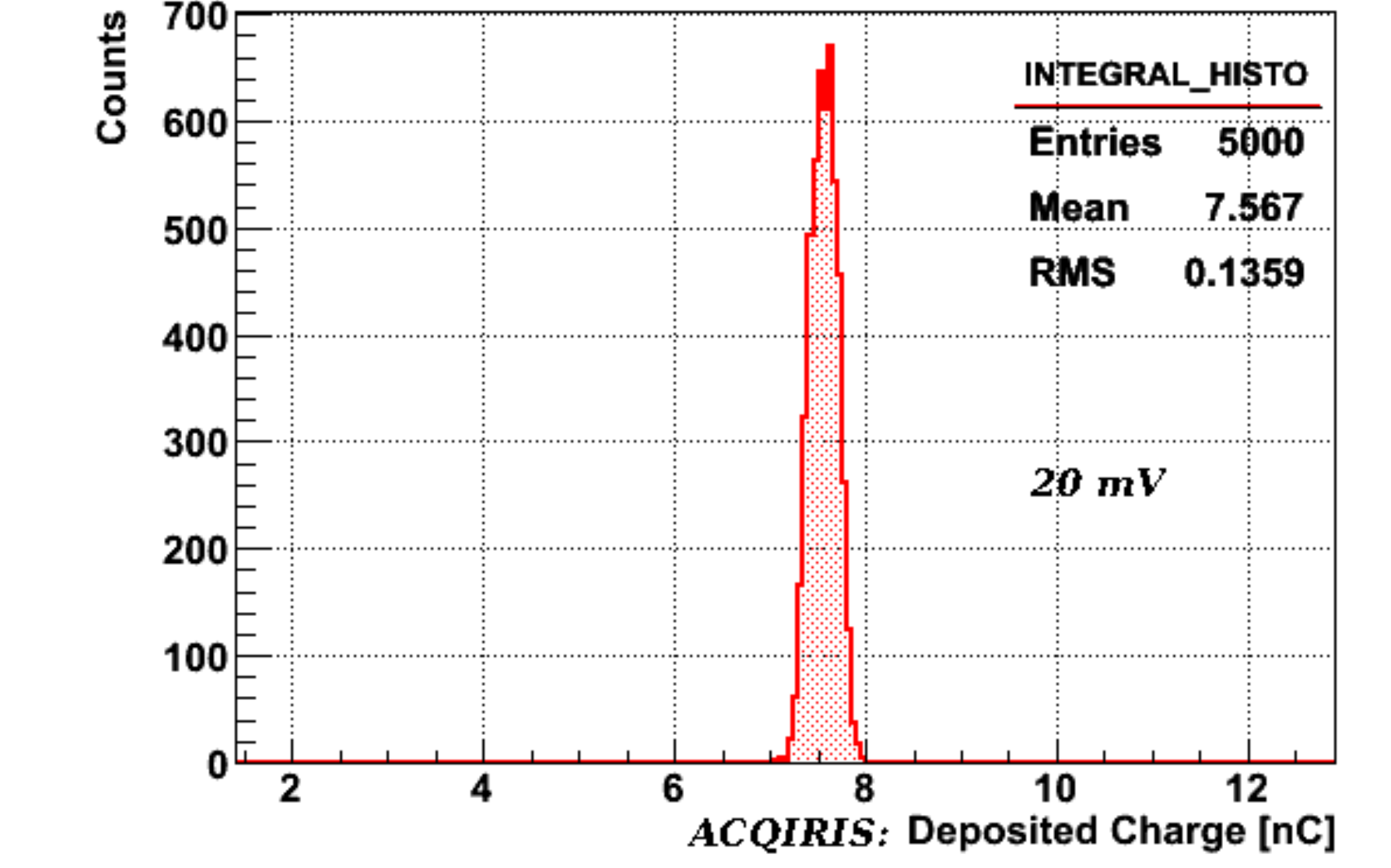}\\

  \vspace{-0.7cm}
  \caption{Integrated charge injected in the WASA and in the ACQIRIS 
           readout systems with the new transimpedance amplifier  
           {\bf with frequency filtering} 
           for the value $20$ mV of the voltage pulse amplitude.}
  \label{fig:NEWAMPL_FILTER_COMPARE_CHARGE_PULSER_20mV}
\end{figure}

\begin{figure}[h!]
 \hspace{-2.cm}
 \includegraphics[height=5.4cm,width=7.5cm]
         {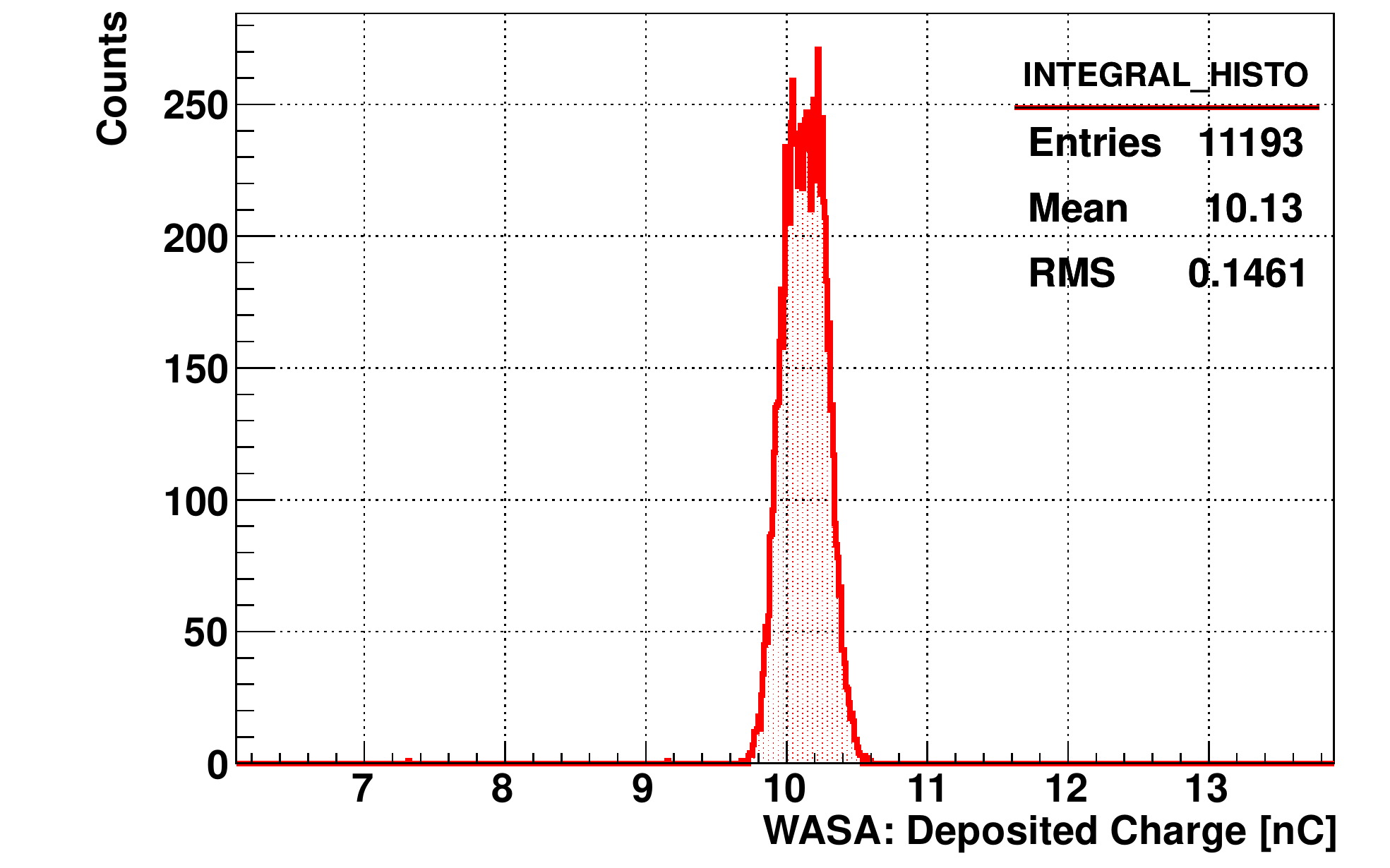}
 \includegraphics[height=5.4cm,width=7.5cm]
         {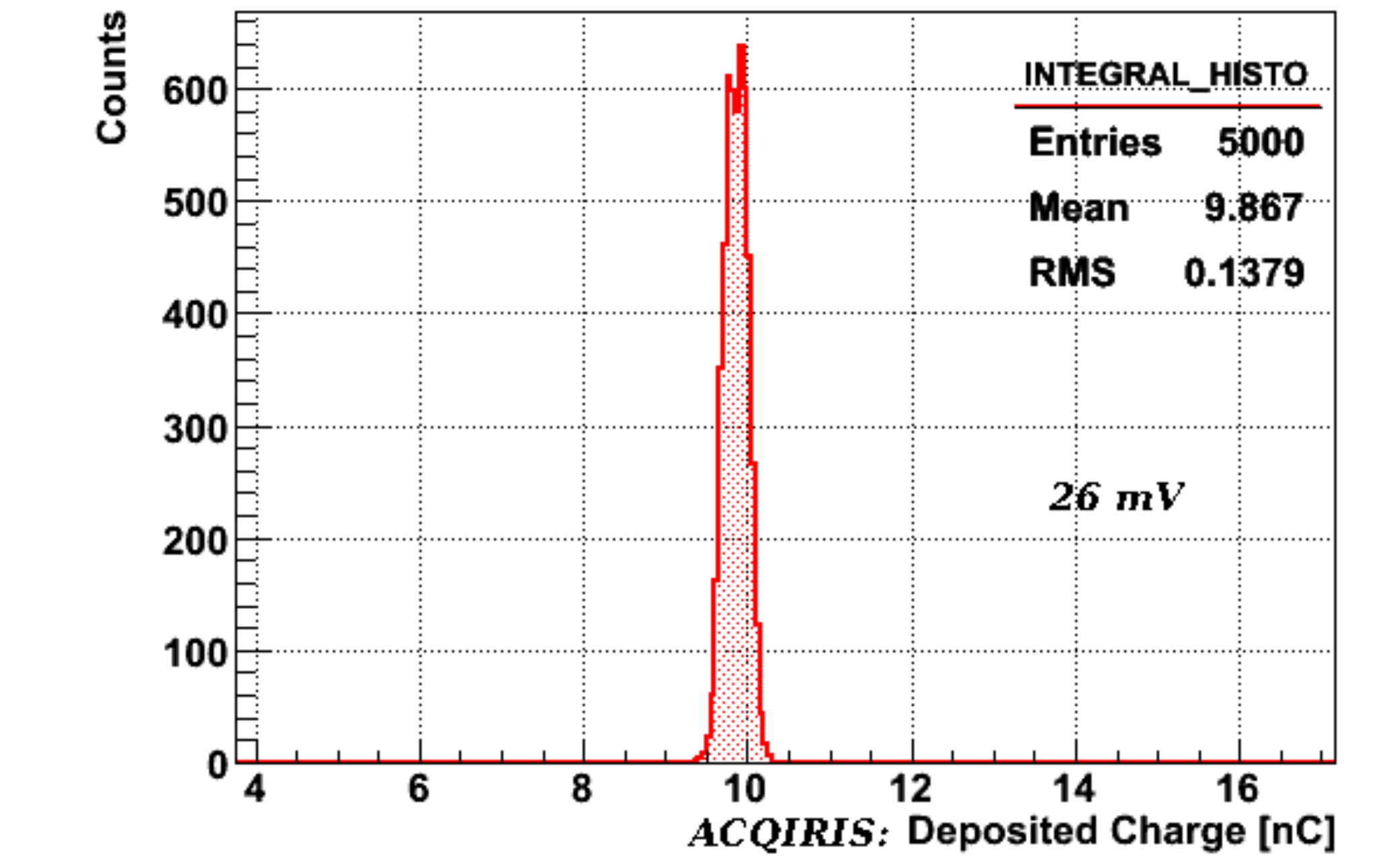}\\

  \vspace{-0.7cm}
  \caption{Integrated charge injected in the WASA and in the ACQIRIS 
           readout systems with the new transimpedance amplifier  
           {\bf with frequency filtering} 
           for the value $26$ mV of the voltage pulse amplitude.}
  \label{fig:NEWAMPL_FILTER_COMPARE_CHARGE_PULSER_26mV}
\end{figure}

The comparison of the integrated charge as measured by the two systems is 
presented in 
Fig.~\ref{fig:NEWAMPL_FILTER_COMPARE_CHARGE_PULSER_10mV}~-~\ref{fig:NEWAMPL_FILTER_COMPARE_CHARGE_PULSER_26mV}
for the 
three values of the voltage pulse amplitude used with the ACQIRIS 
electronics. 


For all three experimental setups a relative $1-3\%$ difference appears, 
better as in the previous setup.
Preliminary, for the moment this value can be considered as the systematical
uncertainty of the measurement.

\subsection{Measurements with WASA System:\\ Differential Signal}
%
The last test done with the new amplifier was performed using the complete 
differential signal at its output (as shown in Fig.~\ref{fig:TRANSIMPEDENCE}),
connecting both poles directly to the ADC in the QDC board. When located
close to the current source this signal transmission technique should 
help to limit deformation of the signal along its way to the ADC module.

\begin{figure}[t!]
\hspace{-1cm} 
      \includegraphics[height=5cm,width=7.cm]{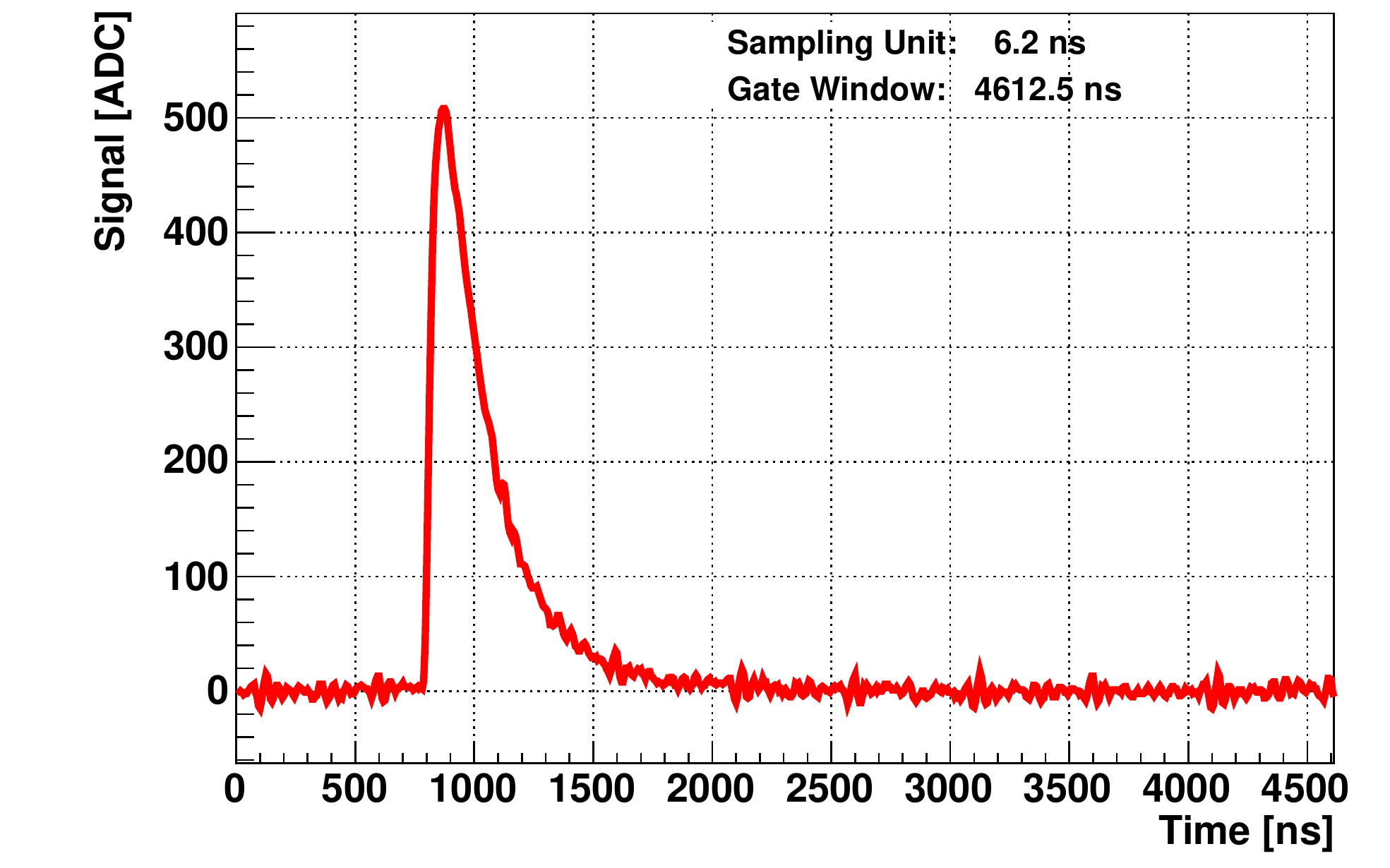}
      \includegraphics[height=5cm,width=7.cm]{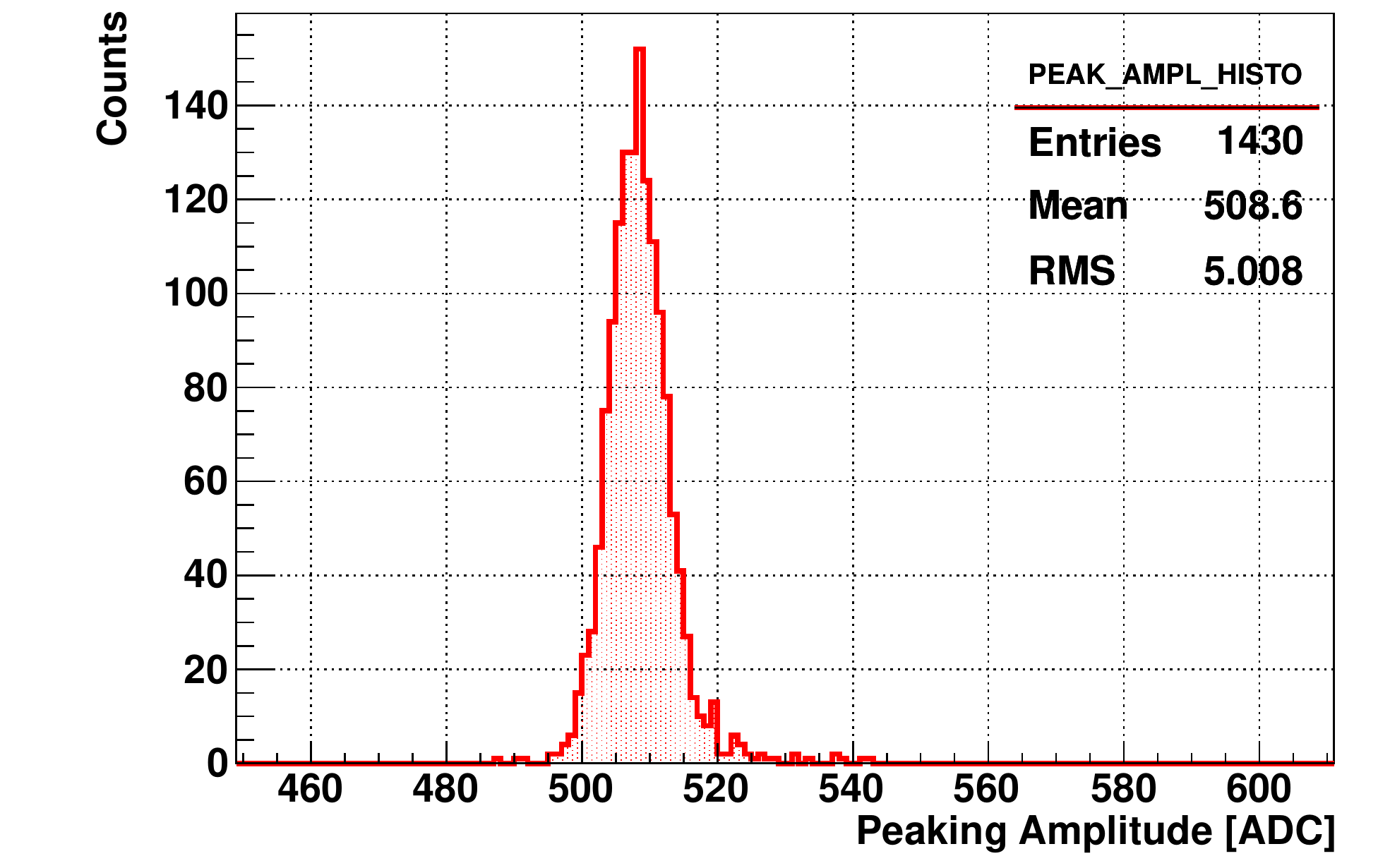}\\

\hspace{-1cm} 
      \includegraphics[height=5cm,width=7.cm]{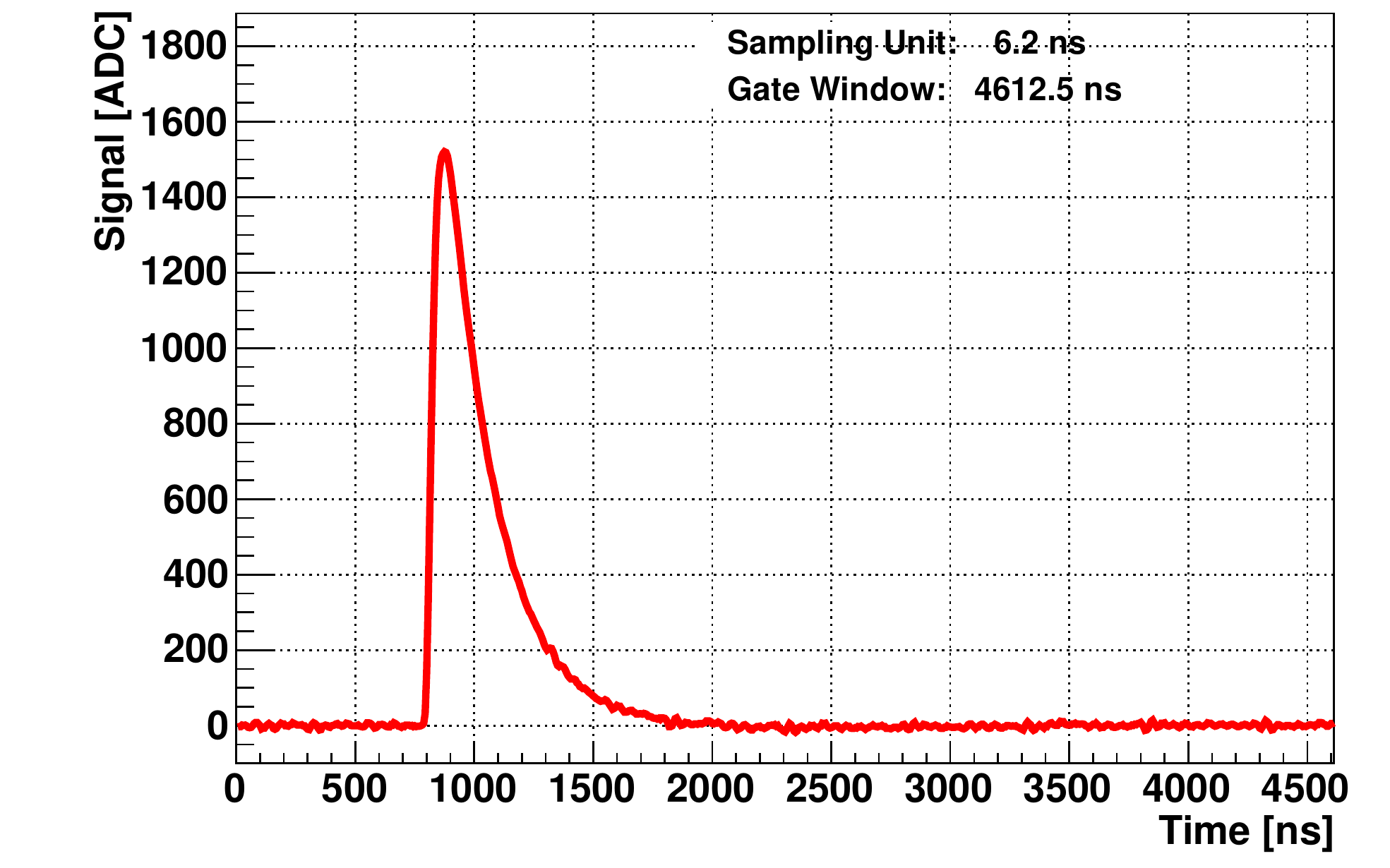}
      \includegraphics[height=5cm,width=7.cm]{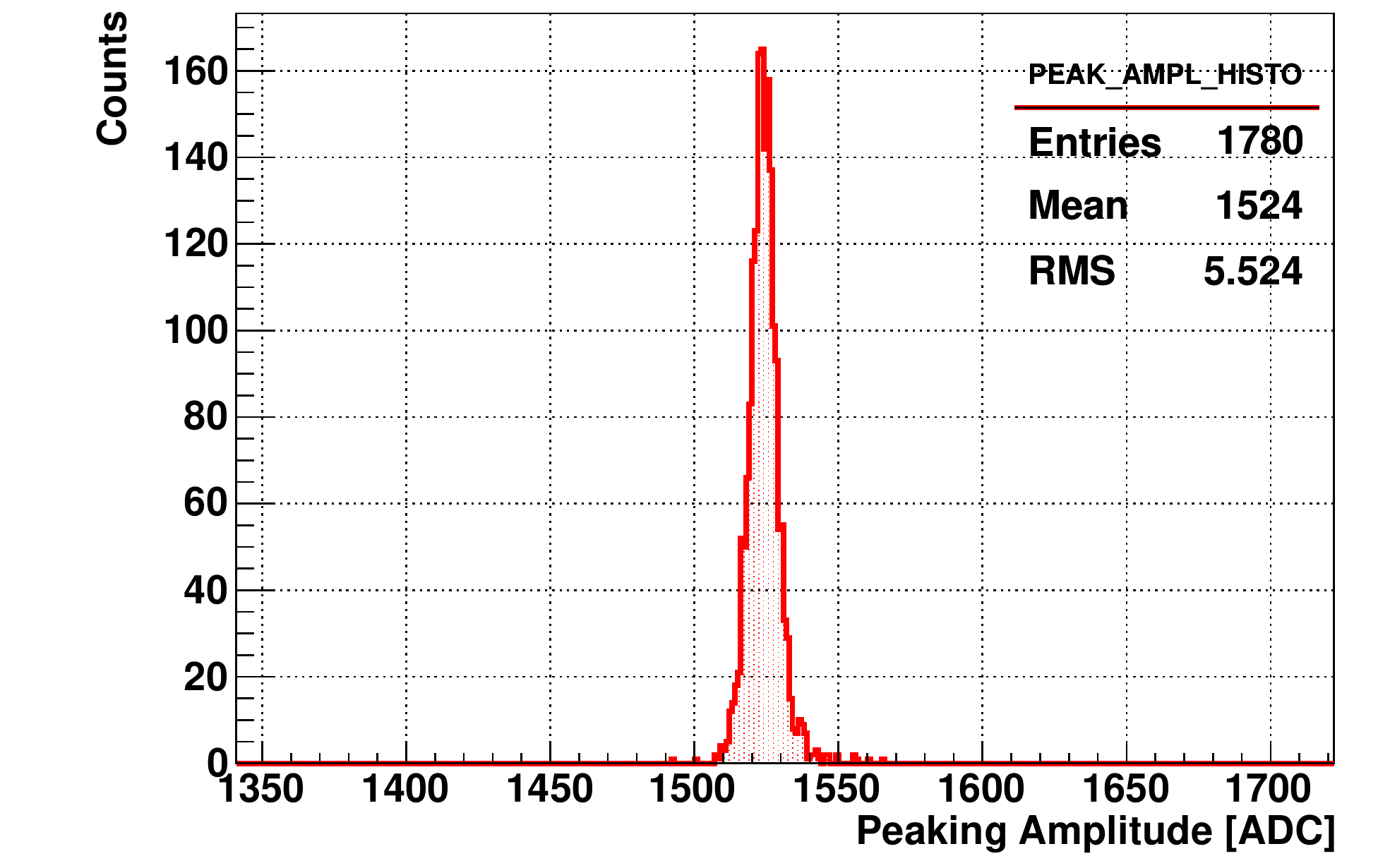}\\

  \vspace{-0.8cm}
  \caption{Waveform 
           analysis with the WASA electronics using the developed 
           transimpedence amplifier above described. 
           Measurements done {\bf with differential output}.
           These examples were obtained 
           with $5$ mV (upper panels) and \mbox{$15$ mV}
           (bottom panels) of voltage 
           amplitude provided by the pulse generator.}
  \label{fig:NEWAMPL_DIFF_WASA_PULSER_EXAMPLE}
\end{figure}

Examples of these measurements are presented in the upper and bottom 
panels of Fig.~\ref{fig:NEWAMPL_DIFF_WASA_PULSER_EXAMPLE} 
for the input voltage 
amplitude of $5$ and $15$ mV, respectively.
It is visible the higher gain of the amplifier with the 
differential signal output, which is expected to be 
approximately double than when coupling the negative 
output pin to the ground, as in the previous section 
with a single-ended output cable of the amplifier.
%

\subsubsection{Linearity of the Experimental Setup System}
%
Using a differential signal no direct comparison could be 
done with the ACQIRIS readout system, which has single-ended
input connectors.  
Nevertheless it is interesting to verify the linearity of the
experimental system, after the applied modifications.
To investigate the linearity of the system a scan is performed, 
injecting in the WASA system waveforms with different 
amplitude values of the input voltage 
pulse, as shown in the Tab.~\ref{tab:NEWAMPL_DIFF_WASA_SCAN_PULSER}.
\begin{table}[t!]
  \begin{center}
\begin{tabular}{||c|c|c||}
\hline 
\hline
\multicolumn{2}{|c|}{ \textbf{WASA System: Linear Scan} }\\
\hline
\hline
   \bf{Output Pulser [mV]:} & \bf{Measured Peaking Amplitude [ADC]:} \\
\hline 
    \small  3 &   \small  306 \\
    \small  5 &   \small  509 \\
    \small  7 &   \small  711 \\
    \small  9 &   \small  914 \\
    \small 11 &   \small 1117 \\
    \small 15 &   \small 1524 \\
    \small 17 &   \small 1726 \\
    \small 19 &   \small 1930 \\
\hline
\end{tabular}\\

\vspace{0.5cm}
\caption{Scan performed with the WASA readout system
         using the new transimpedence amplifier and 
         {\bf with differential output}. 
         The total error in the measured peaking amplitude is 
         conservatively considered as being unity.}
\label{tab:NEWAMPL_DIFF_WASA_SCAN_PULSER}
  \end{center}
\end{table}

When comparing the current results with what obtained with the
amplifier in single-ended
output mode (and the frequency filtering active in both case) 
shown in Tab.~\ref{tab:NEWAMPL_FILTER_WASA_SCAN_PULSER}, 
the signal appears to have the expected double amplitude.

Without any information from the ACQIRIS system, we cannot have a
precise measurement of the system gain (the input voltage from the 
pulse generator 
is in millivolt, while the measurement of the peaking amplitude
performed by WASA after the amplification is in terms of ADC). 
Nevertheless the linearity can be investigated comparing the measurements
by WASA with respect to the voltage amplitude provided by the input
pulser (the use of the gain factor would result infact 
simply in a rescaling of the reference input).  

\begin{figure}[t!]
  \hspace{-2cm}
  \includegraphics[height=6.cm,width=8cm]
     {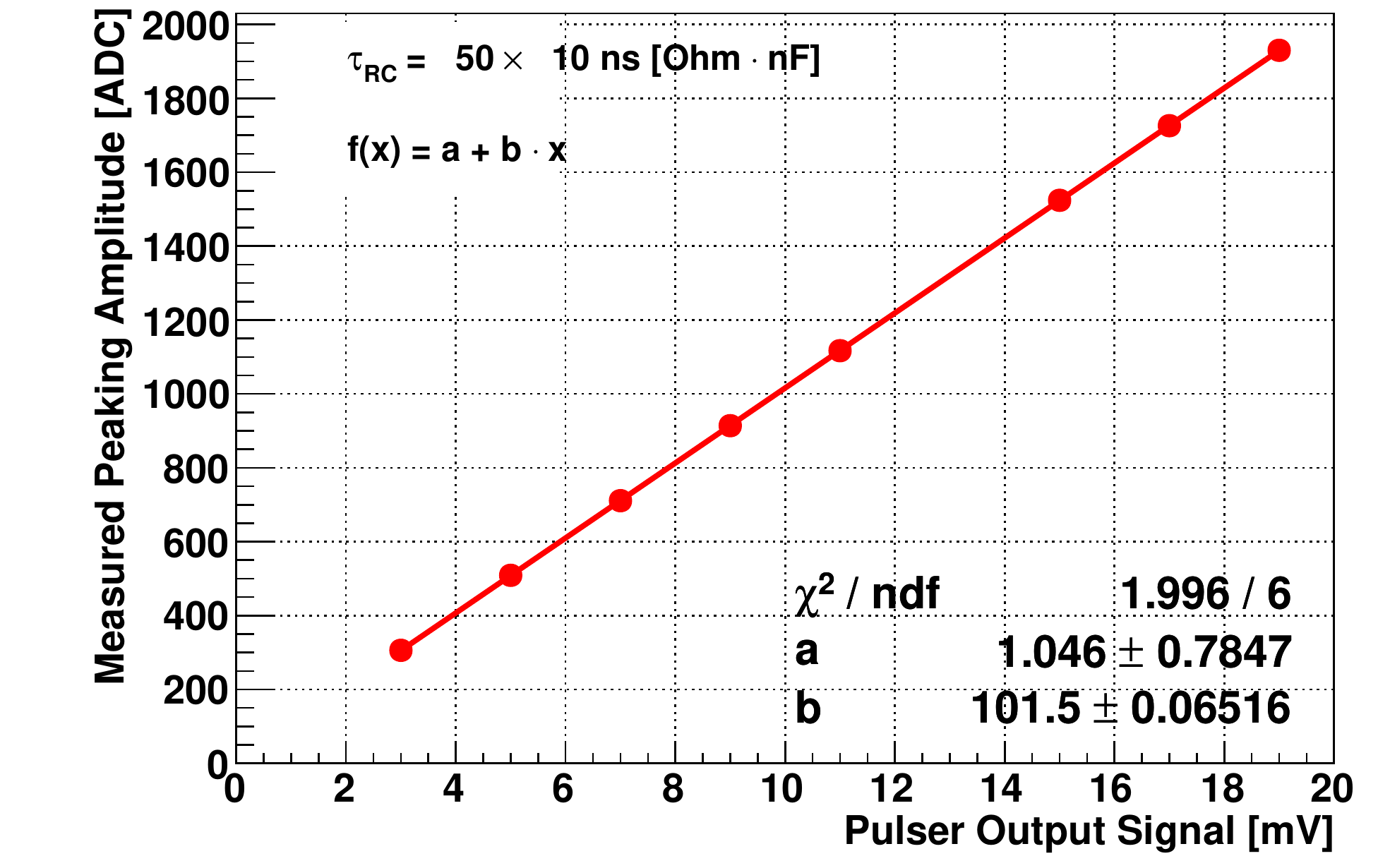}
  \includegraphics[height=6.cm,width=8cm]
    {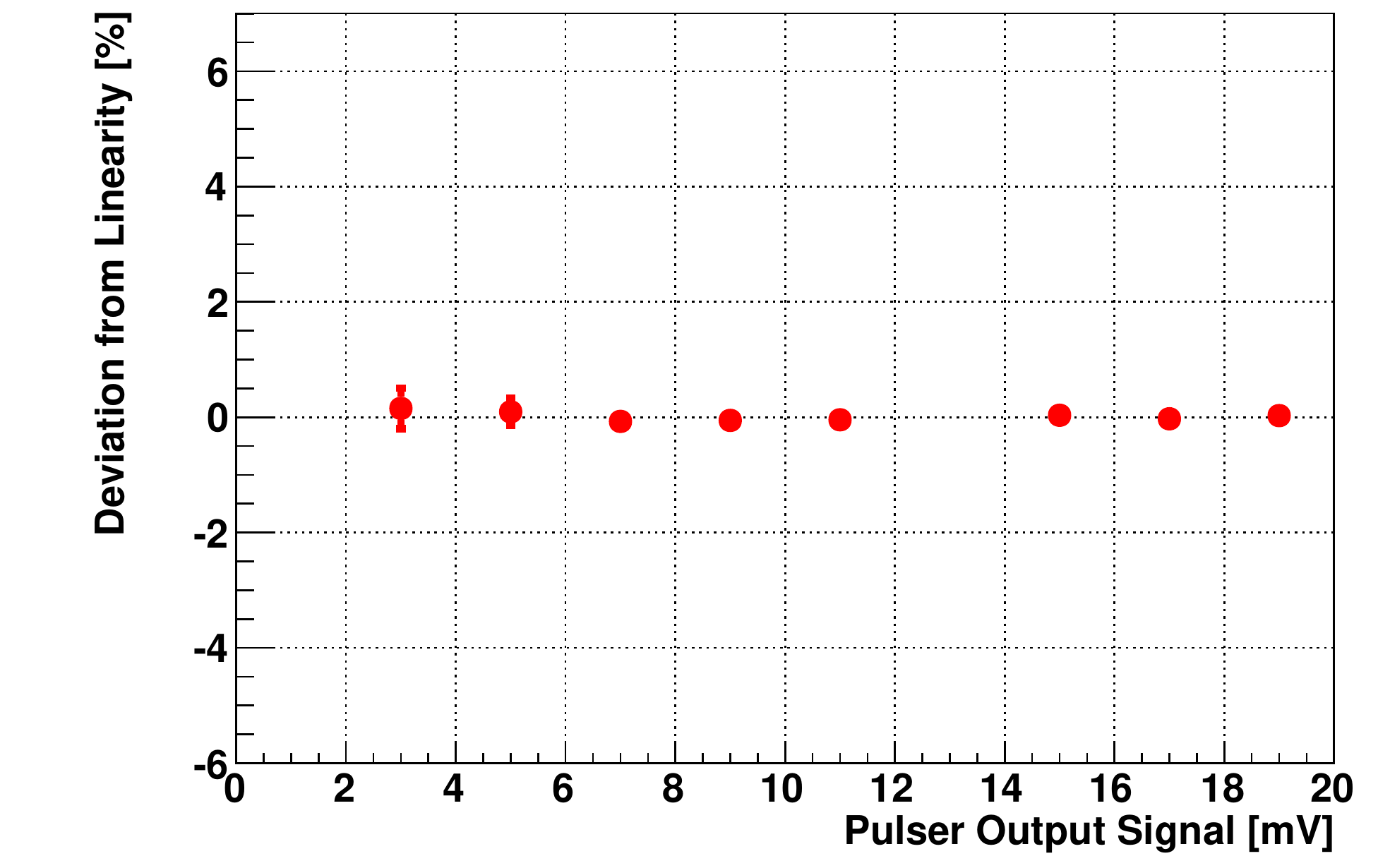}\\

  \vspace{-0.5cm}
  \caption{The linearity of the experimental system 
           is here extracted via a linear fit 
           to the data of the scan w.r.t. the amplitude of the 
           input signal sent by the pulser to the WASA board. 
           The deviation to linearity shows that the system is linear within 
           $1\%$ through all available ADC range (right panel).
           This measurement was performed with the new 
           transimpedence amplifier and {\bf with differential output}.}
  \label{fig:NEWAMPL_DIFF_WASA_LINEARITY_PULSER}
\end{figure}

The WASA scan data are then presented, without the above mentioned rescaling,
in Fig.~\ref{fig:NEWAMPL_DIFF_WASA_LINEARITY_PULSER}.
A two-parameter linear function is fitted to the data, and the 
deviation to linearity, calculated in percent, is presented in the 
left panel of the picture. The deviation is well below $1\%$.
 
\begin{figure}[t!]
  \hspace{-2cm}
  \includegraphics[height=6.cm,width=8cm]
    {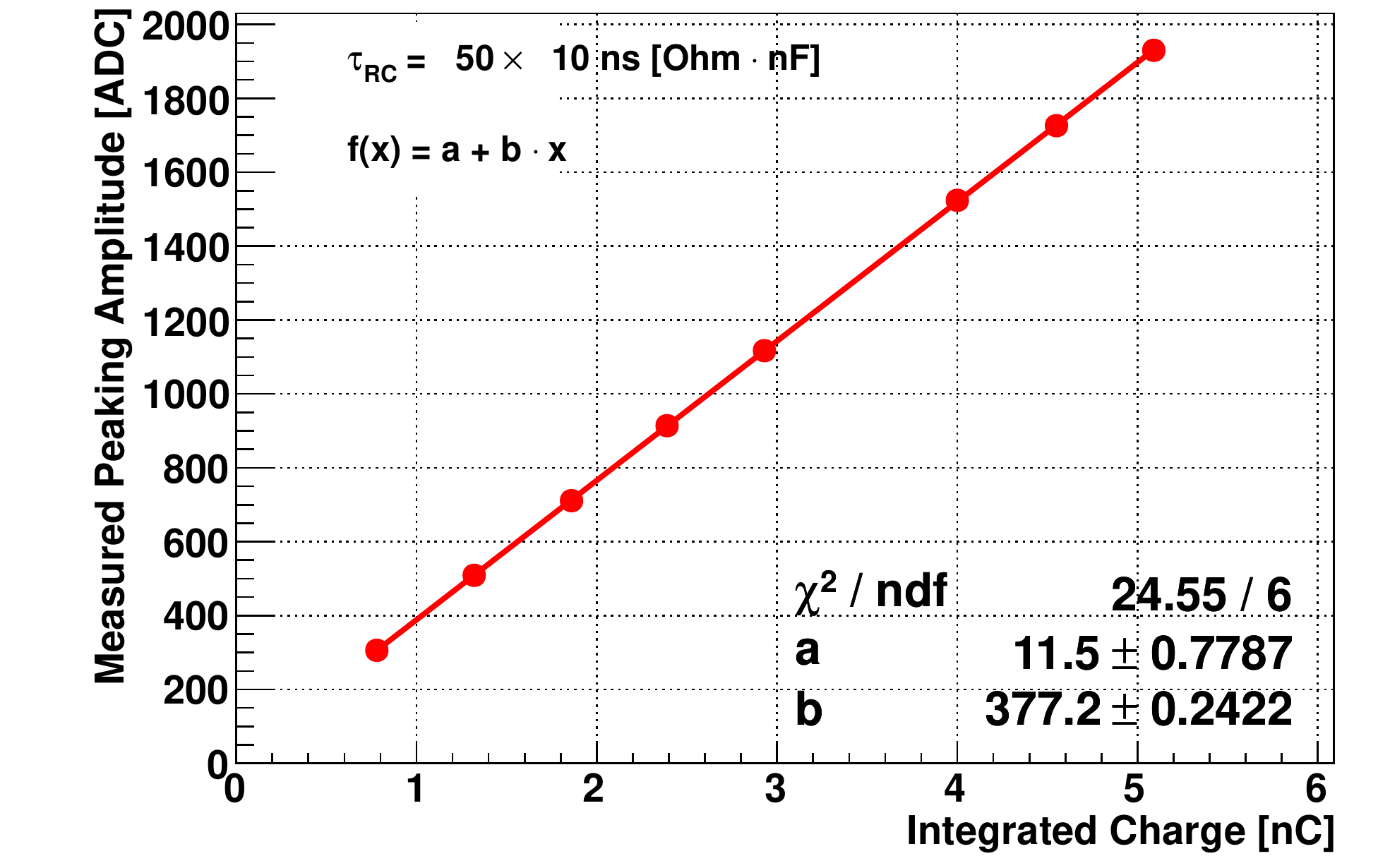}
  \includegraphics[height=6.cm,width=8cm]
    {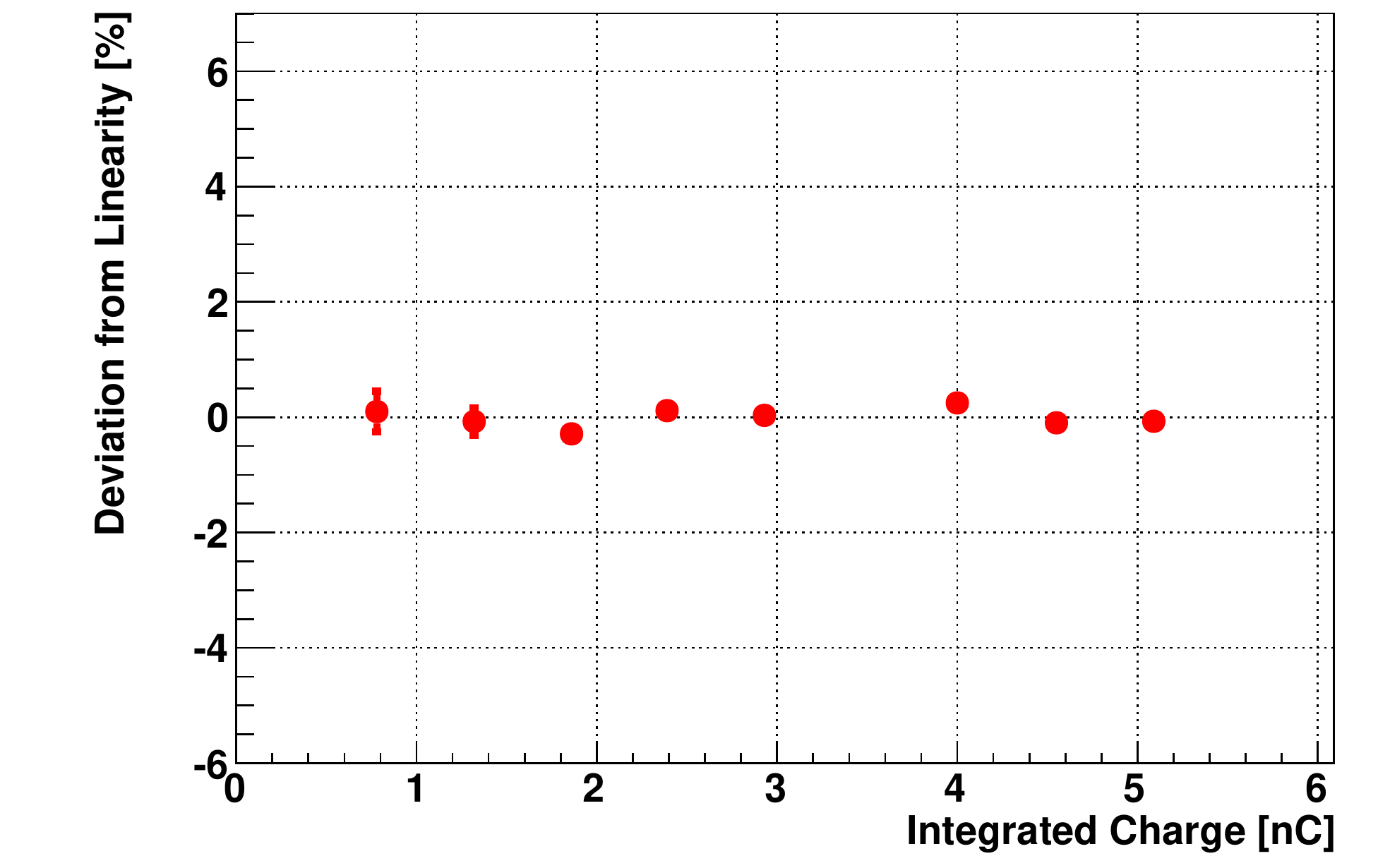}\\

  \vspace{-0.5cm}
  \caption{The peaking amplitude is presented with respect 
           to the corresponding integrated charge showing 
           a good linearity. 
           The deviation to linearity shows that the used analysis algorithms
           provides results linear within 
           $1\%$ through all available ADC range (right panel).
           This measurement was performed with the new 
           transimpedence amplifier and {\bf with differential output}.}
  \label{fig:NEWAMPL_DIFF_CHARGE_VS_PEAK}
\end{figure}

Using the mean mV/ADC conversion factor $0.46$ found in the
previous analyses the integrated injected charge can be calculated, 
and compared with the peaking amplitude (also converted in millivolts) 
of the waveform. The results are presented in 
Fig.~\ref{fig:NEWAMPL_DIFF_CHARGE_VS_PEAK} for all 
the values of the pulse amplitude used in the scan.

The system and the analysis algorithm provide results linear well 
within $1\%$. The slightly high value of the $\chi^2$ is possibly 
given by the underestimation of the uncertainty of the measurements.
Following the recommendations in the Particle Data Book~\cite{PDG}, 
the uncertainty could be enlarged with a common factor bringing 
the $\chi^2$ down to a reasonable value. The "extra" uncertainty 
should be considered as a contribution to the systematical uncertainty 
of the measurements.

\section{Conclusions}
%
During summer 2013 a preliminary waveform analysis was 
performed using two readout electronics systems, 
ACQIRIS DC282 and WASA, to investigate their precision 
in measuring the collected charge by the devices
using several types of amplifier configurations and 
different amount of injected charge. The goal of the 
analysis was to verify the possibility to use in 
future applications for neutron detection one of the two 
systems without experiencing unexpected biases in the 
measurement. It is clear that it would be desirable 
in the future to perform additional measurements, as 
on the rate efficiency, gain adjustment, pedestal 
uniformity and cross-talk. 

For each experimental configuration the results 
show a relative agreement between the two devices within 
few percents in the charge measurement. Also, some
discrepancy in the extracted ADC resolution of few 
percents was observed when using different experimental 
setups.
Being the results preliminary, no further investigation 
was done on the source of this systematical uncertainty, 
although the measurements suggest a dependence on 
the correct ADC calibration in the WASA; this feature should, 
in principle, be a minor issue, due to the possibility to
perform that calibration in a more controlled and stable
experimental setup.

Quite interesting is the use of a new transimpedance 
amplifier with differential output in order to minimize, 
when locating the device close to the current source,
the signal distorption when picking up noise from 
neighbouring instrumentation in an experimental hall 
like FRM-II. This device in the used experimental test-bench 
setup showed a linear response well below $1\%$.

As next evaluation, it is forseen to use the 
new transimpedence amplifier in single-ended mode 
in combination with the available 16 input channel 
ACQIRIS DAQ crate (which processes signals in 
single-ended mode) to run a new Anger Camera prototype
(already built at ZEA-2), based on scintillating plates, 
and made of 4x4 vacuum photomultipliers R268 by Hamamatsu. 
The measurements will be made initially with the neutron 
$Cf^{252}$ source availabe in our laboratory, and the
results will be published as soon as they will be available.   

The results presented in this work should ensure the 
possibility to use the WASA system in the waveform 
analysis, expecially in the case of need to build
and run the Anger Camera with a larger spatial coverage and input 
channels, being the WASA crate (by WIENER~\cite{WIENER}) 
capable to host up to $15$ readout modules (plus the
interface module to the external computer), resulting 
theoretically in $240$ readout channels.

%
\begin{quotation}
      \vspace{0.5cm}
  \begin{center}
      {\bf Acknowledgments}
      \vspace{0.25cm}
  \end{center}
 The authors acknowledge A.\ Erven, J.\ Heggen and 
 \mbox{C.\ Wesolek} for their technical 
 contribution to some results here presented.
 We are also grateful to the ZEA-2 Management for its 
 efforts in supporting this project. 
\end{quotation}



\end{document}